\documentclass{article}

\usepackage{PRIMEarxiv}

\usepackage[utf8]{inputenc} 
\usepackage[T1]{fontenc}    
\usepackage{hyperref}       
\usepackage{url}            
\usepackage{booktabs}       
\usepackage{amsfonts}       
\usepackage{nicefrac}       
\usepackage{lipsum}
\usepackage{fancyhdr}       
\usepackage{graphicx}       
\graphicspath{{media/}}     

\usepackage{tfrupee}
\usepackage[most]{tcolorbox}
\tcbset{
  Cbox/.style={
    colback=blue!5!white,
    colframe=blue!70!black,
    boxrule=0.8pt,
    arc=3pt,
    left=6pt,
    right=6pt,
    top=6pt,
    bottom=6pt
  }
}

\title{Carbon Farming: An Expository, Inter-Disciplinary Survey
}

\author{
    V. Priyanka, Geetha Charan, Rohit P. Suresh, Thandava Sunkara, Manojkumar Patil, Kartik Sagar \\
    Indian Institute of Science, Bangalore \\
    \texttt{\{priyankav, geethacharan, rohitpsuresh, thandavas, pmanojkumar,  kartikcsagar\}@iisc.ac.in} \\
    \And
    Aashman Trivedi \\
    University of Chicago, Chicago \\
    \texttt{atrivedi@cisb.org.in} \\
    \And
    K. Soumya, Subir Paul, Parashuram Hadimani, Ganesh Babu \\
    Boomitra, Bangalore \\
    \texttt{\{soumya, subir, parashuram, ganesh\}@boomitra.com} \\
    \And
    Ravi Trivedi \\
    Piramal Foundation, Bangalore \\
    \texttt{ravi.trivedi@alumni.duke.edu} \\
    \And
    Yadati Narahari \\
    Indian Institute of Science, Bangalore \\
    \texttt{narahari@iisc.ac.in} \\
}

\begin{document}
\maketitle

\begin{abstract}
Carbon farming is the collection of agricultural best practices specifically designed to maximize the capture and long-term storage of atmospheric carbon dioxide in soils and plant biomass, while simultaneously reducing greenhouse gas emissions from cultivation practices. Carbon farming can be viewed as a promising pathway to simultaneously address climate change mitigation, soil degradation, and farmer welfare. For example, if the entire agricultural cropland in India practices carbon farming, this will spectacularly offset about 50\% of emissions from  the country's annual transport-sector emissions. However,  practical deployment of carbon farming is constrained by scientific challenges, inherent complexity, and fragmented understanding across disciplines. As a result, in India, for example,  fewer than 1\% of farmers participate in carbon credit programs. This inter-disciplilinary, expository survey offers the first unified treatment of carbon farming for practitioners, policymakers, and researchers. The survey integrates insights from agronomy, soil science, climate science, measurement, reporting, and verification (MRV), economics, carbon markets, and policy design.  We begin by establishing the conceptual foundations of soil organic carbon dynamics and agricultural carbon sequestration, and compare carbon farming with the paradigms of sustainable, regenerative, and organic agriculture. We then present a comprehensive landscape analysis of carbon-farming best practices, including both generic and crop-specific interventions, and systematically examine their co-benefits and trade-offs. The paper offers a rigorous review of MRV frameworks, emerging digital MRV technologies, and the carbon-credit project life cycle, followed by a structured analysis of voluntary and compliance carbon markets. Drawing on six representative case studies, we synthesize implementation models, successes, and failure modes. Building on this integrated analysis, we highlight key scientific, economic, institutional, and adoption challenges, and propose potential remedies to make carbon farming a credible, scalable, and attractive proposition for global agriculture. We finally highlight the important role that artificial intelligence, game theory, and computation can play in improving various dimensions of carbon farming.
\end{abstract}


    \section{Introduction}\label{sec1}
Carbon farming has emerged as an important theme in the global agricultural landscape. It offers the promise of enhancing soil health, building climate resilience in farming systems, as well as generating new income opportunities by trading carbon credits through carbon markets. To provide an example of the significant benefits of carbon farming, India has approximately 180 million hectares of agricultural land \cite{PIB_CroppedArea}, which, at a sequestration rate of 1 metric ton of carbon dioxide equivalent (CO$_2$e) per hectare annually, could sequester about 180 million tons of CO$_2$e each year. This potential sequestration can offset nearly 50\% of the country's annual transport-sector emissions (368.2 MtCO$_2$e) \cite{jain2023analysing}. 

The carbon farming system is a complex conglomeration that lies at the intersection of agronomy, soil science, climate science, MRV (Monitoring, Reporting, Verification), behavioural economics, computation, and carbon markets. The objective of this survey is to provide an informed understanding of the various building blocks of carbon farming and examine the promises and challenges associated with it. 

This section introduces carbon farming in its contemporary context and motivates the need for a unified treatment. It first outlines the climate and agricultural drivers that have brought carbon farming to the forefront. The section then describes carbon farming relative to sustainable, regenerative, and organic farming paradigms, clarifying conceptual overlaps and distinctions. Finally, the section positions the present survey within the existing literature, identifies gaps in prior work, and outlines the scope and organization of the paper.

\subsection{Background}
\begin{figure}[ht]
    \centering    
    \includegraphics[width=\linewidth]{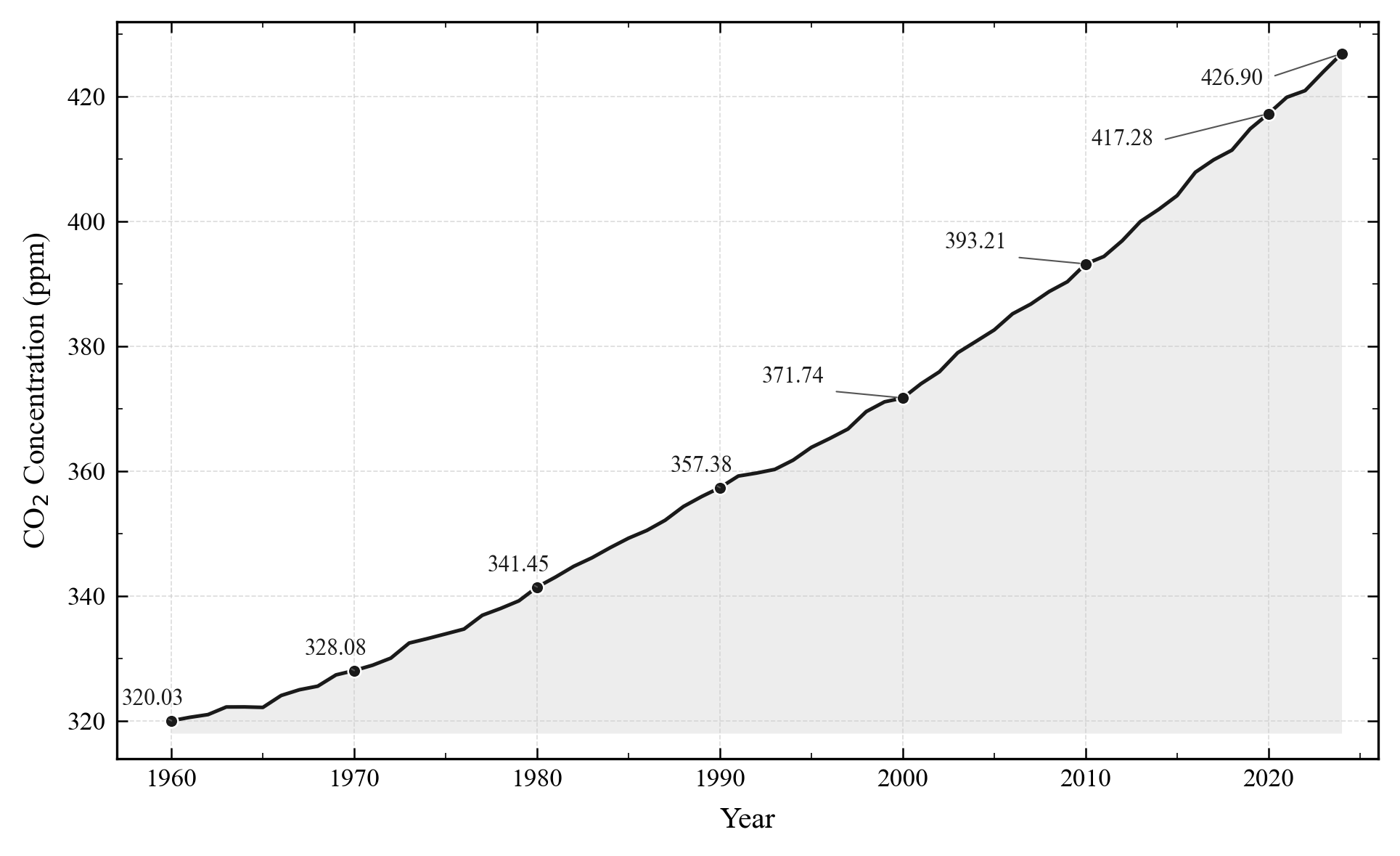}
    \caption{Atmospheric carbon dioxide levels from 1960 to 2024, recorded at the end of May each year \cite{NOAA24a}}
    \label{fig:co2_levels}
\end{figure}

\begin{figure}[ht]
    \centering
    \includegraphics[width=\linewidth]{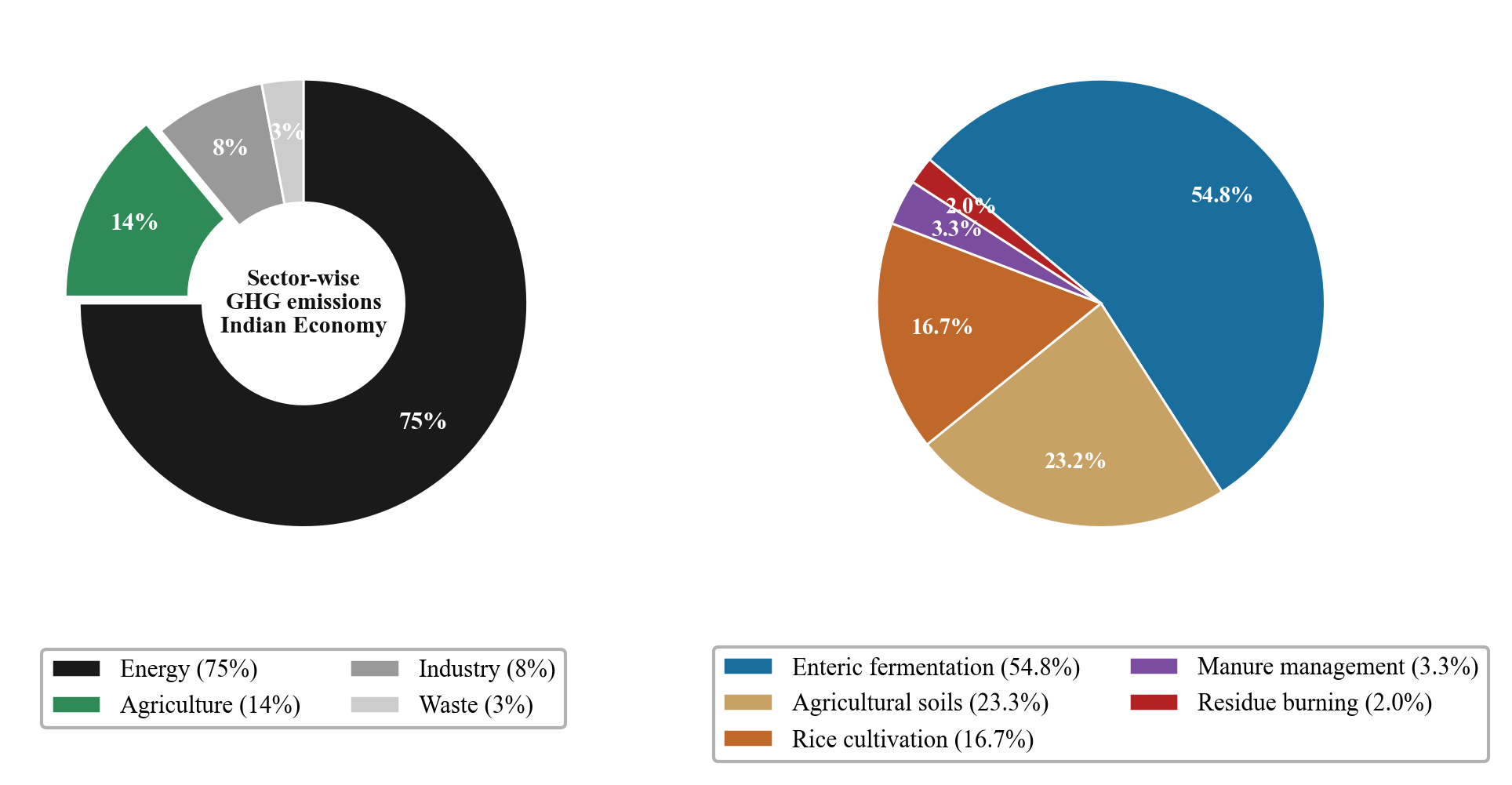}
    \caption{Breakdown of industry-wise emissions and agriculture sector emissions
    }
    \label{fig:emissions_breakdown}
\end{figure}

Global warming represents one of humanity's most significant challenges in the current century, primarily driven by unprecedented increases in atmospheric {greenhouse gas} (GHG) concentrations. Since 1959, atmospheric carbon dioxide concentration has risen sharply from 315.54 parts per million (ppm) in 1959 to 426.90 ppm in 2024, reflecting a 34\% increase \cite{NOAA24a, NOAA24b} (See Fig.~\ref{fig:co2_levels}). This marked elevation in GHG levels has contributed to a global temperature rise of approximately 1.1$^{\circ}$C above pre-industrial values \cite{EARTH24}, precipitating an environmental and socioeconomic crisis. Notably, emissions from agriculture account for 13.44\% of total GHG emissions. Within the agriculture sector and its related activities, enteric fermentation is the largest source, contributing 54.84\% of emissions, followed by agricultural soils (23.26\%), rice cultivation (16.68\%), manure management (3.33\%), and the burning of agricultural residues in fields (2.00\%) \cite{NITIAAYOG20}, which is depicted in Fig.~\ref{fig:emissions_breakdown}.


In the backdrop of escalating concerns about global warming, several best practices in agricultural farming have gained prominence. According to \cite{CACHO13}, effective land management and forest conservation represent practical and cost-effective measures to mitigate climate change. Three primary approaches to support climate mitigation have been identified in this context: (1) generate {carbon offsets} through the sequestration of {soil organic carbon} (SOC) in soils and plant biomass; (2) reduce emissions of methane and other greenhouse gases; and (3) replace fossil fuels by producing green energy.

\subsection{Paradigms of Agriculture}  
In the context of emerging best practices, several paradigms of agriculture, such as sustainable agriculture, regenerative agriculture, organic farming, and carbon farming, are widely discussed. 
These four agricultural paradigms represent overlapping rather than mutually exclusive approaches, sharing common best practices such as reduced tillage, crop diversification, and soil health preservation. The boundaries between them are fluid, with even experts disagreeing about precise definitions. What distinguishes them is primarily their framing, objectives, and measurement emphasis rather than any fundamentally different practices.

\begin{figure}[ht]
    \centering
    \includegraphics[width=0.9\linewidth]{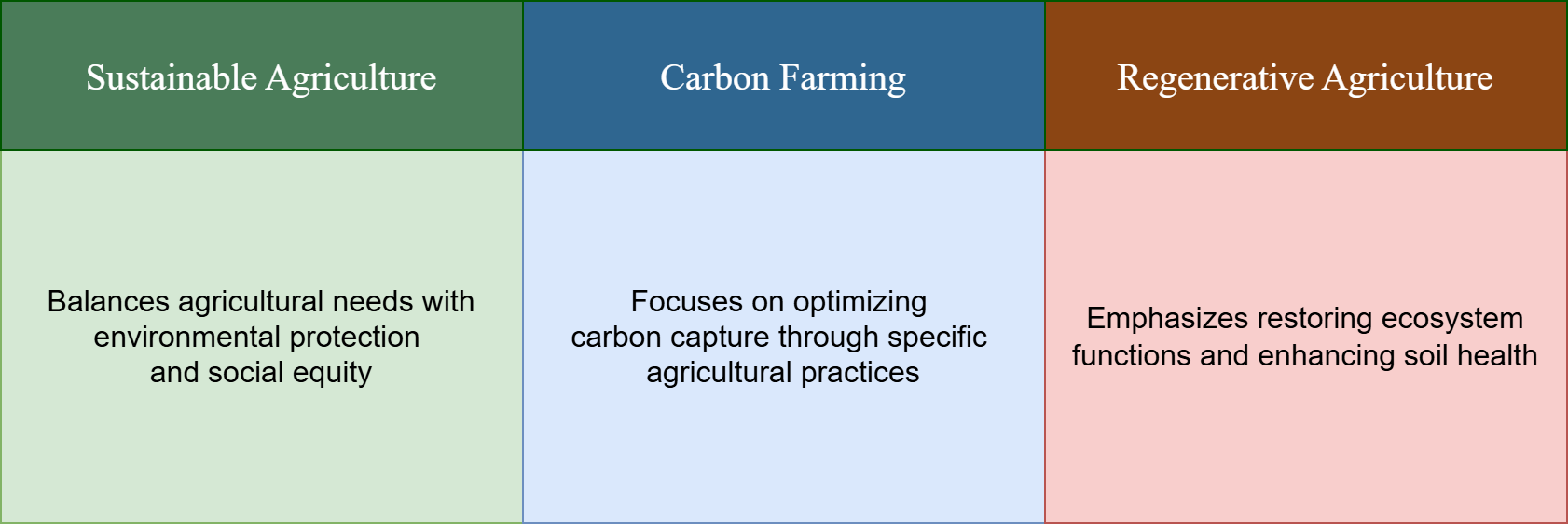}
    \caption{Climate-resilient farming systems}
    \label{fig:approaches_in_agriculture}
\end{figure}

Although terms are used differently, sustainable agriculture, regenerative agriculture, organic farming, and carbon farming collectively seek to create farming systems that balance social responsibility, economic viability, and environmental health while focusing on ecosystem benefits such as soil health, biodiversity, and climate resilience (Fig.~\ref{fig:approaches_in_agriculture}). What matters most are the practices employed: techniques that enhance soil health, conserve biodiversity, reduce emissions, and support farming communities.

\subsubsection*{Sustainable Agriculture}
\begin{figure}[ht]
    \centering
    \includegraphics[width=\linewidth]{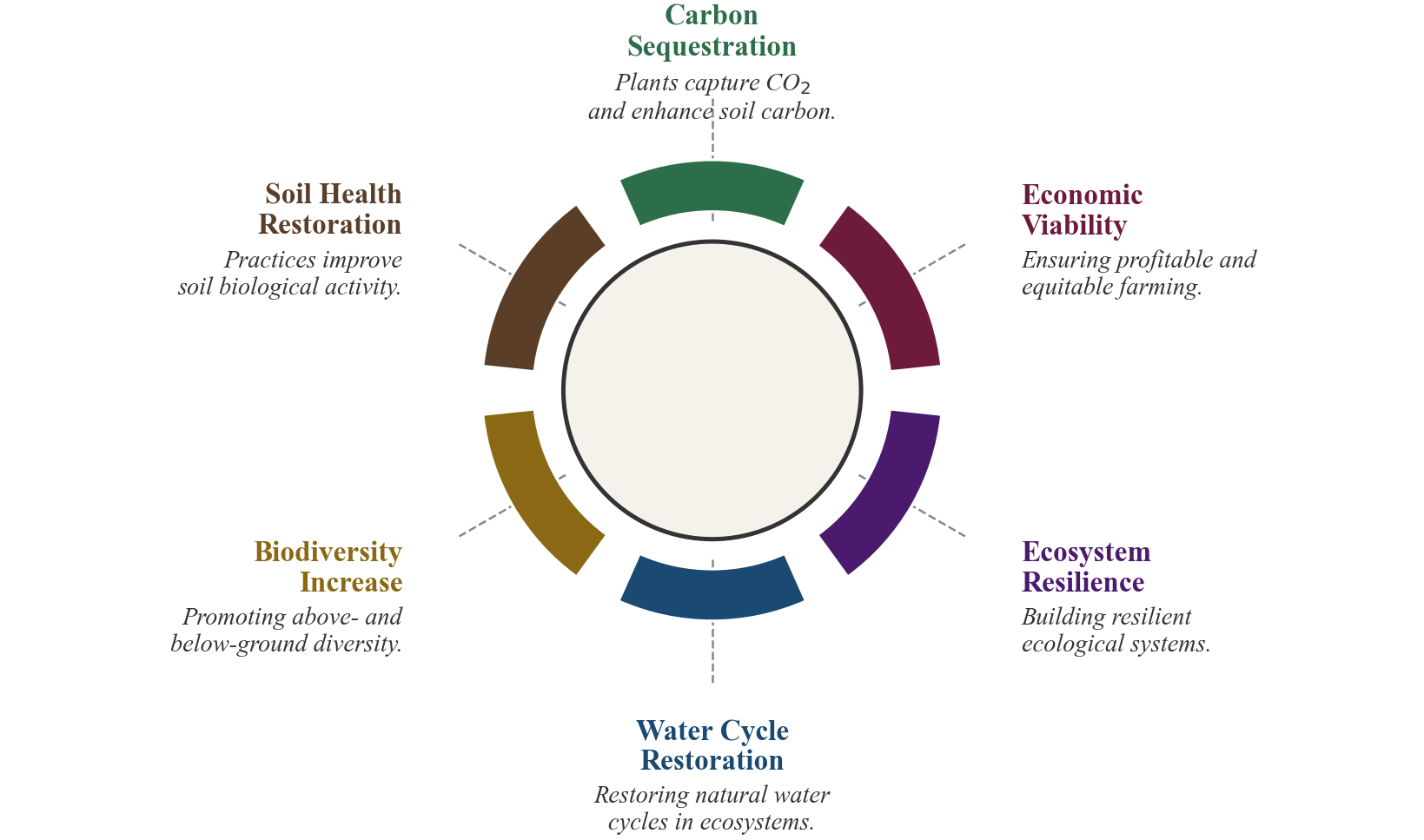}
    \caption{Cycle of sustainable agricultural practices}
    \label{fig:cycle_of_sustainable_practices}
\end{figure}

Sustainable agriculture is a holistic agricultural paradigm that seeks to balance environmental protection, economic viability, and social equity in a mutually reinforcing manner. Rather than optimizing a single objective, it integrates soil health, biodiversity conservation, water quality, climate resilience, farmer livelihoods, and long-term food security (Fig.~\ref {fig:cycle_of_sustainable_practices}). 

Best practices of sustainable agriculture \cite{KIBBLEWHITE08} include diversified cropping systems, crop rotations, integrated nutrient and pest management, efficient water use, and reduced dependence on external chemical inputs. A central focus is on improving soil health by enhancing organic matter, microbial activity, and nutrient cycling, thereby increasing productivity and resilience over time. By safeguarding natural resources while supporting viable farming systems, sustainable agriculture ensures that present needs are met without compromising the capacity of future generations to farm and thrive.

\subsubsection*{Regenerative Agriculture}
The focus in {regenerative agriculture} is on restoring and enhancing ecosystem function rather than merely sustaining current conditions. Its defining characteristic is improvement over time - rebuilding degraded soil organic matter, increasing biodiversity, and strengthening natural water cycles through biological approaches. It is outcome-oriented but does not necessarily quantify the improvements.

Regenerative agriculture is a farming model that works in synchronization with natural systems to restore and improve soil health, going beyond the goal of mere sustainability. It was first introduced by Robert Rodale in the 1980s, and the concept has gained substantial prominence in recent years in response to escalating concerns about climate change \cite{GILLER21}. 
Regenerative agriculture helps farmers and their communities deal with ecological, economic, and social challenges, while also strengthening global ecosystems by making agricultural systems more diverse and resilient \cite{ELEVITCH18, GORDON22}. 

The defining characteristic of regenerative agriculture is its emphasis on restoring degraded ecosystem functions through agricultural land management practices that support and improve natural processes. Such practices improve soil biological activity, promote above- and below-ground biodiversity, and restore natural water cycles. Regenerative agriculture views agricultural land as an interconnected ecosystem where productivity is derived from healthy ecological relationships, rather than dependency on external inputs. Although it shares principles with carbon farming and other sustainable approaches, regenerative agriculture is distinguished by its focus on ecological restoration and building resilience in farming systems \cite{TITTONELL22}.

\subsubsection*{Organic Farming}
{Organic farming} is an ecologically grounded cultivation approach that minimizes or completely avoids synthetic fertilizers, pesticides, genetically modified organisms, and growth regulators \cite{BritannicaOrganicFarming2025}. It emphasizes the use of natural processes and organic resources to produce crops and manage livestock. Its defining characteristics include the use of organic manures, composting, green manures, crop rotations, biological pest control, and soil health management aimed at improving long-term agro-ecosystem resilience. Scientifically, organic systems emphasize soil organic matter, microbial activity, and nutrient cycling as the main drivers of productivity.

The benefits of organic farming include improved soil structure, higher soil carbon stocks, reduced chemical residues in food, lower water body pollution, and enhanced on-farm biodiversity. In several contexts, especially rainfed and low-input systems, organic practices improve yield stability under climatic stress and reduce dependence on external inputs. However, organic farming faces significant challenges. Average yields are often significantly lower than conventional systems, particularly in crops that require nutrients, and nitrogen availability remains a key biophysical constraint. Certification costs, transition periods, and market access barriers further limit farmer adoption.

Empirical evidence shows that organic farming has succeeded in niche markets, high-value crops, and smallholder systems with strong institutional support. Failures invariably occur when transitions to organic farming are promoted without adequate nutrient sources, extension support, or market premiums, resulting in a decline in yields, farmer distress, and even disillusionment. However, there is no doubt that the best practices in organic agriculture have a great deal to contribute to carbon farming, which is described next.

\subsubsection*{Carbon Farming}
Carbon farming overlaps substantially with both sustainable agriculture and regenerative agriculture, but distinguishes itself through explicit measurement, verification, and monetization of {carbon sequestration} and emissions reductions. While employing similar best practices, carbon farming requires quantifiable carbon accounting to generate tradable credits, making it the most precisely defined among the four paradigms.


For the purposes of this survey paper, we define carbon farming as the collection of agricultural management practices specifically designed to maximize the capture and long-term storage of atmospheric carbon dioxide in soils and plant biomass, while simultaneously reducing greenhouse gas emissions from cultivation practices. Unlike conventional agriculture, which focuses primarily on yield maximization, carbon farming employs a dual mandate: maintaining agricultural productivity while quantifiably enhancing the land's capacity to sequester carbon and mitigate climate change.


Agricultural practices that are currently prevalent have contributed significantly to the GHG emissions and depletion of {soil carbon stocks}, largely reflected in losses of SOC. Much of this decline occurs when natural ecosystems such as forests and grasslands are converted to cropland, leading to substantial reductions in SOC. For example, converting primary forests to croplands results in a 25\% decrease in SOC, while conversion to perennial cropping systems can cause losses of up to 30\%. Even the conversion of primary forests to grasslands leads to a 12\% reduction in soil carbon \cite{DON11}.

Therefore, restoring soil carbon stocks through carbon farming presents an attractive strategy that enhances carbon sequestration, improves {soil health}, increases water retention, and increases the resilience of the crop to environmental stress. The potential of this approach is substantial: agricultural soils could sequester between 0.4 and 1.2 gigatons of carbon annually at the global level. In India alone, the annual potential for soil carbon sequestration ranges from 39 to 52 million tons \cite{NAIR15}. Yet for this opportunity to materialize, three ecosystem prerequisites must be established: functional markets that provide liquidity and fair pricing; trusted Measurement, Reporting, and Verification (MRV) systems to ensure credit integrity; and supportive policy frameworks that define carbon rights and reduce transaction costs.

An important byproduct of carbon farming is the accumulation of {\em carbon credits\/} by farmers. By trading these carbon credits in carbon markets, an attractive economic opportunity arises for the farmers.

Carbon farming generates multiple benefits that usually outweigh or operate independently of carbon-credit revenue. These co-benefits influence soil health, farm productivity, risk reduction, biodiversity, water security, and farmer livelihoods. It is worth noting that these co-benefits come with some caveats.

\subsubsection*{A Comparative Perspective}
Together, sustainable agriculture, regenerative agriculture, organic farming, and carbon farming should be viewed as complementary lenses rather than competing paradigms. They are based on a large and widely shared pool of best practices in agronomy, but differ in their main objectives, narratives, and success metrics. Sustainable agriculture provides an overarching normative framework that balances environmental, economic, and social goals. Regenerative agriculture emphasizes ecological restoration and building resilience over time. Organic farming is distinguished by clearly codified input restrictions and certification standards. Carbon farming introduces a quantitative, market-oriented dimension by explicitly measuring and monetizing climate mitigation outcomes. In practice, effective agricultural transitions often blend elements from all four paradigms, adapting them to local biophysical, socio-economic, and institutional contexts. From a policy and implementation standpoint, focusing on verifiable outcomes, namely, soil health, ecosystem services, resilience, and farmer welfare, rather than rigid labels, is likely to yield more scalable and durable impacts.

\subsection{Positioning of this Survey Paper}
Recent studies have extensively examined carbon farming as a prominent strategy to promote sustainable agriculture and mitigate climate change. This approach encompasses the adoption of agricultural practices that not only reduce GHG emissions but also enhance carbon sequestration, particularly within soils. SOC has become a central focus, and research has evaluated the effects of various practices, such as improved agricultural land management (e.g., conservation tillage, residue incorporation, cover cropping, improved crop planting), {agroforestry}, and improved grazing, on SOC sequestration. \cite{devideen_2025} highlights the considerable potential for SOC sequestration in agriculture, while \cite{Sharma2021} emphasizes the broader contribution of agriculture to the global {carbon cycle}. Regenerative agricultural practices are gaining significant attention in climate mitigation because they increase SOC in temperate arable lands without reducing crop yields \cite{Jared24}. Because of this dual potential to address climate change and strengthen food security, regenerative agriculture has attracted considerable interest from investors, farmers, and governments across Southeast Asia. To meet their nationally determined contributions (NDCs), policymakers in the region are increasingly exploring the use of agricultural lands to enhance carbon sequestration and directly reduce total GHG emissions from the agricultural sector.


The development of {carbon markets} and {carbon credits} has introduced financial incentives for adopting carbon farming as part of global efforts to reduce GHG emissions and promote sustainable agriculture. \cite{sainath2025} provides a comprehensive analysis of the opportunities, challenges, and policy considerations in the Indian context, stressing the importance of high-integrity carbon credits and robust {Monitoring, Reporting, and Verification (MRV) frameworks}. 



Several survey articles have appeared covering different aspects of carbon farming. The paper by Zheng et al. (2025) \cite{Zheng2024} presents a synthesis of carbon farming and climate-smart agriculture practices, summarizing evidence on soil organic carbon (SOC) responses, regenerative agriculture methods, and policy pathways. The emphasis is on India and other developing regions and on agronomic interventions and mitigation potential. The paper does not get into details on monitoring, reporting, and verification (MRV) systems, digital measurement technologies, carbon crediting methodologies, or carbon market architectures. The report by Bhattacharyya et al. (2011) \cite{bhattacharyya2011_carbon_status_indian_soils} provides a soil-science-centred assessment of SOC in Indian agricultural systems, drawing upon long-term field experiments, soil datasets, and agro-ecological analyses. This paper also does not get into much detail on carbon markets, MRV requirements, project development processes, or institutional and economic considerations. The paper by Petropoulos et al. (2025) \cite{Petropoulos2025} provides an in-depth global review of remote-sensing and machine-learning approaches for SOC estimation, encompassing satellite and UAV platforms, digital soil mapping techniques, model accuracies, and methodological trends. However, their paper does not discuss carbon-farming practices, farmer-level adoption issues, or the functioning of carbon markets and policies in depth. 

The paper by Durrer et al. \cite{DURRER2024} systematically reviews six crucial concepts: permanence, additionality, leakage, uncertainty, transaction costs, and heat-trapping ability of gases. This paper is very complementary to our paper.
Buck and Palumbo-Compton (2022) \cite{BUCK2022} present a fascinating study of 37 empirical social science studies on the adoption of carbon farming, specifically soil carbon sequestration, by farmers. The paper also highlights many adoption barriers. 
The paper by Newton et al. (2020) \cite{NEWTON2020} is a definitive review that analyzes 229 peer-reviewed articles and 25 practitioner websites on definitions of regenerative agriculture. This paper finds that around 86\% of the articles defined it as outcome-based and 64\% mentioned carbon sequestration. 
The survey article by Dev and Krishna (2025) \cite{DEV2025} is a systematic review of 69 empirical studies on VCMs worldwide. The report by the World Bank (2023) \cite{WORLDBANK2023} provides a comprehensive technical guidance on MRV technologies for soil organic carbon. 

Together, the above papers cover important but isolated components of the carbon-farming landscape. However, these papers do not provide a unified treatment that connects farming practices, SOC processes, MRV technologies, carbon-crediting methodologies, market mechanisms, policy frameworks, and India-specific implementation models within a single coherent narrative. Our paper strives to fill this gap. It is essential to note that the above papers are significantly more detailed than our paper in terms of the specific aspects they address.

\subsection{Outline of the Paper}
We now provide a section-by-section outline of this survey paper.

\subsubsection*{Section 2: Conceptual Preliminaries}
This section lays the conceptual foundations for carbon farming. It introduces the global carbon cycle, soil organic carbon dynamics, and mechanisms of carbon sequestration in agro-ecosystems. The section explains how agriculture interacts with carbon flows and clarifies the relationship between sustainable agriculture, regenerative agriculture, and carbon farming as overlapping, practice-based paradigms.

\subsubsection*{Section 3: The Carbon Farming Ecosystem}
Section 3 provides a big picture of the carbon farming ecosystems, presenting the various interacting components or building blocks. This section traces all key activities from best practices to carbon credit generation and monetization. The section also describes the end-to-end carbon project development cycle. 

\subsubsection*{Section 4: Best Practices in Carbon Farming}
The section surveys best practices in carbon farming. It first discusses generic agronomic interventions such as minimal tillage, crop rotation, cover cropping, efficient water management, and biochar application. It then presents crop-specific practices for major systems like rice and sugarcane, highlighting their operational features, scientific rationale, and documented impacts on soil carbon and greenhouse gas emissions.

\subsubsection*{Section 5: Co-Benefits and Trade-offs}
Section 5 analyses the co-benefits and trade-offs of carbon farming beyond carbon credit revenues. It synthesizes evidence on soil health improvement, water infiltration and retention, erosion control, biodiversity enhancement, resilience to droughts and floods, and reduced input costs. The section also discusses transition risks, context-specific limitations, and potential negative side effects that must be managed carefully.

\subsubsection*{Section 6: Measurement, Reporting, Verification, and Certification}
This section provides details of Measurement, Reporting, and Verification (MRV) frameworks for soil carbon initiatives. It covers field-based sampling protocols, empirical and process-based SOC models, remote sensing and digital MRV innovations, and the role of verification bodies. Particular attention is paid to uncertainty, baseline setting, and how modern digital tools enable scalable, lower-cost monitoring in smallholder-dominated landscapes.

\subsubsection*{Section 7: Carbon Markets and Trading}
Section 7 introduces carbon markets and trading mechanisms. It traces the evolution from early market-based environmental instruments through the Kyoto Protocol to the Paris Agreement. The section distinguishes compliance and voluntary markets, explains how carbon credits are generated and priced, and outlines key actors, governance structures, and integrity concerns that shape market performance and credibility.

\subsubsection*{Section 8: Representative Case Studies of Carbon Farming}
This section presents a representative set of six case studies from agricultural carbon projects and major carbon market systems. It highlights implementation models, farmer engagement strategies, technological innovations, and measured environmental and economic outcomes. Of the six case studies presented, three are from India.

\subsubsection*{Section 9: Carbon Farming in India: Current Status}
The next section examines India’s evolving carbon farming landscape. It discusses national and state-level policies, the emergence of India’s carbon market architecture, and the role of initiatives promoting regenerative agriculture and soil carbon enhancement. Specific attention is given to smallholder constraints, aggregation models such as Farmer-Producer Organizations, and India-specific opportunities and implementation challenges.

\subsubsection*{Section 10: Challenges and Potential Remedies}
This section presents the key challenges in carbon farming and suggests possible remedies. It discusses scientific, technical, financial, institutional, and social barriers limiting large-scale adoption, especially among small and marginal farmers. 

\subsubsection*{Section 11: AI, Game Theory, and Computation in Carbon Farming}
Section 11 presents the applications of artificial intelligence, game theory, and computation to carbon farming. These applications include digital MRV, farm management, farmer advisories, cooperative carbon farming, and carbon market design. 

\subsubsection*{Section 12: Conclusions and Future Directions}
The final section is a crisp summary of the paper and also presents broad directions for future work.

\subsubsection*{Glossary}
The {Glossary} provides concise definitions of key technical terms, acronyms, and concepts used throughout the paper, including greenhouse gases, SOC, MRV, ALM methodologies, and major carbon standards. It is intended as a quick reference for readers from diverse backgrounds, helping to ensure clarity and consistent interpretation of specialist terminology.

\subsubsection*{Appendix} 
The appendix offers supporting material that complements the main text. Appendix~A summarizes major carbon sequestration approaches in agriculture and land use. Appendix~B lists best practices in carbon farming, including crop-specific options. Appendix~C documents representative carbon projects in India, while Appendix~D reviews the startup ecosystem and major global stakeholders active in carbon markets.

\section{Conceptual Preliminaries}\label{sec2} 
This section attempts to establish the scientific basis for carbon farming by outlining the global carbon cycle, soil organic carbon dynamics, and sequestration mechanisms in agricultural systems. The section clarifies how farming practices influence carbon fluxes.

\subsection{The Carbon Cycle}
	The carbon cycle is the biogeochemical process of exchanging carbon among the main reservoirs of the Earth: the atmosphere, oceans, land, and living organisms. This cycle is fundamental to maintaining Earth’s climate and supporting life. 

\subsubsection*{Role of the Carbon Cycle in Agriculture}
The atmospheric concentration of $\text{CO}_2$ demonstrates a pronounced seasonal cycle, the amplitude of which has increased by up to 50\% in the Northern Hemisphere over the past 50 years \cite{MacBean2014}. While several factors, such as increased fossil fuel emissions, changes in oceanic $\text{CO}_2$ fluxes, and atmospheric transport processes, have been suggested to explain this trend, recent research indicates that the intensification of agricultural productivity is a major contributor.

Using crop production statistics from the Food and Agriculture Organization (FAO), \cite{MacBean2014} employed a carbon-accounting method to estimate the annual carbon uptake of major crops like maize, wheat, rice, and soybeans in the northern extratropics. This analysis revealed that increased productivity of these crops has resulted in an additional annual carbon exchange of approximately 0.33 petagrams, accounting for roughly 17--25\% of the enhanced seasonal carbon flux, with maize alone responsible for up to 66\% of this increase in certain regions.

In a complementary study, \cite{zeng2014agricultural} modified the terrestrial biosphere model VEGAS to incorporate changes in agricultural management practices. Their analysis suggested that increased agricultural productivity in mid-latitude regions could explain about 45\% of the rise in the amplitude of global net surface carbon fluxes between 1961 and 2010, comparable to contributions from climate change and $\text{CO}_2$ fertilization.

These findings demonstrate that current-day agricultural practices are significantly altering the terrestrial carbon cycle. The associated carbon uptake and release changes have important implications for regional agricultural productivity, global climate regulation, and carbon budgeting.

As agricultural intensification continues, its influence on the global carbon cycle is expected to become even more significant. Understanding these dynamics is essential for developing effective strategies to enhance soil carbon sequestration and support climate change mitigation efforts.

\subsubsection*{Carbon Cycle Models}
Various models have been developed to simulate the carbon cycle at different spatial and temporal scales. Process-based models such as \textbf{CENTURY} \cite{Parton1996} and its daily time-step counterpart, \textbf{DayCent} \cite{delgrasso2012}, simulate soil organic matter dynamics by incorporating interactions among climate variables, soil properties, and management practices. The \textbf{RothC} model \cite{colemanrothc1996} is also widely utilized to estimate soil carbon turnover, particularly in non-waterlogged soils, by modeling the decomposition of organic matter under diverse environmental conditions.

At larger spatial scales, Dynamic Global Vegetation Models (DGVMs) such as \textbf{LPJ} (Lund-Potsdam-Jena) \cite{sitch2003}, \textbf{ORCHIDEE} \cite{bowring2019}. The \textbf{Community Land Model (CLM)} \cite{lawrence2019} integrates vegetation dynamics with biogeochemical processes to predict global carbon fluxes. Additionally, empirical approaches draw on remote sensing data (e.g., MODIS) and flux measurements from observation networks such as \textbf{FLUXNET} \cite{pastorello2015}, providing estimates of carbon uptake and release across various ecosystem types and geographical regions.

\subsection{Carbon Sequestration} 
\begin{figure}[ht]
    \centering
    \includegraphics[width=\linewidth]{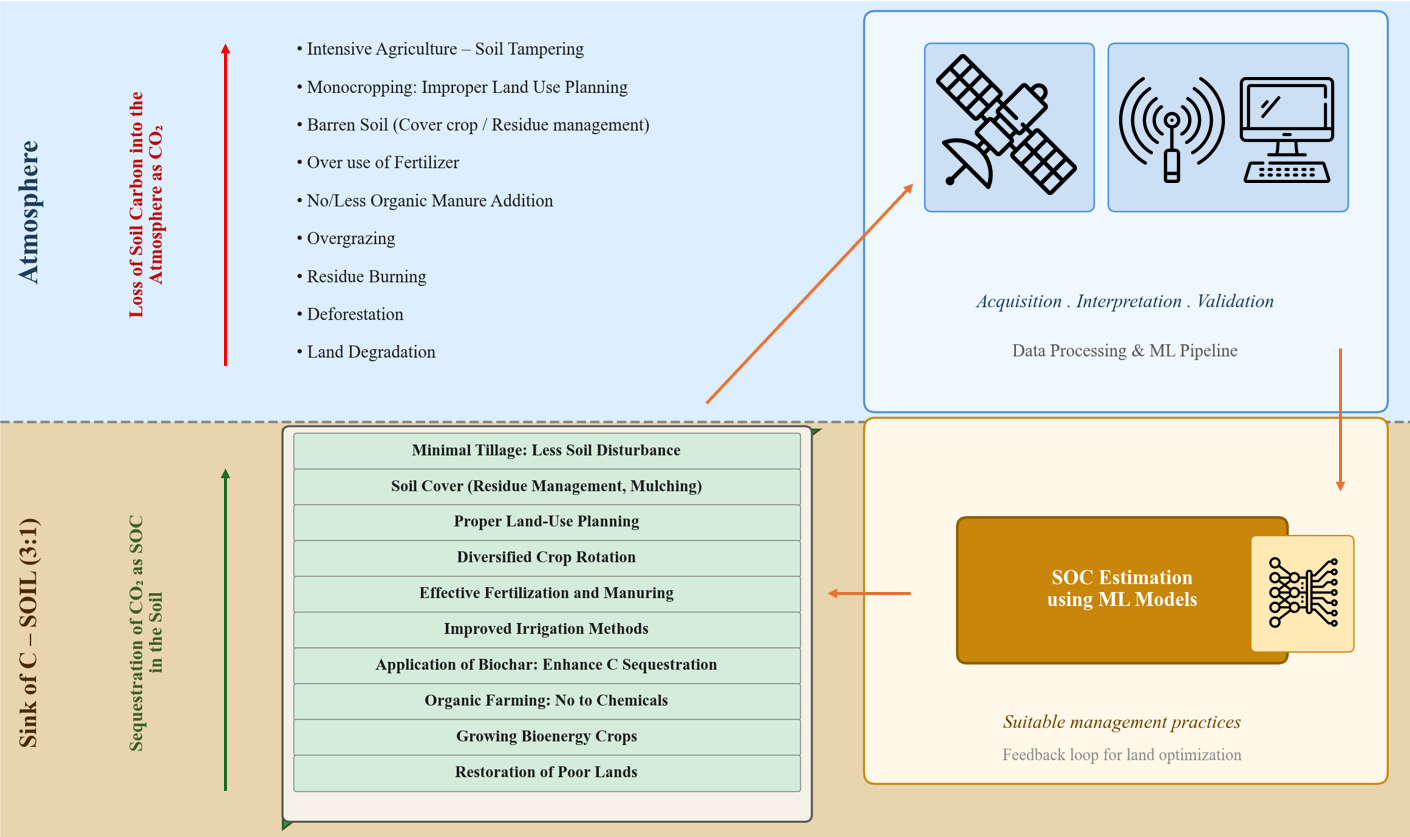}
    \caption{Carbon sequestration and monitoring for sustainable land management}
    \label{fig:Sinkc}
\end{figure}

Carbon sequestration is a critical process that involves the capture and long-term storage of atmospheric carbon dioxide, accomplished through natural ecosystem functions and human-engineered interventions. In agricultural systems, carbon sequestration occurs primarily through two principal mechanisms: the fixation of atmospheric carbon through plant photosynthesis and its subsequent incorporation and storage as organic matter in the soil, as shown in Fig.~\ref{fig:Sinkc}. The effectiveness of carbon sequestration depends on two fundamental factors: the rate of carbon uptake from the atmosphere and the stability of the storage mechanisms that retain the captured carbon. This dual requirement highlights the necessity for scientists and engineers to develop efficient carbon capture processes and establish stable, long-term storage solutions, ensuring that sequestered carbon dioxide does not return prematurely to the atmosphere. 
SOC is fundamental for soil health and constitutes the largest terrestrial {carbon sink} \cite{LAL16}. The relationship between SOC and soil health operates through multiple interconnected pathways crucial to soil functionality and agricultural productivity.

\subsection{Soil Organic Carbon and Soil Health}
SOC profoundly influences soil physical properties by increasing soil porosity, water retention capacity, and aggregate stability \cite{CHENU00}. These enhancements improve soil tilth and cultivability, directly supporting higher agricultural yields. A landmark paper by Lal (2004) \cite{LAL2004} establishes the scientific basis for soil as a carbon sink, quantifying its sequestration potential and highlighting the co-benefits for food security. This is one of the most cited papers on SOC sequestration. 
 In addition to its impact on physical properties, SOC plays a pivotal role in soil chemical processes by promoting the availability of nutrients and facilitating the exchange of ions, including the retention and mobilization of metal ions. Moreover, SOC is the main source of energy for soil organisms, which supports essential biological processes \cite{REEVES97}. The paper by Paustian et al. \cite{PAUSTIAN2016} defines “climate-smart soils,” provides SOC sequestration pathways, discusses barriers, and quantifies mitigation potential. This is very relevant to explaining why SOC matters and framing global mitigation estimates.

Soil health is defined by its ability to sustain agricultural productivity and provide ecosystem services. Soil health is governed by several fundamental functions: (1) carbon transformations, (2) nutrient cycling, (3) maintenance of soil structure, and (4) regulation of pests and diseases \cite{KIBBLEWHITE08}. These functions are shaped by factors such as soil type, the presence and activity of soil organisms, carbon and energy flows, and nutrient availability. The synergistic relationship between SOC and soil health is evident in improved soil structure, enhanced water retention, optimized nutrient cycling, enriched microbial habitats, and increased resistance to erosion.

This interaction creates a dynamic system in which higher levels of SOC are generally associated with improved soil structure, increased nutrient availability, and greater microbial diversity \cite{DEB15}. These interconnected benefits underscore the fundamental importance of SOC in maintaining and improving soil health and agricultural productivity.

\section{The Carbon Farming Ecosystem}
This section frames carbon farming as an interconnected socio-technical system and identifies key actors, data flows, and institutional components or building blocks. Together, these building blocks form a closed loop: farmers follow best practices and generate climate benefits; data is collected to quantify the benefits; aggregators orchestrate the relevant processes; certifiers and registries validate and issue credits; and buyers consume the credits \cite{cmi2024_key_stakeholders,climatedrive_key_stakeholders}. The section links farm practices to verified carbon credits, providing a system-level understanding of how biophysical results are translated into economic value.

\subsection{Building Blocks of Carbon Farming}

\begin{figure}[ht]
    \centering
    \includegraphics[width=0.9\linewidth]{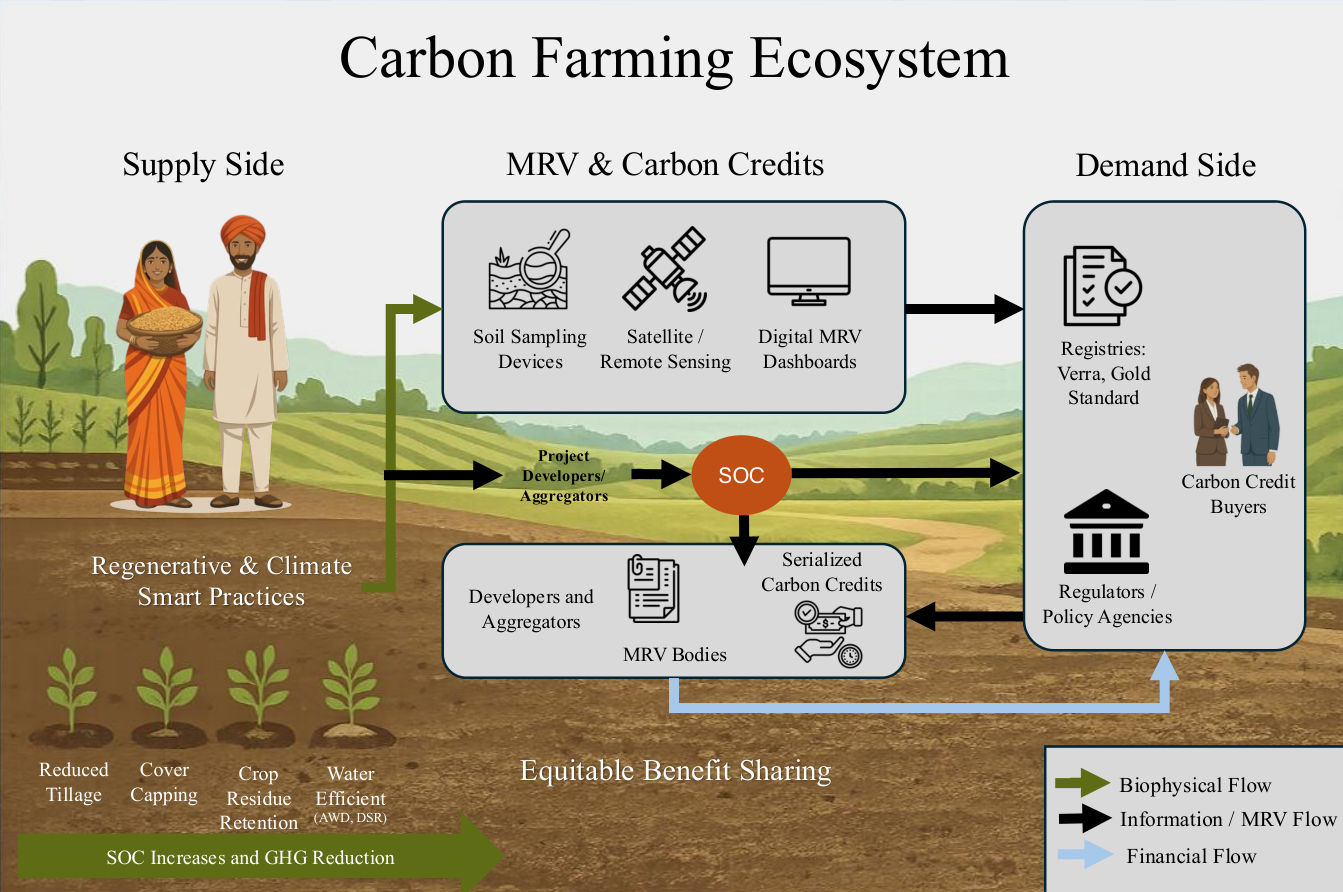}
    \caption{Carbon farming ecosystem linking best practices to carbon markets}
    \label{fig:CFEco}
\end{figure}

We now briefly describe the building blocks of the carbon farming ecosystem. Fig.~\ref{fig:CFEco} presents these building blocks.

    \subsubsection*{Source of Sequestration}
    Farmers adopt practices such as reduced tillage, cover cropping, residue retention, and water-efficient cultivation (all described in the next section) that increase soil organic carbon (SOC) levels and reduce greenhouse gas emissions. These on-farm actions constitute the biophysical foundation of carbon farming.

    \subsubsection*{Data Layer}
    High-frequency, georeferenced measurements from soil-sampling devices, satellite and remote-sensing platforms, yield maps, and other machine-readable sources provide the empirical basis for establishing baselines and quantifying changes in SOC. This data infrastructure enables transparent, intervention-based accounting rather than static activity reporting.

    \subsubsection*{Aggregator}
    Project developers and co-operatives bundle many dispersed smallholdings into coherent projects, standardise practice protocols, manage implementation, and shoulder administrative tasks. Aggregators also act as intermediaries, ensuring that information and financial flows reach farmers.

    \subsubsection*{Certification, MRV, and Credit Issuance}
    Independent MRV (measurement, Reporting, and Verification) bodies measure, report, and verify net CO$_2$-equivalent gains obtained. Following verification, recognised registries (such as Verra and Gold Standard) issue carbon credits and ensure they are uniquely tracked to avoid double-counting. This building block provides credibility and transparency to the system.

    \subsubsection*{Carbon-Credit Buyers}
    Corporations and other entities purchase verified credits to offset residual emissions or fulfill sustainability commitments. Their expenditure ultimately finances the system, with revenue flowing back through project developers to farmers via equitable benefit-sharing arrangements.

Fig.\ref{fig:CFEco} illustrates the end-to-end value chain through which carbon farming practices are translated into verified carbon credits. On the supply side, farmers adopt best practices that increase soil organic carbon and reduce greenhouse gas emissions. These biophysical changes are captured through soil-sampling devices, satellite and remote-sensing technologies, and digital MRV platforms, forming the data backbone required for transparent quantification. Project developers and aggregators organise multiple farms into unified projects and channel data to independent MRV bodies, which verify SOC gains. Verified outcomes are then issued as carbon credits by recognised registries (such as Verra and Gold Standard). On the demand side, corporate buyers purchase these credits to meet climate and ESG (environmental, Social, and Governance) commitments, with the resulting financial gains flowing back to farmers through structured, equitable benefit-sharing mechanisms. Distinct flows (biophysical, informational, and financial) are shown in the picture to highlight how carbon, data, and value circulate through the system.

\subsection{Development Life Cycle for Carbon Farming Projects}\label{sec6}
\begin{figure}[htbp]
    \centering
    \includegraphics[width=\linewidth]{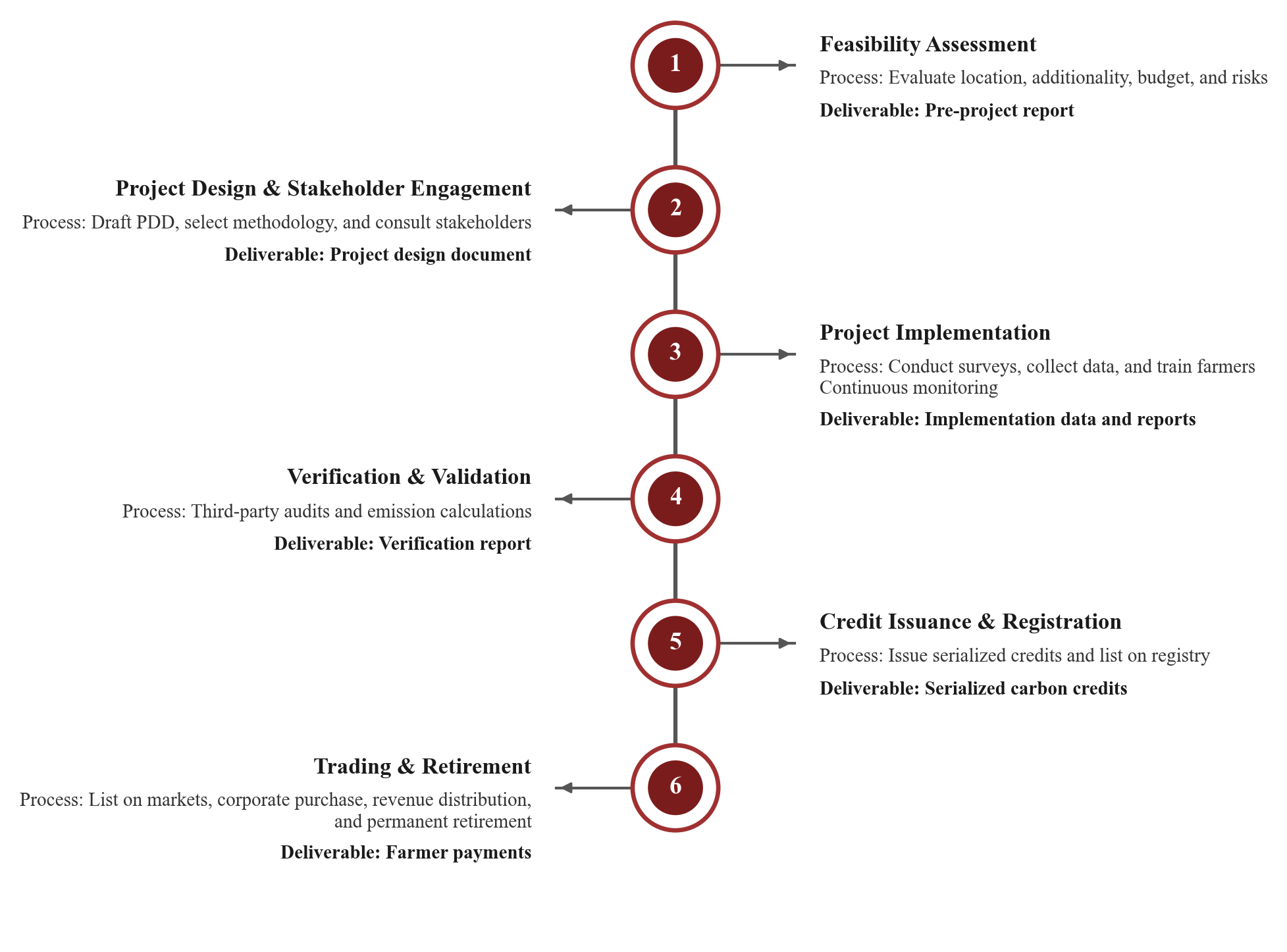}
    \caption{Overview of the development life cycle for carbon farming projects}
    \label{fig:lifecycle}
\end{figure}

Generating carbon credits involves a rigorous, multi-stage process focused on accurately measuring, reducing, and verifying reductions in greenhouse gas emissions and/or increases in the stock of soil organic carbon, as illustrated in Fig. ~\ref{fig:lifecycle}.
\subsubsection*{Project Feasibility Assessment} 
The feasibility assessment \cite{VERRAprojectfeasibility} of the project begins with evaluating the location and conditions of the project and identifying the activities and carbon pools of the project affected by the project. This activity examines the project's {additionality criterion} (best practices in carbon farming beyond traditional best practices already being followed in the project location) and the baseline scenario, ensuring that the project would not occur without intervention, in accordance with carbon methodology and standards. The baseline scenario helps assess the environmental impact without the project, serving as a reference for carbon reduction. A pre-project report is prepared detailing the budget, including resources, costs, funding sources, and risk mitigation strategies.

\subsubsection*{Project Design and Stakeholder Engagement}
Project developers initiate the process by designing the project, drafting a Project Design Document (PDD) in accordance with the selected methodology and carbon standard (e.g., popular standards such as the Verified Carbon Standard or Gold Standard), and engaging local stakeholders. This stage includes the onboarding of farmers, consultations with various stakeholders, promotion of the project and its activities, including emission reduction and SOC stock improvement opportunities, as well as the implementation and development of robust monitoring plans. The project is expected to undergo a public comment period and resolve all comments raised on the carbon standard platform before it is listed on the registry as a project under development. 

\subsubsection*{Project Implementation} 
Project developers implement project activities according to the methodology and standard selected for measurement and verification. This involves conducting detailed surveys, collecting data, implementing systematic monitoring, and providing transparent reporting as outlined in the project design document to ensure accountability. The project must also establish additionality by demonstrating that the emission reductions or removals would not have occurred without the carbon finance incentive, thereby validating the project's eligibility under the chosen standard.

\subsubsection*{Verification and Validation} 
Accredited Verification and Validation Bodies (VVBs), in collaboration with the selected carbon standard, conduct third-party audits to validate the project design and verify the implementation of regenerative or sustainable practices, as well as the accuracy of claimed emission reductions. This is achieved through a combination of site visits, stakeholder interviews, and a detailed review of the documentation. To ensure methodological accuracy, VVBs can also engage independent modeling experts to assess and approve credit calculation processes. This rigorous validation process safeguards the credibility, transparency, and environmental integrity of carbon credits issued.

\subsubsection*{Issuance of Carbon Credits} 
After successful verification, recognized carbon registries (for example, Verra, Gold Standard) issue unique serialized credits (1 credit = 1 ton of $CO_2$ eq). Then, these credits enter carbon markets for sale to entities that offset their emissions. Once used for offsetting, credits are permanently retired from registry accounts to prevent double-counting. 
Carbon projects in AFOLU (Agriculture, Forestry, and Other Land Use) must demonstrate {permanence} (that is, long-term impact) and avoidance of leaks (that is, no change of emissions elsewhere). Documentations (project design documents, audits) are made publicly available via registries.

\section{Best Practices in Carbon Farming}
\label{app:best_practices}
\begin{figure}[htbp]
    \centering
    \includegraphics[width=\linewidth]{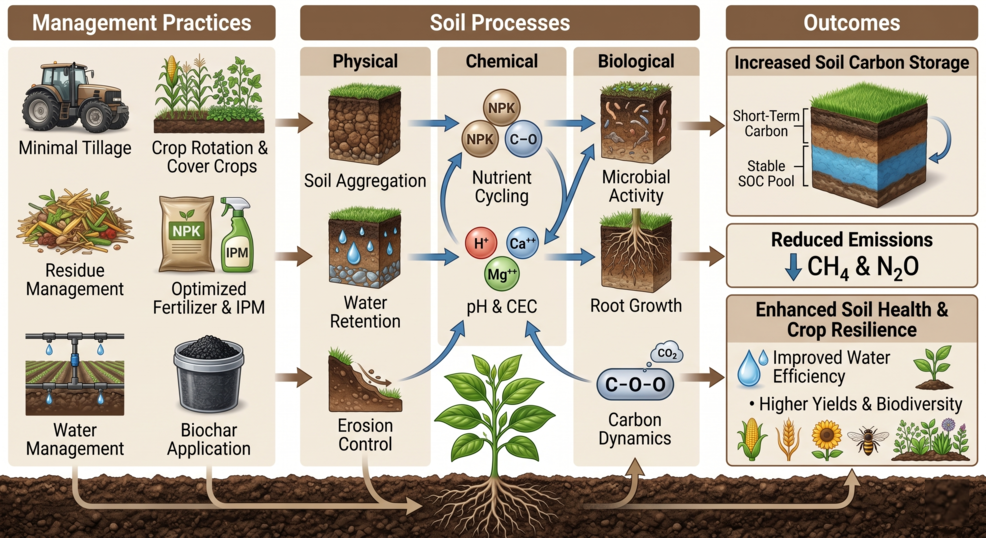} 
    \caption{Carbon farming mechanisms in agricultural soils}
    \label{fig:practices_for_agriculture}
\end{figure}

Carbon farming encompasses agricultural practices scientifically designed to maximize soil carbon sequestration while maintaining or enhancing crop productivity. These practices increase soil carbon stocks and improve overall soil health through synergistic interactions among physical, chemical, and biological soil processes. The effectiveness of carbon farming operations is highly dependent on local conditions, including climate, soil properties, and crop selection. 
This section reviews best practices in such agronomic interventions. The section presents evidence on both generic and crop-specific practices, highlighting their documented impacts on productivity, emissions, and long-term soil health. The benefits can be seen in Fig. \ref{fig:practices_for_agriculture}.

\subsection{Generic Best Practices}
\label{app:generic_practices}
 First, we cover operational best practices that are generic in nature and are applicable across all crops.


    \subsubsection*{Minimal Tillage} 
    {Minimal tillage} practices, including {no-till} and strip-till systems, are fundamental to preserving soil structure and improving carbon sequestration. By minimizing soil disturbance, these practices support the natural formation of the soil and the stability of soil aggregates, making the soil substantially less vulnerable to wind and water erosion \cite{WOODYARD17}. The preservation of soil structure is vital for long-term carbon storage and overall soil health.

    Reduced mechanical disturbance in minimal tillage systems slows the decomposition of organic residues by limiting the breakdown of soil aggregates and reducing the exposure of organic matter to rapid degradation \cite{ADEKIYA17}. Additionally, minimal tillage moderates soil temperature fluctuations, further contributing to organic matter accumulation by slowing decomposition rates and creating conditions conducive to increased carbon sequestration. Minimal tillage also improves the biological activity and diversity of the soil. These systems protect essential soil organisms by preserving earthworm burrows and maintaining mycorrhizal fungal networks that form symbiotic relationships with host plants \cite{WOODYARD17}. In contrast, intensive tillage disrupts these beneficial biological structures and destroys critical microhabitats necessary for robust soil organism communities \cite{COHEN19}.

    \subsubsection*{Crop Rotation} 
    Strategic crop rotation is a highly effective agricultural practice for enhancing soil carbon sequestration. By introducing a diverse sequence of crops throughout the rotation cycle, farmers provide the soil with varied inputs of organic matter, as different species contribute different residues and root exudates \cite{Mohler09}. This diversity strengthens the carbon storage capacity of the soil and improves the efficiency of the nutrient cycle. Variant patterns of uptake of nutrients among crops help prevent the depletion of specific soil nutrients that often accompany continuous monoculture. An effective crop rotation system typically incorporates three core elements: deep-rooted perennial plants that access nutrients and store carbon in deeper soil layers; leguminous crops capable of fixing atmospheric nitrogen, thereby reducing dependence on synthetic fertilizers; and high-biomass crops that contribute substantial amounts of organic matter. This combination maximizes carbon sequestration, improves soil fertility \cite{agronomy21}, and naturally suppresses pest and disease cycles, reducing dependence on chemical interventions that may otherwise disrupt soil ecosystems.

    \subsubsection*{Cover Crops} 
    {Cover crops} are plant species that are cultivated between main crop seasons primarily to protect and improve soil health, rather than for commercial harvest. Species commonly utilized as cover crops, including legumes, grasses, and brassicas, provide a range of ecological services that can enhance the agronomic performance of subsequent cash crops. The presence of living vegetation during fallow periods maintains continuous carbon assimilation through photosynthesis, facilitating sustained allocation of carbon below the ground. Cover crops also contribute to improving soil quality by providing organic matter through both aboveground residues and root biomass \cite{agronomy21}. As cover crop roots grow, senesce, and decompose, they create biopores and deposit organic substrates throughout the soil profile, thereby enhancing soil structural integrity. This improved structure increases the water retention capacity of the soil and confers greater resistance to erosion caused by wind and water.


\subsubsection*{Optimal Usage of Fertilizers and Pesticides} Reducing the dependence on synthetic fertilizers and pesticides is essential for enhancing carbon sequestration and directly reducing agricultural greenhouse gas emissions \cite{agronomy21}. Synthetic nitrogen fertilizers, in particular, are the largest source of nitrous oxide (N$_2$O) emissions from agriculture -- a gas with nearly 300 times \cite{griffis2017nitrous} the warming potential of CO$_2$.

By optimizing chemical inputs through precision nutrient management and Integrated Pest Management (IPM), farmers can stimulate natural soil biology, reduce soil acidification, and promote microbial communities vital for building stable soil organic carbon \cite{Mohler09}. Importantly, lowering fertilizer use also reduces the substantial fossil fuel emissions associated with fertilizer manufacturing.

Complementing reduced chemical inputs, the use of green manure crops - such as clover, vetch, peas, and beans - further strengthens soil carbon dynamics. As these crops decompose, they add organic matter to the soil, improving moisture retention, nutrient availability, and soil structure, all of which facilitate carbon capture and storage. Leguminous green manure species additionally fix atmospheric nitrogen through symbiosis with nitrogen-fixing microorganisms, thereby reducing the need for synthetic fertilizers.

Taken together, precision nutrient strategies, reduced reliance on synthetic inputs, and the integration of biological amendments create a balanced approach that reinforces the soil’s intrinsic capacity to store carbon. This synergy contributes to the long-term sustainability, productivity, and resilience of agricultural systems. 

\subsubsection*{Efficient Water Management and Irrigation}
Effective water management and advanced irrigation techniques are crucial for maximizing carbon sequestration in agricultural soils \cite{GANGOPADHYAY2022}. Maintaining optimal soil moisture levels creates favorable conditions for microbial activity, which is integral to the transformation of organic matter and carbon stabilization. Improved water management practices also reduce soil erosion, promote healthy plant growth, and increase biomass production, directly supporting higher levels of carbon storage. In addition, appropriate irrigation techniques help prevent waterlogging and the development of anaerobic soil conditions that could restrict carbon sequestration processes \cite{Chaudhary23}. The application of modern technologies, such as drip irrigation and sprinkler systems, soil moisture conservation measures, and monitoring devices, allows precise regulation of water distribution and supports efficient soil moisture management, thus optimizing the conditions for carbon accumulation in soils \cite{Mboyerwa22}. 

\subsubsection*{Use of Biochar} 
{Biochar} is a stable form of carbon produced from biomass - such as wood chips, crop residues, or manure - through pyrolysis, a thermochemical process that converts organic material into long-lasting carbon. Its application in agricultural soils represents a promising strategy for enhancing soil fertility while simultaneously sequestering carbon. Because biochar is highly resistant to decomposition, a substantial portion of its carbon remains in the soil for decades to centuries, contributing directly to long-term carbon storage.

The use of biochar also improves nutrient use efficiency by utilizing locally available biomass and converting it into a soil amendment that reduces nutrient losses and enhances soil–plant interactions. When incorporated into soils, biochar influences their physical, chemical, and biological properties, affecting soil structure, pH, cation-exchange capacity, microbial communities, and root-soil interactions \cite{LEHMANN09}. The magnitude of these effects depends on factors such as feedstock type, pyrolysis conditions, the physical characteristics of the resulting biochar, and soil conditions, including texture, temperature, and moisture.


\subsection{Crop-Specific Best Practices}
\label{app:crop_specific}
As examples of crop-specific practices, we cover two crops below: rice and sugarcane.
\subsubsection{Rice-Specific Practices}
\label{app:rice_practices}
Rice cultivation is a significant source of GHG emissions, mainly due to methane ($CH_4$) released from continuously flooded paddy areas and nitrous oxide ($N_2O$) emissions associated with nitrogen fertilizer use \cite{Mboyerwa22}. Flooded paddy cultivation not only contributes to GHG emissions but also places a heavy stress on water resources. To grow just 1 kilogram of rice, the water requirement typically ranges from 1,000 to 3,000 liters. This significant amount of water is needed for irrigation, evapotranspiration, and other losses during the cultivation process, according to the International Rice Research Institute (IRRI). Farmers burn rice stubble because it is the quickest and cheapest way to clear fields after harvesting; however, this practice significantly contributes to the GHG footprint. While it helps prepare fields for the next crop, stubble burning negatively impacts soil productivity for subsequent crops. It destroys beneficial microorganisms, depletes soil nutrients, and releases carbon, all of which increase air pollution and environmental degradation. These emissions represent a major portion of global agricultural GHG emissions, making rice management a critical area for sustainable intervention. Some specific carbon sequestration practices for rice include:

\subsubsection*{Rice Crop Residue Management (CRM)}
CRM is not only crucial for addressing environmental, agronomic, and health challenges but also plays an important role in carbon farming. In the Indo-Gangetic Plains, where rice residue burning is widespread, it is estimated that India produces about 165.8 million tons (Mt) of rice straw annually, with nearly 50 Mt of this being burned \cite{kaur2022rice}. This contributes significantly to air pollution, respiratory diseases, and soil degradation. However, through effective CRM practices, the environmental impact can be greatly reduced, and carbon sequestration can be enhanced. In-situ CRM practices, such as incorporation, mulching, and the use of advanced technologies like the Happy Seeder, Super Straw Management System (Super SMS), and straw chopper-spreader, help in directly increasing soil carbon storage. By incorporating rice stubble into the soil or using it as mulch, carbon is sequestered in the soil, improving soil organic matter (SOM) and enhancing long-term soil fertility. For instance, the Happy Seeder has been shown to reduce particulate pollution by over 98\%, GHG emissions by approximately 80\%, and, importantly, support the sequestration of carbon by retaining organic material in the soil. In addition to in-situ methods, ex-situ uses of rice straw, such as baling, biochar production, and bioenergy generation, also contribute to carbon sequestration and provide valuable resources for sustainable agricultural practices. Biochar, produced by pyrolyzing rice straw, locks carbon into a stable form that can be used to enrich the soil, further enhancing carbon storage.

\subsubsection*{System of Rice Intensification (SRI)} 
The System of Rice Intensification adopts optimized plant spacing, reduced water application, and intensive soil organic matter management. This approach reduces methane emissions by 40--60\% and increases soil carbon sequestration by 12.58 to 49.51 tons of $CO_2$ equivalent per hectare \cite{GANGOPADHYAY2022}. SRI further promotes environmental stewardship with 20--30\% water savings and yield improvements of 15--25\%.

\subsubsection*{Alternate Wetting and Drying (AWD)} 
Alternative Wetting and Drying is a practice of water management involving intermittent field flooding. This strategy achieves a 30--50\% reduction in methane emissions and increases carbon sequestration \cite{GANGOPADHYAY2022}. The emission of $CH_4$ is reduced by alternate wetting and drying irrigation, which significantly increases the absorption of atmospheric oxygen ($O_2$) into the soil \cite{Mboyerwa22}. Although increased nitrification of $\mathrm{NH_4^+}$ during the dry episode and subsequent denitrification of $NO_3^-$ during rewetting of dry soils have been reported to cause a slight increase in $N_2O$ emissions during AWD irrigation, it still lowers total GHG emissions from rice fields mainly due to decreased $CH_4$ emissions.

\subsubsection*{Direct Seeded Rice (DSR)} 
Direct seed rice involves direct seeding, bypassing traditional transplantation, and minimizing standing water during the early stages of growth \cite{Chaudhary23}. There are two principal types of DSR, namely, dry DSR and wet DSR. In case of dry DSR, dry seeds are seeded in a well-prepared non-puddled seed bed either through broadcast or using seed drills (often practiced in rain-fed and deep-water ecosystems), whereas, in the case of wet DSR, pre-germinated seeds are seeded in puddled (wet) soil using specialized equipment such as drum seeders \cite{Anjani24}. DSR reduces GHG emissions by 69.9\% and increases carbon sequestration \cite{REDDY2025100238}.

\subsubsection*{The Saguna Rice Cultivation Technique (SRT)} 
This is a climate-smart, zero-tillage system that eliminates soil puddling. It involves directly sowing seeds in unploughed soil, often incorporating efficient drip irrigation and mulching with crop residue. This approach reduces water consumption by up to 60\%, reduces methane ($CH_4$) emissions, and saves fuel and labor. Furthermore, by promoting legume crop rotation in the following season, SRT enhances natural soil fertility   \cite{Wadghane24}. This holistic method builds long-term organic soil carbon, offering a highly scalable and profitable solution for sustainable, resilient agriculture.

\subsubsection{Best Practices Specific to Sugarcane}
Sugarcane is a C4 plant (C4 plants are a group of plants, primarily grasses, that are characterized by a specialized pathway for photosynthesis to enhance their efficiency in hot, dry environments). C4 plants efficiently sequester atmospheric CO$_2$ across roots, shoots, and leaves. In India, sugarcane monoculture is depleting soil health and increasing the dependence on chemicals. The burning of sugarcane stubble exacerbates air pollution and damages soil microorganisms. Adopting crop diversification through practices like mulching and inter-cropping offers a sustainable solution. Mulching helps retain moisture, improve soil organic carbon (SOC), and reduce weed growth, while inter-cropping enhances soil fertility and biodiversity. \cite{KUMAR24a} comprehensively analyzes the carbon storage capacity of sugarcane, revealing that the leaves are the main carbon reservoir, accumulating approximately 877.08 kg/ha. In general, sugarcane cultivation stores approximately 14,439.6 kg of $CO_2$ per hectare per crop, of which 7293.4 kg are conserved in the soil, and 7146.2 kg are stored in plant biomass. This research underscores the significant potential of sugarcane cultivation for carbon sequestration and its important role in climate change mitigation strategies. Some specific carbon sequestration practices for sugarcane are described below.

\subsubsection*{Green Cane Trash Blanketing (GCTB)} 
This is a sugarcane residue management practice, a sustainable alternative to burning sugarcane fields after harvest \cite{Kingston05}. In GCTB, the crop is harvested in green and the leafy residue (trash) is left on the surface of the soil as thick mulch instead of being burned. This practice completely eliminates significant methane ($CH_4$), $N_2O$, and black carbon emissions from burning. Furthermore, the trash blanket decomposes over time, sequestering 1-2 tons of carbon per hectare annually, conserving soil moisture by up to 30\%, and suppressing weed growth, thus reducing herbicide use.

\subsubsection*{Sub-surface Drip Irrigation (SDI) and Fertigation} 
Sugarcane is a water-intensive crop, traditionally grown with flood irrigation. SDI involves delivering water and liquid fertilizers (fertigation) directly to the root zone through buried drip lines. This method can reduce water consumption by more than 50\% and pump energy use. By preventing waterlogged soil conditions, it reduces potential $CH_4$ emissions. Fertigation is a form of precision agriculture that minimizes $N_2O$ emissions and nutrient runoff, ensuring efficient uptake by the plant. 

\subsubsection*{Biomass Management } 
Biomass management in sugarcane systems plays a crucial role in strengthening soil carbon storage, particularly when residues are retained rather than removed or burned. Long-term trials have shown that residue retention can substantially increase soil organic carbon, with total SOC improving by 21\% and stable carbon pools forming a dominant share of soil carbon stocks \cite{pradhan2023impact}. These gains arise from continuous biomass inputs that promote carbon stabilization and support soil biological functioning. Enhanced biological activity under residue retention—evidenced by higher enzyme activities—further accelerates carbon transformation and the incorporation of carbon into stable pools. As a result, systems that retain biomass are able to sequester around $0.68~\text{Mg C/ha/year}$ while simultaneously improving carbon retention efficiency and overall crop productivity. These findings highlight that practices designed to maximize biomass return to the soil provide a strong foundation for long-term carbon sequestration and soil health improvement in sugarcane cultivation.

\subsubsection*{Sustainable Utilization of Sugarcane Bagasse} 
Sugarcane bagasse, the fibrous residue remaining after juice extraction, plays a central role in enhancing the carbon efficiency of sugarcane production systems. When used as a renewable fuel in cogeneration units, bagasse can displace fossil energy sources, thereby lowering greenhouse gas emissions and improving energy self-sufficiency in sugar mills. Recent assessments of bioenergy pathways incorporating post-combustion carbon capture indicate that bagasse-based systems can deliver meaningful mitigation benefits, with carbon capture costs estimated at approximately 262 USD per ton of $CO_2$, potentially decreasing to 17.2 USD per ton under optimized operating conditions. These findings highlight the feasibility of integrating bioenergy with carbon capture and storage (BECCS) in sugarcane biorefineries, positioning bagasse as a viable negative-emission resource \cite{WIESBERG2021111486}. Beyond energy production, the application of bagasse-derived ash to soils can enhance soil organic carbon, improve aggregate stability, and increase nutrient availability \cite{KUMAR24b}. Together, these uses underscore the importance of sustainably managing bagasse to close carbon loops and strengthen the overall carbon sequestration potential of sugarcane systems.

\section{Carbon Farming: Co-Benefits and Trade-offs}
\label{app:co_benefits}
\begin{figure}[htbp]
    \centering
    \includegraphics[width=\linewidth]{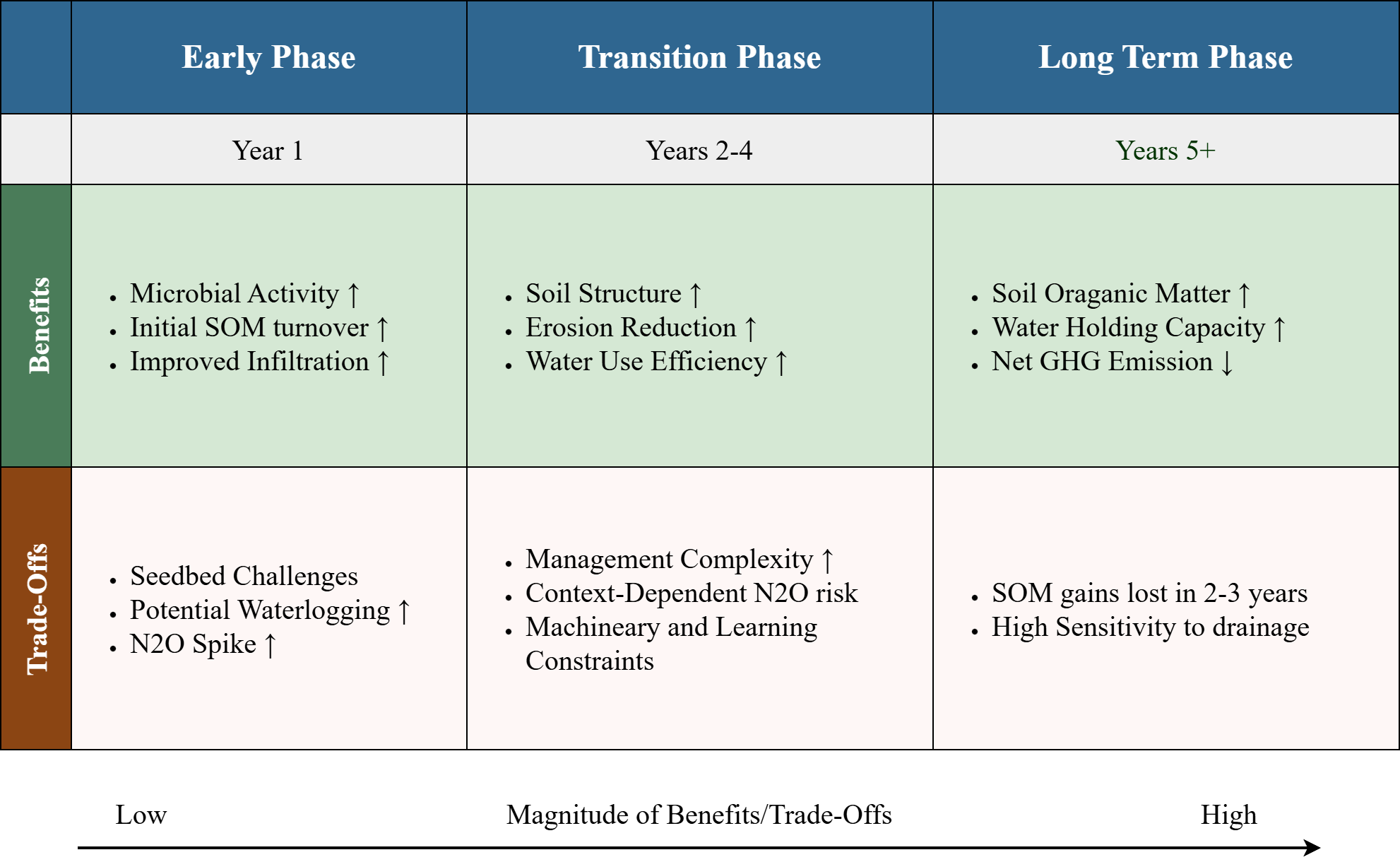} 
    \caption{Temporal Benefit–Risk Matrix for Carbon Farming Practices}
    \label{fig:cobenefits_tradeoffs}
\end{figure}

The true benefit of carbon farming lies in generating multiple benefits that usually outweigh any revenue generated by the sale of carbon credits. These co-benefits influence soil health, farm productivity, risk reduction, biodiversity, water security, and farmer livelihoods \cite{Zheng2024,PAUSTIAN2016}.
This section examines the broader ecological and socio-economic consequences of carbon farming. The section presents co-benefits related to soil health, water regulation, biodiversity, and resilience, while also identifying trade-offs, transition risks, and conditions under which anticipated benefits may not fully materialize. Fig. \ref{fig:cobenefits_tradeoffs} shows the benefits and risks associated with practices over the duration of the practices used for carbon farming.
Wherever possible, we have tried to quantify the benefits available from different sources or provided educated estimates.

\subsubsection*{Additional Revenue Generation}
It is the direct benefit a farmer obtain from adopting these carbon farming practices. Carbon sequestration is widely regarded as a “win–win” approach to sustainable agriculture, as it delivers environmental and productivity benefits at relatively low abatement costs \cite{Dutta21}, making it a cost-effective and eco-friendly strategy for enhancing soil health and crop yields \cite{Yousra24}. Based on the Aadi project (VCS 2590) by Grow indigo the economics of carbon farming \cite{GrowindigoProject} has been quantified and compiled in the  Table~\ref{tab:carbon_credits}.

\begin{table}[htbp]
\centering
\caption{Carbon Credit Earnings per Acre}
\label{tab:carbon_credits}
\begin{tabular}{l l r r}
\hline
\textbf{Category} & \textbf{Units} & \textbf{Value (\$)} & \textbf{Value (INR)} \\
\hline
Credits generated & Credit / acre & 1  & -- \\
Minimum earnings  & Rupees / acre & 10 & 916.73 \\
Average earnings  & Rupees / acre & 25 & 2291.81 \\
Maximum earnings  & Rupees / acre & 40 & 3666.90 \\
\hline
\end{tabular}
\end{table}

\subsubsection*{Enhanced Soil Organic Matter (SOM)}
Increased SOM improves soil structure, creating stable aggregates that resist erosion and improve tilth. SOM acts as the {\em glue\/} binding soil particles, creating pore spaces for air and water movement while providing habitat for soil organisms. It is known that an approximately 1\% increase in SOM can increase water-holding capacity by 20000 gallons per hectare \cite{SOM2015}. 
However, it takes 5-7 years to see a perceptible increase in SOM at 0-30 cm soil layer \cite{soilsystems9030103}. Benefits could be lost rapidly (within 2-3 years) if the best practices are discontinued. Initial transition may temporarily reduce SOM due to increased microbial activity before the new inputs stabilize. In very sandy soils or extreme climates, SOM accumulation may be minimal despite best efforts.

\subsubsection*{Improved Water Infiltration and Retention}
Carbon farming practices dramatically improve the ability of the soil to capture and store water, reducing both waterlogging and drought stress. This is often more valuable than carbon credits in water-stressed regions. Conservation systems can increase infiltration rates by 50--100\% compared to conventional tillage. Cover crops typically reduce surface runoff by 30--70\% \cite{Madgoff2021, USDA_NRCS_2022_Infiltration_Runoff}. Each 1\% increase in soil organic carbon increases available water capacity by 15-20 mm per meter of soil depth. Alternate wetting and drying rice saves 25--35\% of irrigation water while maintaining yields.
The initial implementation may temporarily disrupt drainage in heavy clay soils. The practice requires a learning curve for irrigation management. The benefits could take 2-3 years to fully materialize as soil structure improves. In extremely arid regions, water benefits may be insufficient without supplemental irrigation \cite{Murtaza2025}.

\subsubsection*{Enhanced Nutrient Cycling and Reduced Fertilizer Dependence}
Biological processes in healthier soils mobilize nutrients from organic matter and mineral sources, reducing synthetic fertilizer requirements while maintaining or improving crop nutrition. For example, Legumes are highly valued in sustainable agriculture for their ability to conserve resources by fixing atmospheric nitrogen, a process that naturally enriches the soil with an estimated 20--60 kg of nitrogen for both the current crop and subsequent plantings in a farming system. Legume-based intensification, including rotation, intercropping, cover cropping, green manuring, pasture management, hedgerow planting, and other legume-based land practices, is promoted to improve crop yields by about 10--25\% \cite{LALOTRA202227}. Improved biological activity increases phosphorus and micronutrient availability, where biochar application boosts phosphate activity by 32\% \cite{soilsystems8020054}. Long-term organic matter accumulation creates a slow-release nutrient reservoir.
The transition period (2-4 years) may require maintaining or even slightly increasing fertilizer use while soil biology develops. Initial Carbon:Nitrogen ratio of crop residues (especially cereal straw) can cause temporary Nitrogen immobilization, requiring supplemental Nitrogen. The legume cover crops could occupy land/time that could grow cash crops.

\subsubsection*{Increased Microbial Biomass and Diversity}
Carbon farming practices nurture diverse soil microbial communities that drive nutrient cycling, disease suppression, and soil structure formation. For example, Microbial biomass can increase 30--100\% under conservation agriculture \cite{chen2020global,LI201850}. Fungal: bacterial ratios shift toward fungi, which create more stable carbon compounds. Mycorrhizal colonization increases 20--50\% \cite{nthebere2025conservation}, improving Phosphoros uptake and drought tolerance. Beneficial organisms (earthworms, beetles) could increase 50--200\% \cite{briones2017conventional}.
It could take 3-5 years for microbial communities to fully equilibrate. It may increase some pest organisms alongside beneficials (requires integrated pest management). Microbial activity can temporarily increase N$_2$O emissions during transition. Benefits are fragile – a single tillage event can undo years of microbial community development.

\subsubsection*{Reduced Soil Erosion}
Reducing soil erosion remains a central objective in the adoption of cover crops, as these systems prolong the period during which the soil surface is shielded by living vegetation or plant residues. Fields that maintain moderate levels of crop residues experience substantially lower water runoff and soil loss, with reductions of roughly half for runoff and up to four-fifths for erosion compared with bare soil, and cover cropping further strengthens this effect by markedly decreasing surface runoff and erosion, with global assessments reporting average reductions of about 67\% and 80\% under conservation systems \cite{du2022conservation} and, in some cases, erosion-control improvements approaching 90\% under well-managed cover-crop practices \cite{USDA_NRCS_2022_Infiltration_Runoff}. Ground cover from crop residues, cover crops, and improved soil structure dramatically reduces both water and wind erosion, preserving topsoil and preventing sedimentation of water bodies. 
On the flip side, residue retention can create seedbed challenges (requires specialized planters). Heavy residue loads may harbor pests and diseases. In high-rainfall areas with poor drainage, residue can cause waterlogging.

\subsubsection*{Enhanced Drought Resilience}
Cover crop adoption consistently enhances drought resilience, with multi-year use shown to improve yields even under moisture stress; for instance, farmers reported gains of 9.6\% in corn and 11.6\% in soybeans during the 2012 drought \cite{Clark2015}. Improved water-holding capacity, deeper root systems, and better soil structure help crops tolerate dry spells with reduced yield loss, as soils with higher organic matter can retain plant-available water for 7--14 additional days, and cover crops can increase rooting depth by 20 to 50\%, enabling access to deeper moisture reserves. These advantages become most visible in stressful years and are realized fully only when residue retention, minimal tillage, and cover cropping are implemented together, as partial adoption yields relatively modest benefits.

\subsubsection*{Temperature Moderation (Microclimate)}
Surface residue mulch moderates soil temperature extremes by reducing heat transfer into the soil, and studies show that mulched plots can be 5–8$^{\circ}$C cooler than bare or un-mulched soils during hot periods, as illustrated by \cite{Kamara1986} who observed temperatures of 40--44$^{\circ}$C at 5 cm depth in bare and un-mulched plots compared with about 35$^{\circ}$C under mulched conditions; this thermal buffering also lowers daily temperature fluctuations by 50--70\% and helps reduce heat stress on germinating seedlings in summer crops.
However, in cool seasons/regions, mulch may delay soil warming, slowing germination, and can create a favorable habitat for pests (rodents, slugs) in some contexts.

\subsubsection*{Enhanced Above-Ground Biodiversity}
Diversified cropping systems, reduced chemical inputs, and permanent ground cover create habitat for beneficial insects, birds, pollinators, and other wildlife. Flower-rich cover crops and vegetative strips contribute to greater pollinator activity by offering alternative foraging and nesting sites, particularly during periods when floral resources in the main crop are limited \cite{Haaland2011, Kremen2015}. Similarly, cover crops and reduced-tillage practices create favourable microhabitats for predatory arthropods such as ground beetles, spiders, and parasitoids, thereby strengthening natural pest-regulation processes \cite{Rodriguez2020, Frasier2016}. Agroforestry and other structurally diversified systems also tend to support richer bird communities by supplying more complex vegetation layers and stable foraging opportunities than monoculture fields \cite{Bhagwat2008}. While these ecological enhancements contribute positively to ecosystem functioning, increased vegetation complexity may also provide refuge for certain pests, underscoring the need for carefully designed and well-managed diversification strategies \cite{Landis2000}.

\subsubsection*{Enhanced Below-Ground Biodiversity}
Soil biodiversity underpins core ecosystem processes, with microbial and faunal communities driving nutrient cycling, soil aggregation, and plant health. Conservation practices such as diversified rotations, reduced tillage, residue retention, and cover crops generally enhance below-ground biological activity, and a recent global meta-analysis reports an average 21\% increase in microbial biomass under such systems \cite{PerezSanchez2023}. These practices also promote richer soil-faunal communities and improved mycorrhizal development, although increases may include some pathogens or parasites. Because soil biota are highly sensitive to disturbance and require several years to stabilise, even a single disruptive event can degrade community structure, underscoring the need for long-term, consistent management \cite{gomez2008}.

\subsubsection*{Reduced Input Usage}
Carbon farming substantially reduces dependence on purchased agricultural inputs such as fuel, synthetic fertilizers, pesticides, herbicides, and irrigation water, thereby improving farm profitability even in the absence of carbon payments. Studies \cite{soilsystems9030103, nikolic2021effect, fogliatto2021cover} show that reduced tillage alone can lower fuel expenditure by approximately \rupee 1,500--3,000 per hectare, while transitions to residue retention, improved nutrient cycling, and legume-based rotations can decrease synthetic fertilizer use by 20--50\% over time. Enhanced ecological regulation within carbon farming systems also contributes to lower chemical use, with pesticide requirements declining by as much as 50\% and natural weed suppression reducing herbicide use by 20--40\%. In water-intensive crops such as rice, improved soil structure and moisture retention translate into irrigation cost savings of roughly \rupee 5,000--15,000 per hectare. Collectively, these reductions highlight the capacity of carbon farming to minimize external input dependency while supporting more resilient and cost-effective production systems.

\subsubsection*{Reduced Methane Emissions}
Water management practices (AWD, DSR, SRT) dramatically reduce methane emissions from rice paddies, providing climate benefits beyond carbon sequestration. For example, AWD reduces methane emissions by 30--50\% \cite{GANGOPADHYAY2022}. DSR reduces methane emissions by 69.9\% \cite{REDDY2025100238}. The climate benefit translates to 0.5--2.0 tCO$_2$eq/ha/season from methane reduction alone.
However, it may slightly increase $N_2O$ emissions (less potent but still important). It will require careful water management, which is labor-intensive and knowledge-intensive.

\subsubsection*{Reduced Nitrous Oxide Emissions}
Precision fertilizer management and organic amendments reduce Nitrous oxide emissions, a potent greenhouse gas 273 times more powerful than CO$_2$ \cite{cui2024global}. The climate benefit can be as much as 0.2--0.8 tCO$_2$eq/ha/year from Nitrous oxide reduction.

\subsubsection*{Improved Farmer Health and Safety}
Reducing agrochemical use and increasing mechanization lowers both toxic pesticide exposure and injuries associated with manual labor. Conservation agriculture combined with integrated pest management (IPM) can achieve comparable reductions, while avoiding the recurrent incidence of tens of thousands of acute poisoning cases reported annually among cotton and vegetable producers.

\subsubsection*{Reduced Crop Residue Burning}
Retaining residues instead of burning eliminates massive air pollution, improving public health, especially in northern India. Studies attribute 25--30\% of New Delhi’s winter air pollution to crop residue burning \cite{PTI2023StubbleBurning}. Eliminating burning would significantly improve air quality and reduce the incidence of respiratory illnesses.

\subsubsection*{Climate Change Mitigation (Aggregate Impact)}
Beyond soil carbon sequestration, the combined effect of reduced emissions (methane, nitrous oxide, fossil fuels) and sequestration provides substantial climate benefit. Scientific studies show that carbon farming provides 2-4 times the climate benefit of soil carbon sequestration alone when the impact of all greenhouse gases is included.
\subsubsection*{Labor Savings}
No-till and reduced-till systems, along with direct seeding, substantially lower labor requirements—an important advantage in regions facing chronic rural labor shortages. In conventional rice cultivation, transplanting typically requires 60--80 person-days per hectare, whereas direct-seeded rice uses only 10--15 person-days per hectare, resulting in labor savings of 28--39 person-days per hectare \cite{soilsystems9030103}. At an average agricultural wage rate of \rupee 330 per person-day in India, this reduction translates to savings of roughly \rupee 9,240--12,870 per hectare. Although these efficiencies improve farm profitability, they may also contribute to labor displacement and reduce employment opportunities in rural areas. Moreover, the successful adoption of such practices often depends on access to mechanization, which introduces additional capital costs for smallholder farmers.

\section{Carbon Measurement, Monitoring, Verification (MRV), and Certification Frameworks}
\label{app:mrv}Accurate measurement and monitoring of carbon stocks and GHG emissions form the foundation of credible soil carbon initiatives and carbon farming programs.
This section presents the methodological foundations of credible carbon accounting. The section reviews MRV frameworks, measurement techniques, modeling approaches, and certification processes, with particular emphasis on uncertainty, spatial heterogeneity, cost–accuracy trade-offs, and scalability.

\subsection{Measurement, Reporting, and Verification} 
MRV in carbon farming is fundamentally a problem of estimating small, slow, and spatially heterogeneous changes in soil organic carbon under tight cost and credibility constraints.

MRV protocols typically follow a structured sequence that involves establishing applicability conditions, defining spatial and temporal project boundaries, delineating baseline and intervention scenarios, assessing additionality, and implementing systematic monitoring and reporting cycles \cite{fao2020protocol}.

 These MRV procedures ensure the credibility, transparency, and traceability of soil carbon initiatives. In agricultural landscapes, MRV systems must also capture spatial variability, management heterogeneity, and the inherently slow rate of change of SOC, while maintaining a balance between measurement accuracy, cost, and operational feasibility.

MRV systems, in the context of Carbon Projects, typically integrate three components:
\begin{enumerate}
  \item \textbf{Measurement}: quantification of SOC stocks and fluxes through direct measurement (soil sampling) or indirect measurement (remote sensing, modeling). 
  \item \textbf{Reporting}: Structured documentation of methods, assumptions, and results, aligned with IPCC (Inter Governmental Plan on Climate Change) or voluntary carbon standard requirements.
  \item \textbf{Verification}: Independent third-party assessment of the accuracy and conformity of the data with the standards.
\end{enumerate}


Depending on the purpose of the project, donor reporting, NDC (Nationally Determined Contribution) alignment, or participation in carbon markets, MRV systems vary in their methodological complexity, with higher-tier approaches such as direct measurement combined with process-based modeling typically required for the generation of certified carbon credits \cite{WORLDBANK2023}. 

Petropoulos et al \cite{Petropoulos2025}, in their systematic review of 86 studies on SOC assessment methods, report substantial methodological heterogeneity between sampling protocols, laboratory analytical techniques, and modeling approaches, which complicates cross-study comparability. Their review highlights that robust SOC assessment requires an integrated framework that combines field sampling, laboratory analysis, and modeling tools, while remote sensing provides valuable but inherently limited complementary information. 
The challenge of measurement uncertainty, further intensified by spatial and temporal variability, underscores the need for harmonized, transparent, and scientifically robust MRV frameworks, particularly in agricultural landscapes that exhibit heterogeneous management conditions.

\subsubsection*{Additionality}
Additionality is a foundational concept that ensures that credited emission reductions or removals would not have occurred in the absence of carbon finance. In carbon farming, demonstrating additionality is particularly challenging because many practices, such as reduced tillage or improved water management, may already be adopted for agronomic or economic reasons. Establishing additionality, therefore, requires careful definition of baselines, assessment of prevailing regional practices, and evaluation of financial, technological, or behavioral barriers. Precise computation of additionality is fundamental to high-integrity carbon farming projects.

 Additionality assessment, often supported by SOC models and standardized GHG estimation methods, is conducted prior to investing in resource-intensive field monitoring to ensure that projected climate benefits exceed business-as-usual outcomes. Once additionality is established, the monitoring is carried out through field-based soil sampling, laboratory analysis, and model-based projections, with periodic reporting used to document activity data, SOC measurements, and estimated emission changes.

\subsubsection*{Field-Based Soil Carbon Measurement}
Field-based measurement provides the most direct and credible approach to quantifying SOC stocks \cite{poeplau2017soil}. Standard protocols involve soil sampling at defined depths (commonly 0-30 cm and 30-60 cm), followed by laboratory analysis using methods such as dry combustion (e.g., elemental analyzers) to determine the concentration of SOC. The SOC stock is calculated using the following equation:

\begin{equation}
\text{SOC stock (Mg C ha}^{-1}) = \frac{C \times \rho_b \times d \times 10}{1000}
\end{equation}
\noindent
where $C$ is the SOC concentration (\%), $\rho_b$ is the bulk density (g cm$^{-3}$), and $d$ is the sampling depth (cm). Stratified sampling, random grids, or transects are used to account for spatial heterogeneity, and the sample size is determined based on desired confidence and variability. Periodic resampling at consistent locations supports temporal monitoring of SOC change.

\subsubsection*{Modeling Approaches}
Modeling approaches are critical for estimating SOC stocks and changes when direct measurements are unavailable, infrequent, or expensive. These models are broadly categorized into two types: \textbf{empirical} and \textbf{process-based}, each with distinct data requirements and applications.

\subsubsection*{Empirical Models}
Empirical models use statistical or machine learning models to predict SOC from site-specific or remotely sensed covariates.
\begin{equation}
\hat{Y}_{SOC} = f(X_1, X_2, \ldots, X_n) + \varepsilon
\end{equation}
where $\hat{Y}_{SOC}$ is the predicted SOC, $X_1, X_2, \ldots, X_n$ are predictors such as the NDVI (normalized difference vegetation index), topographic indices, or climate variables, and $\varepsilon$ is the residual of the model. These models are used in IPCC Tier 1/2, FAO EX-ACT, and digital soil mapping tools. Tier 1 uses default emission factors, while Tier 2 incorporates empirical models and country-specific factors, providing a lower-cost, large-area assessment. These models often require calibration using local field data~\cite{brungard2015}.

\subsubsection*{Process-based Models}
\textbf{Process-based models} simulate the turnover of SOC by representing carbon inputs, decomposition, microbial activity, and environmental interactions \cite{colemanrothc1996}:
\begin{equation}
SOC_{t+1} = SOC_t + C_{\text{input}} - (k_d \cdot SOC_t)
\end{equation}
where $C_{\text{input}}$ is the annual carbon input (for example, crop residues, manure) and $k_d$ is the decomposition rate constant. Process-based models like \textit{RothC} and \textit{CENTURY} divide SOC into active, slow, and passive pools, each governed by first-order kinetics ~\cite{colemanrothc1996}. These models are integrated in certified MRV protocols such as \textit{Verra VM0042} under the Measure-and-Model approach.

The selection of models depends on the desired spatial scale, temporal resolution, and data availability. Empirical models are easier to implement but limited in extrapolation, while process models capture SOC dynamics more accurately, but require robust input datasets and calibration.

\subsubsection*{Remote Sensing and Digital MRV}

Remote sensing and digital MRV (dMRV) provide scalable frameworks for soil SOC monitoring, especially in fragmented smallholder systems. SOC is commonly estimated through spectral reflectance, where soil organic matter absorbs strongly in the visible and near-infrared (VNIR) spectrum. A generalized estimator is the following.
\[
\hat{SOC} = f(R_{\lambda_1}, R_{\lambda_2}, \ldots, R_{\lambda_n}) + \varepsilon,
\]
with $R_{\lambda}$ denoting reflectance at wavelength $\lambda$, $f(\cdot)$ a regression or machine learning function (e.g., PLSR, random forests), and $\varepsilon$ model residuals. Performance is typically evaluated using the root mean square error (RMSE) and the coefficient of determination $R^2$. Fusion of VNIR/SWIR hyperspectral data with synthetic aperture radar (SAR) backscatter improves predictive accuracy under varying soil moisture and canopy conditions \cite{gomez2008}. 

Calibration is achieved through stratified soil sampling and laboratory analysis. Model calibration requires regression between measured $SOC_{stock}$ and predicted $\hat{SOC}$, with cross-validation applied to minimize overfitting. Eddy covariance towers and in-situ sensors provide complementary high-frequency data streams for model refinement.

Digital MRV integrates these calibrated models with automated data pipelines \cite{verra2023dMRV}. Core functions include parcel-level identifiers to prevent double-counting, automated detection of canopy change and land-use conversion, and geotagged farmer-submitted records. Blockchain-enabled registries and digital carbon wallets allow tamper-proof issuance and direct benefit transfer. Verra and Gold Standard are piloting digital submissions, reflecting a shift toward automated verification \cite{verra2024vcs}.

The World Bank's (2023) comprehensive MRV Sourcebook for Agricultural Landscapes \cite{WORLDBANK2023} acknowledges that while technological advances in remote sensing and modeling have reduced per-hectare monitoring costs, verification of carbon stock changes still requires significant investment in producing ground-truth data and in quality assurance, particularly for smallholder systems with high spatial heterogeneity.

\subsection{Verification and Certification}

Verification and certification anchor the credibility of carbon projects. Independent Validation and Verification Bodies (VVBs) audit projects to ensure compliance with approved methodologies. Reported SOC changes are quantified as follows:

\[
\Delta SOC = (SOC_{t} - SOC_{0}) - (SOC_{t}^{BL} - SOC_{0}^{BL}),
\]
where $SOC_{t}$ and $SOC_{0}$ denote project stocks at time $t$ and baseline, and $SOC^{BL}$ represents the counterfactual baseline scenario. Leakage, the shift of emissions from the project area to outside the project boundary, must also be prevented and accounted for.

The uncertainty assessment is integral. A common approach propagates errors from bulk density ($\sigma_{\rho_b}$), carbon concentration ($\sigma_C$), and depth ($\sigma_d$):

\[
\sigma_{SOC}^2 \approx \left( \frac{\partial SOC}{\partial \rho_b} \sigma_{\rho_b} \right)^2 + 
\left( \frac{\partial SOC}{\partial C} \sigma_{C} \right)^2 +
\left( \frac{\partial SOC}{\partial d} \sigma_{d} \right)^2.
\]
This is crucial because certification standards often require conservative uncertainty deductions from the total credits issued, ensuring the buyer receives only reliably quantified climate benefits. Certification standards require uncertainty deductions where the 90\% confidence interval exceeds a defined threshold.

\subsubsection*{Certification Standards: Verra and Gold Standard}
Verra and Gold Standard are two
leading organizations that develop and manage certification standards for climate
action projects, ensuring the credibility and transparency of GHG emission reductions and removals. Verra's verified carbon standard (VCS) provides methodologies such as VM0042 (Improved Agricultural Land Management) and VM0044 (Biochar Use), emphasizing rigorous baselines and transparent monitoring \cite{verra2024vcs}. It is the world’s largest voluntary GHG crediting program, with more than 4,000 registered projects in 95 countries. The Gold Standard focuses on high-integrity credits with strong links to Sustainable Development Goals (SDGs) \cite{goldstandard2020}. Its methodologies cover diverse areas such as the reduction of methane in rice cultivation (through alternate wetting and drying, shortened flooding, aerobic rice, and direct seeding) and SOC enhancement (through zero tillage, cover cropping, and managed pastures). Both require adherence to principles of additionality, permanence, leakage prevention, and transparency. Upon successful verification, serialized credits are issued to registries, ensuring traceability and avoiding double-counting. Other emerging standards, such as the Social Carbon Nature Standard and SCM0005, align closely with Verra’s VCS by emphasizing regenerative agriculture and soil organic carbon monitoring through measured and modeled approaches.

\subsubsection*{Permanence}
Permanence refers to the durability of carbon sequestration achieved through carbon farming best practices, such as soil carbon enhancement, agroforestry, or reforestation. For carbon credits to have climate value, the sequestered carbon must remain stored for long time horizons, perhaps at least 25 to 50 years. However, biological carbon pools are inherently reversible. Episodes such as drought, wildfire, pest outbreaks, land-use change, or even management practices can potentially release previously stored carbon back into the atmosphere. This is called the permanence risk.

Permanence risk quantification seeks to estimate the likelihood and magnitude of carbon reversal over a given period. This  involves assessing risks of various types: (1) biophysical risks (climate variability, soil type, vegetation);  (2) management risks (farmer behavior, etc.); and (3) external risks (policy changes, market pressures).  The different carbon standards usually address permanence risk by applying buffer pools or discount factors, where a portion of credited carbon is set aside to insure against future reversals.

Effective permanence risk quantification improves the credibility of carbon farming by aligning credited sequestration with realistic long-term storage expectations. Risk quantification is important for transparency, comparability across projects, and investor confidence. As uncertainties of various kinds increase,  robust methods for permanence risk quantification  are becoming a central requirement for credible carbon farming programs.

\subsubsection*{Leakage}
Leakage occurs when emissions reductions or carbon sequestration achieved within a carbon farming project boundary cause increased emissions outside that boundary. This could partially or even fully offset the climate benefit. In carbon farming, the main reasons for leakage could be: (a) commonly  activity shifting (for example, agricultural production displaced to other land); (b) market leakage (for example,  reduced supply driving production elsewhere). For example, converting cropland to regenerative practices could push food production to new areas, potentially leading to land-use change emissions.

Leakage quantification is important and seeks to estimate the scale of such indirect emissions. Approaches for leakage quantification typically use  models that use location specific land-use data and  production statistics.  Some carbon standards apply conservative leakage deductions. Other carbon standards  define project eligibility rules to minimize leakage risk upfront, such as restricting participation to degraded lands or maintaining baseline production levels.

In any case, accurate leakage quantification is critical to ensuring net climate benefits. Overlooking leakage can result in over-crediting and undermining the credibility of carbon farming. While precise measurement is difficult, transparent assumptions, conservative accounting, and continuous monitoring help manage leakage risk and uphold the integrity of carbon farming initiatives.

\subsection{Carbon Crediting Methodologies for Agricultural Land Management (ALM)}

ALM methodologies, also known as “regenerative agriculture carbon credit methodologies,” are a subcategory within Agriculture, Forestry, and Other Land Use (AFOLU)
methodologies. These methodologies specifically address emissions reductions and
carbon sequestration from changes in agricultural practices, such as soil carbon
enhancement, improved crop management, and livestock-related interventions. 

They are designed to reward farmers and land managers for adopting regenerative practices that build organic matter in the soil, reduce the reliance on synthetic inputs, and improve resilience to climate change. The emphasis is on generating measurable, reportable, and verifiable climate benefits while also enabling sustainable agricultural transitions.

\subsubsection*{VM0042 Methodology} 
VM0042 (Improved Agricultural Land Management) of Verra offers three main routes: (i) \emph{measure-and-model}, combining soil sampling with calibrated process-based models such as RothC, CENTURY, and DNDC; (ii) \emph{measure-and-re-measure}, relying solely on repeated field sampling and laboratory analysis; and (iii) \emph{default emission factors}, which apply IPCC-derived coefficients. Recent updates to VM0042 have strengthened guidance on baseline determination, model calibration, and leakage accounting, improving applicability across diverse agro-ecosystems \cite{verra2024vm0042}. Baselines can be set using either static approaches (historical land-use data, typically covering 5 years) or dynamic approaches (periodically updated with remote sensing, surveys, or national statistics). This flexibility helps balance rigor with practicality, particularly in regions with evolving farming practices.

The methodology also emphasizes uncertainty deductions, stratification of project areas by soil type, climate, and management system, and requires Independent Modeling Experts (IMEs) to review model calibration and parameterization. Together, these measures safeguard against over-crediting and enhance scientific credibility.

\subsubsection*{VM0044 Methodology} 
VM0044 (Biochar Utilization) of Verra quantifies durable removals from stable carbon stored in biochar, avoided CH$_4$ and N$_2$O emissions from biomass diversion, and soil flux adjustments following application. Biochar projects require monitoring of feedstock type, pyrolysis conditions, and soil incorporation, ensuring permanence over 100+ years \cite{verra2025vm0044}.

Two project components are explicitly recognized:
\begin{itemize}
    \item \textbf{Biochar production:} conversion of agricultural residues or other waste biomass into biochar through controlled pyrolysis, preventing emissions from open burning or decomposition.
    \item \textbf{Biochar application:} incorporation of biochar into soils where it provides long-term carbon storage and may also reduce fertilizer needs.
\end{itemize}
The accounting of GHG under VM0044 is multifaceted, capturing both removals and avoided emissions. It also accounts for potential changes in soil GHG fluxes and requires conservative deductions wherever uncertainty exists. By integrating waste management with long-term carbon sequestration, VM0044 delivers a dual climate benefit.

\subsubsection*{Gold Standard SOC Framework} The Gold Standard SOC framework provides a flexible pathway where direct soil sampling can be combined with peer-reviewed models and conservative defaults. It explicitly integrates SDG co-benefits, making it attractive for projects prioritizing livelihood and biodiversity outcomes alongside carbon \cite{goldstandard2020}.

It requires setting baselines based on the continuation of the past five years of management practices, with stratification of fields by soil type, climate zone, cropping system, and input levels. Quantification may rely on (a) direct SOC measurement through statistically robust sampling, (b) peer-reviewed models or datasets validated for the local context, or (c) conservative default factors where data is lacking.

Unlike Verra’s more rigid frameworks, Gold Standard explicitly recognizes and credits social and ecological co-benefits, ensuring projects contribute to poverty reduction, food security, gender equality, and biodiversity. This makes it highly attractive for development-focused agricultural projects, particularly in smallholder systems.

For smallholder-dominated contexts, hybrid approaches are practical: default factor methods for broad areas, with calibration plots intensively monitored to improve credibility. Grouped projects, where many farmers join under a single methodology, further reduce transaction costs and allow standardized verification.

\subsection{Toward Scalable and Inclusive Monitoring}

Moving from methodological rigor to practical deployment, the final challenge for soil carbon MRV is the transition to scalable and inclusive monitoring that integrates smallholders into the market.

\textbf{Cost Efficiency.} dMRV platforms reduce per-hectare monitoring costs by automating remote sensing analysis, integrating machine learning, and minimizing dependence on extensive field sampling. They also enable near real-time detection of land-use change, canopy cover dynamics, and crop rotations, thereby improving transparency and responsiveness \cite{verra2023dMRV}.

\textbf{Aggregation Models.} Smallholder plots are typically too fragmented to generate cost-effective credits. Farmer-Producer Organizations (FPOs) and cooperatives enable pooling of land, creating economies of scale, lowering per-farmer costs, and simplifying monitoring. Such aggregation also improves bargaining power in carbon markets and reduces transaction costs in verification.

\textbf{Integration with National Programs.} Embedding MRV within national initiatives such as India’s Soil Health Card program and ISRO’s satellite missions enhances data credibility. The Carbon Credit Trading Scheme (CCTS) provides a regulatory framework to align voluntary MRV systems with compliance markets. This convergence strengthens trust among buyers and ensures compatibility with future climate policy instruments.

\textbf{Equity and Inclusion.} Transparent benefit-sharing mechanisms are essential. Carbon wallets linked to geotagged farmer identities ensure direct transfers, while advisory services and training address digital literacy gaps. Digital interfaces in the local language strengthen inclusivity. The emphasis of the Gold Standard on the alignment of the SDG illustrates how MRV can support social and environmental outcomes \cite{goldstandard2020}. Ensuring gender equity and prioritizing marginalized farmer groups further enhances the legitimacy of these systems.

By coupling advanced digital MRV with aggregation and equity safeguards, MRV systems can evolve into robust national infrastructures that credibly integrate smallholder farmers into global carbon markets. In the long run, such frameworks can serve as the backbone for scalable, transparent, and inclusive climate finance in the agricultural sector.


\section{Carbon Markets and Trading}\label{sec5}
This section positions carbon farming within the evolving architecture of carbon markets. The section describes the generation, valuation, and exchange of carbon credits across voluntary and compliance systems, and discusses governance challenges related to integrity, additionality, permanence, and market credibility.

{Carbon trading} involves a market-based mechanism in which carbon emissions allowances or carbon offset credits are treated as tradable commodities. This approach creates financial incentives for emissions reduction, providing an effective method of achieving environmental goals. By harnessing market dynamics, carbon trading promotes optimal resource allocation, encouraging organizations and businesses to adopt cleaner technologies and sustainable practices while balancing economic growth and environmental stewardship.

The origins of market-based environmental solutions can be traced to the 1960s, but a significant proof of concept emerged in the 1990s with the US acid rain program. This pioneering initiative successfully demonstrated the effectiveness of cap-and-trade systems by achieving substantial reductions in $SO_2$ emissions.

The adoption of the 1997 Kyoto Protocol \cite{UNFCCCKyoto1997} marked the establishment of the first global framework for carbon trading through the introduction of three key mechanisms.

\begin{itemize} 
\item \textbf{Clean Development Mechanism (CDM)}: Facilitates project-based partnerships between developed and developing countries to achieve emission reductions. 
\item \textbf{Joint Implementation (JI)}: Enables collaborative emissions reduction projects between developed countries. 
\item \textbf{International Emissions Trading (IET)}: Allows for trading of surplus emission allowances among countries. 
\end{itemize}
A significant milestone in carbon trading occurred in 2005 with the launch of the European Union Emissions Trading System (EU ETS), which became the world’s first significant carbon market. The EU ETS demonstrated the practical viability of emissions trading by incentivizing the reduction of emissions while stimulating investment in clean technologies.

The 2015 Paris Agreement \cite{ParisAgreement2016} further advanced the role of carbon trading in international climate action. The Paris Agreement allowed countries and other entities to trade emission allowances or carbon credits and established frameworks for market-based approaches, enabling countries to achieve their nationally determined contributions (NDCs) via carbon credit trading and setting guidelines for global cooperation in carbon markets.


Carbon markets can be categorized into two major types: Compliance markets (which are government-administered) and voluntary markets (which are free markets). We describe them next.

\subsection{Compliance Carbon Markets}
{Compliance carbon markets} operate within government-mandated frameworks that require participation from specified sectors or entities. These markets are characterized by rigorous regulation that encompasses standardized procedures for trading, emission monitoring, verification, and comprehensive compliance reporting.

Several compliance markets have been established around the world. At the global level, the European Union Emissions Trading System (EU ETS) is the most mature. Recently, under Article 6 of the Paris Agreement, market-based mechanisms for carbon trading have been expanded, allowing for more flexibility in the international carbon market, particularly through Article 6.2, which introduces cooperative approaches. Article 6.4, known as the Sustainable Development Mechanism (SDM), establishes a framework for trading emission reductions between countries. These changes enable enhanced global cooperation, increasing the potential for international carbon credit transactions and fostering more robust emission reduction efforts.

The EU ETS has successfully reduced emissions in the covered sectors by approximately 50\% since its inception in 2005, meeting and exceeding its initial targets \cite{EUETS23}. The system has been expanded multiple times and now covers approximately 40\% of the total greenhouse gas emissions of the EU. The EU ETS also introduced the auction of allowances, increasing transparency and market efficiency. In 2021, the European Commission proposed a revision to tighten the cap and accelerate emissions reductions, aligning with the EU’s Green Deal and the goal of climate neutrality by 2050. Additionally, the Carbon Border Adjustment Mechanism (CBAM), implemented to prevent {carbon leakage}, is a significant achievement in global climate leadership.

In the USA, California’s Cap-and-Trade Program has demonstrated the viability of emissions trading within the United States. China’s National Emissions Trading Scheme is currently the world’s largest national carbon market by the number of covered emissions. The Regional Greenhouse Gas Initiative (RGGI) exemplifies successful regional cooperation in the USA, while the Korea Emissions Trading Scheme represents a successful implementation in Asia.

\begin{figure}[htbp]
    \centering
    \includegraphics[width=0.9\linewidth]{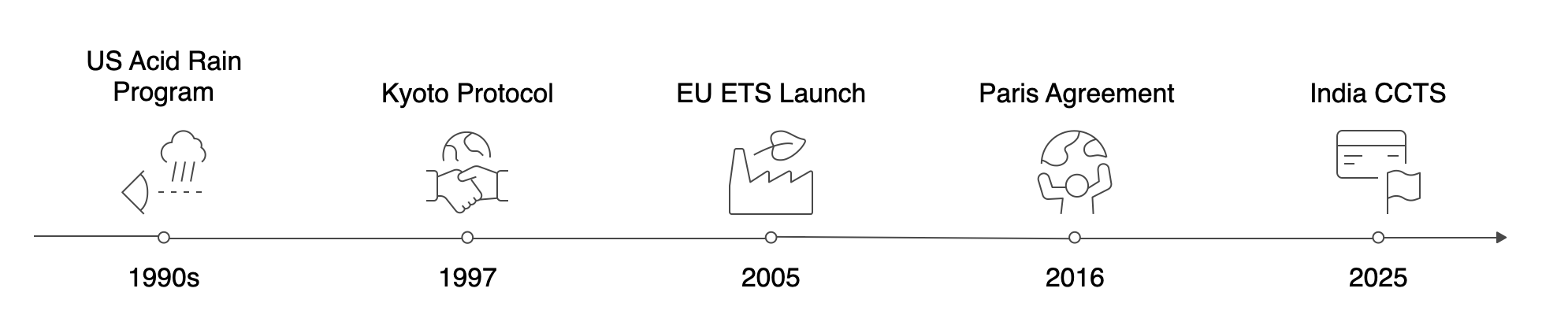}
    \caption{Evolution of carbon markets from the Kyoto Protocol to India's CCTS}
    \label{fig:market_timeline}
\end{figure}

India has introduced its own Emissions Trading System (ETS) under the Carbon Credit Trading Scheme (CCTS), marking a major step toward a national compliance carbon market. Unlike cap-based systems, India’s ETS is rate-based, assigning emission intensity benchmarks to industries rather than absolute caps. Facilities exceeding their targets must purchase credits or face penalties, while those below the benchmarks earn tradable carbon credits. It is crucial to note that agriculture is excluded from this ETS, which currently focuses on industrial sectors. The system is based on the previous Perform, Achieve, and Trade (PAT) scheme and is supervised by the National Steering Committee for the Indian Carbon Market (NSCICM) \cite{PIB25}. Aligning with Article 6 of the Paris Agreement, India's ETS sets the stage for international market links in the future, marking the latest step in the evolution of carbon markets shown in Fig. \ref{fig:market_timeline}.

\subsection{Voluntary Carbon Markets}
Voluntary carbon markets function independently of regulatory oversight, propelled mainly by corporate social responsibility goals and voluntary commitments to sustainability. These markets allow organizations and individuals to offset carbon emissions by purchasing credits from third-party verified emission reduction or removal projects.

Flexible participation criteria and a diversity of project types mark voluntary markets. Market actors may select from various recognized standards, with market supply and demand establishing credit prices. This flexibility has spurred innovation in methodologies and project development, expanding the range and impact of carbon offset projects.

\begin{figure}[ht]
    \centering
    \includegraphics[width=\linewidth]{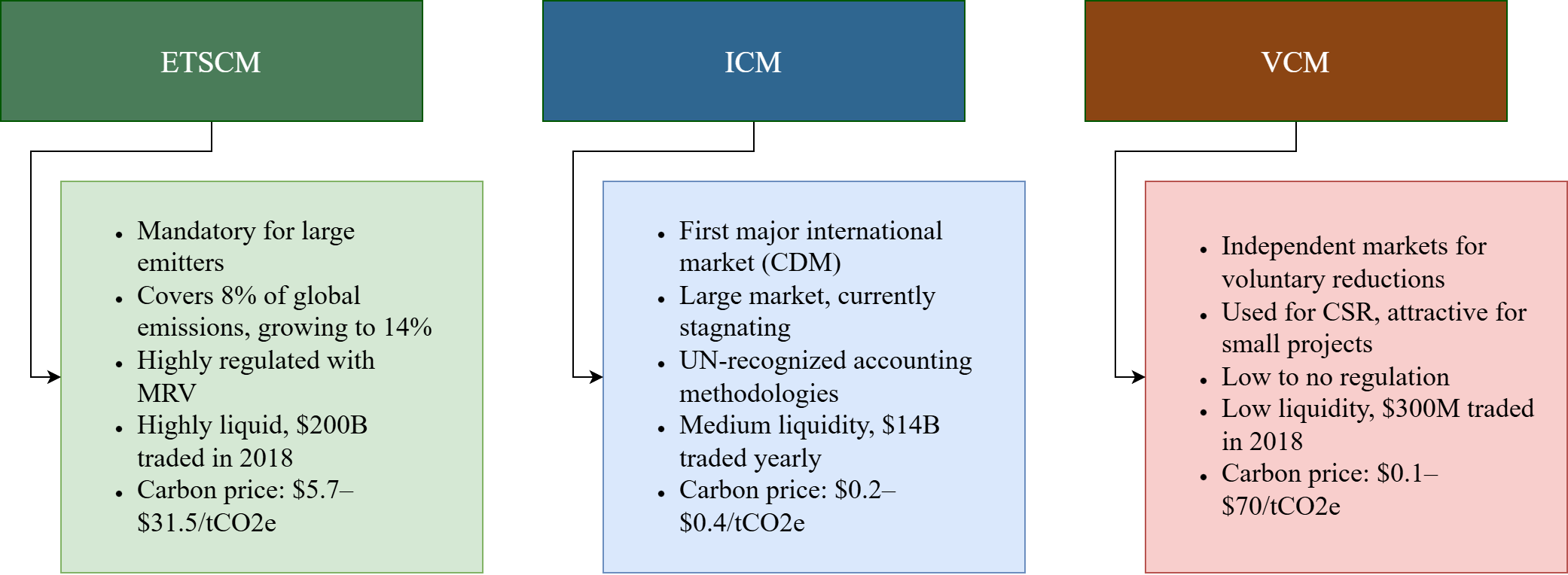}
    \caption{Comparison of emission trading scheme carbon markets (ETSCM), international carbon markets (ICM) and voluntary carbon markets (VCM)}
    \label{fig:enter-label}
\end{figure}

\subsection{Market Participants and Strategies}
The voluntary carbon market ecosystem is composed of two primary actor groups: \textbf{Direct Buyers} (Demand Side) and \textbf{Intermediaries} (Market Infrastructure). The strategies employed by these actors are diverging based on sector, regulatory pressure, and scrutiny regarding credit integrity.

\subsubsection*{Demand Side: Buyer Strategies}
Corporate buyers typically purchase credits to offset residual emissions, meet 'Net Zero' commitments, or manage transitional risks. Current market behavior reveals two distinct strategic archetypes:

\subsubsection*{The Energy and Industrial Strategy (Volume and Compliance)}
Major energy and industrial corporations often prioritize large-scale, cost-effective credits to offset hard-to-abate emissions. Their strategy frequently focuses on nature-based avoidance projects (e.g., forestry) and prepares for future compliance obligations under schemes like CORSIA (aviation) or Paris Agreement Article 6.

\subsubsection*{The Technology Strategy (Quality \& Removals)}
Technology companies, facing different reputational pressures and lower direct emissions intensities, often prioritize 'high-quality' carbon removal credits over avoidance credits, even at a significant price premium.

\subsubsection*{Market Infrastructure: Intermediaries}
Intermediaries facilitate liquidity, price discovery, and quality assurance. They bridge the gap between fragmented project developers and large corporate buyers.

\subsubsection*{Exchanges and Trading Platforms}
These entities provide the infrastructure for standardized spot trading, settlement, and clearing, moving the market away from opaque over-the-counter (OTC) deals.

\subsubsection*{Brokers and Project Advisors}
These firms combine trading with advisory services, often helping corporates design net-zero strategies while sourcing specific credits.

\subsubsection*{Digital and Tokenized Markets}
A new class of intermediaries is leveraging blockchain to enhance transparency and fractionalize ownership of credits.

\subsection{Market Design Considerations}
Several key design considerations determine the efficiency and effectiveness of carbon markets. We briefly describe them below.

\subsubsection*{Price Management and Discovery}
Achieving effective price discovery is critical. The mismatch between the supply of carbon credits from agricultural/land management and the demand from buyers can lead to volatility. Mechanisms such as price floors and liquidity reserves are essential for stability.

\subsubsection*{Quality Assurance and Integrity}
The credibility of carbon markets relies on rigorous MRV (Measurement, Reporting, and Verification). Issues such as 'double counting' (where the same credit is claimed by two entities) and 'permanence' (the risk of reversals via fire or tillage) remain central challenges. High-quality assurance measures and third-party ratings are becoming prerequisites for buyer participation.

\subsubsection*{Buyer Trust}
Buyers often lack knowledge regarding agriculture-based carbon credits. Concerns about changing methodologies and data reliability can dampen demand. Ensuring that credits represent real, additional, and permanent sequestration is vital for maintaining the flow of finance from the demand side (corporates) to the supply side (farmers).


\section{Representative Case Studies of Carbon Farming}\label{sec7}

This section presents six carefully selected, representative case studies of carbon farming initiatives. These case studies provide a comparative lens through which the practical realities of carbon farming can be examined. The cases illustrate a certain configuration of practices, MRV approaches, institutional arrangements, and market linkages, thereby revealing how design choices influence effects on the environment, farmer participation, transaction costs, and scalability.

\subsection{Kenya Agricultural Carbon Project (KACP)}

The Kenya Agricultural Carbon Project (KACP)\cite{WorldBank2012,ViAgro2014,Shames2012} in Western Kenya is one of the first large-scale efforts to convert smallholder Sustainable Agricultural Land Management (SALM) practices into certified soil-carbon credits. Implemented by Vi Agroforestry with the support of the World Bank BioCarbon Fund, the project operates in about 45{,}000 ha in the Lake Victoria basin and involves nearly 60{,}000 smallholders cultivating maize–bean systems in plots typically smaller than 2.5 ha. Farmers are organized into groups whose SALM-related gains in soil organic carbon (SOC) are aggregated and certified under the Verified Carbon Standard (VCS). The project runs over a 20-year crediting period beginning around 2009 and was among the first to pilot the VM0017 SALM methodology, a precursor to the current Verra VM0042 framework.

\subsubsection*{Project Overview}

KACP operates in the Lake Victoria basin in Western Kenya under the leadership of Vi Agroforestry, with technical and financial support from the World Bank BioCarbon Fund. Its scope spans 45{,}000 \,ha and around 60{,}000 farmers who participate through organized groups. The project applies the VCS methodology VM0017 to quantify SOC gains and is designed to generate carbon credits over two decades, starting in 2009.

\subsubsection*{Context and Challenges}

Before the project began, soils in the region were increasingly degraded due to repeated maize–bean cultivation, residue removal, and erosion. Rainfed production systems were exposed to frequent dry spells, producing unstable yields. Low input use, limited credit access, and widespread poverty further constrain productivity. The prevalence of highly fragmented land holdings made coordination, training, and monitoring particularly costly and difficult. Against this backdrop, KACP positioned SALM not only as a carbon-offset mechanism but as an integrated approach to improve both farm productivity and soil-carbon sequestration.

\subsubsection*{SALM Practices Implemented}

The project promotes a standardized SALM package \cite{ViAgro2014} that combines soil-  management, reduced soil disturbance, nutrient recycling, legume-based diversification, agroforestry, and soil-water conservation structures. These practices aim to rebuild soil organic matter, enhance water retention, stabilize yields, and increase on-farm biomass. Farmers adopt different combinations of these measures over time, depending on labor availability, risk preferences, and local agro-ecological conditions.

\begin{table}[ht]
    \centering
    \caption{Key SALM practices in the Kenya agricultural carbon project}
    \label{tab:kacp_salm_practices}
    \begin{tabular}{@{}p{3.8cm}p{9.2cm}@{}}
        \toprule
        \textbf{Practice} & \textbf{Objective / Benefit} \\
        \midrule
        Residue retention and mulching & Return biomass to the soil, protect the soil surface, and reduce erosion. \\
        Reduced / minimum tillage & Limit disturbance, preserve soil structure, and lower fuel requirements. \\
        Composting and improved manure management & recycle nutrients, build SOC, and increase fertility. \\
        Diversified rotations with legumes & Improve soil cover, enhance nitrogen fixation, and break pest and disease cycles. \\
        Agroforestry tree planting & Increase biomass, supply fuelwood and fodder, and add to carbon stocks. \\
        Soil and water conservation (contours, terraces) & Improve infiltration, reduce run-off and stabilise sloping fields. \\
        \bottomrule
    \end{tabular}
\end{table}

\subsubsection*{MRV Approach and Carbon Credits}

KACP introduced an MRV approach that combined farmer-recorded activity data with stratified soil sampling and regionally calibrated biophysical models \cite{Shames2012}. The farmer groups documented the SALM practices implemented in each plot, and the project staff used these records to form management strata from which the sample plots were selected for soil testing. Model simulations, informed by these data, linked observed practice changes to the dynamics of SOC, enabling estimation of emission reductions in a cost-effective manner. In 2014, the project became the first globally to receive soil-carbon credits under the VCS, with an initial issuance of about 25,000 tons of CO$_2$ \cite{WorldBank2012}. Subsequent verifications increased the total credited volume. More importantly, the project demonstrated that smallholder soil-carbon accounting can be conducted without prohibitively expensive sampling on each individual farm.

\subsubsection*{Impact of the Project}

Independent evaluations \cite{WorldBank2012,ViAgro2014} show that the project generated environmental, economic, and institutional benefits. The SALM plots exhibited higher SOC, improved soil structure, better moisture retention, and reduced erosion. Farmers reported more stable and, in many cases, higher yields of maize and beans. Although carbon income contributed only modestly to household revenue, it provided an additional incentive alongside agronomic gains. The project also strengthened farmer organizations, improved local technical capacity in conservation agriculture, and offered practical experience with participatory MRV systems.

\begin{table}[ht]
    \centering
    \caption{Impact of the Kenya agricultural carbon project}
    \label{tab:kacp_impacts}
    \begin{tabular}{@{}p{3cm}p{10cm}@{}}
        \toprule
        \textbf{Dimension} & \textbf{Key Impact} \\
        \midrule
        Environmental & Increases in SOC, improved soil structure, greater water-holding capacity, better ground cover, and reduced erosion. \\
        Economic & More stable and often higher yields of maize and beans, modest carbon income, and opportunities for diversification into trees and dairy. \\
        Social / Institutional & Stronger farmer groups, enhanced extension capacity in SALM, and experience with collective planning and participatory MRV. \\
        \bottomrule
    \end{tabular}
\end{table}

Farmers consistently emphasize that yield stability and reduced climate vulnerability are the most valuable outcomes, while carbon payments are seen as a useful but secondary benefit.

\subsubsection*{Limitations and Lessons}

Although pioneering, KACP also illustrates several challenges \cite{Shames2012}. Transaction and MRV costs remain high relative to the carbon revenue generated, which limits the share reaching individual farmers. The small per-farmer income from credit sales means that SALM adoption is driven primarily by agronomic benefits rather than carbon finance alone. Governance and equity concerns arise because the model relies heavily on an NGO aggregator, external funding, and asymmetries in information and bargaining power between farmers, project developers, and carbon buyers.

For India, the experience shows that smallholder SALM can be organized on a scale and validated through rigorous carbon standards, but it also highlights the importance of public or concessional support to reduce MRV and transaction costs. It further underscores the need for transparent, farmer-centered contract and benefit-sharing arrangements that ensure soil health and livelihood improvements remain the primary goals, with carbon revenue acting only as an additional incentive.

\subsection{The REDD Carbon Project (South-East Asia)}
The Reducing Emissions from Deforestation and Forest Degradation (REDD$^+$) initiative \cite{TransactionCosts} is designed to conserve 300000 hectares of mature tropical, humid, evergreen, and deciduous forests in Southeast Asia threatened by logging and migration. Adherent to the Verified Carbon Standard (VCS) Methodology VM0006, the 20-year project is critical to quantifying emissions reductions from avoided deforestation and anticipates generating 12 million Verified Carbon Units (VCUs) over its lifetime.




\subsubsection*{Transaction Costs}

Transaction costs represent the financial outlays necessary to define, establish, maintain, and transfer carbon credits. These costs are central to the viability and profitability of carbon offset projects and can significantly affect net returns. Understanding the structure and magnitude of these costs is essential for evaluating the economic feasibility of forest conservation through carbon markets.

Transaction costs represent the financial outlays necessary to define, establish, maintain, and transfer carbon credits. These costs are central to the viability and profitability of carbon offset projects. The breakdown of the REDD project, detailed in Table~\ref{tab:costs}, reveals a substantial structural bias.



\begin{table}[ht]
    \centering
    \caption{Transaction costs for the REDD project}
    \label{tab:costs}
    \begin{tabular}{@{}lrr@{}}
        \toprule
        \textbf{Cost Category} & \textbf{Cost (USD)} & \textbf{Percentage of Total Costs} \\
        \midrule
        Search Costs                 & 2,000        & 0.01\%   \\
        Feasibility Studies          & 18,000       & 0.07\%   \\
        Negotiation Costs            & 85,000       & 0.32\%   \\
        Monitoring Costs             & 840,000      & 3.16\%   \\
        Regulatory Approval Costs    & 1,992,000    & 7.50\%   \\
        Insurance Costs              & 23,640,000   & 88.95\%  \\
        USD/ha & 89 & \\ 
        USD/t $CO_2$ & 2.21 & \\
        \midrule
        \textbf{Total Transaction Costs} & \textbf{26,577,000} & \textbf{100\%} \\
        \bottomrule
    \end{tabular}
\end{table}

Insurance costs dominate the transaction cost structure, comprising a remarkable 88.95\% of all transaction costs. This insurance protects against the risk of carbon credit reversals due to unforeseen events (forest fires, pest outbreaks) or policy changes. This high cost directly reflects the need to ensure the long-term stability of the carbon sequestered in forestry projects.

\subsubsection*{Economic Analysis}

The transaction costs were evaluated per hectare and per ton of CO$_2$ sequestered, as tabulated in Table~\ref{tab:costs}. per hectare, transaction costs amount to \$89 per hectare across the 300,000-hectare project area. When normalized by the carbon impact, costs equal \$2.21 per ton of CO$_2$ sequestered over the project lifetime.

\begin{figure}[ht]
    \centering
    \includegraphics[width=0.9\linewidth]{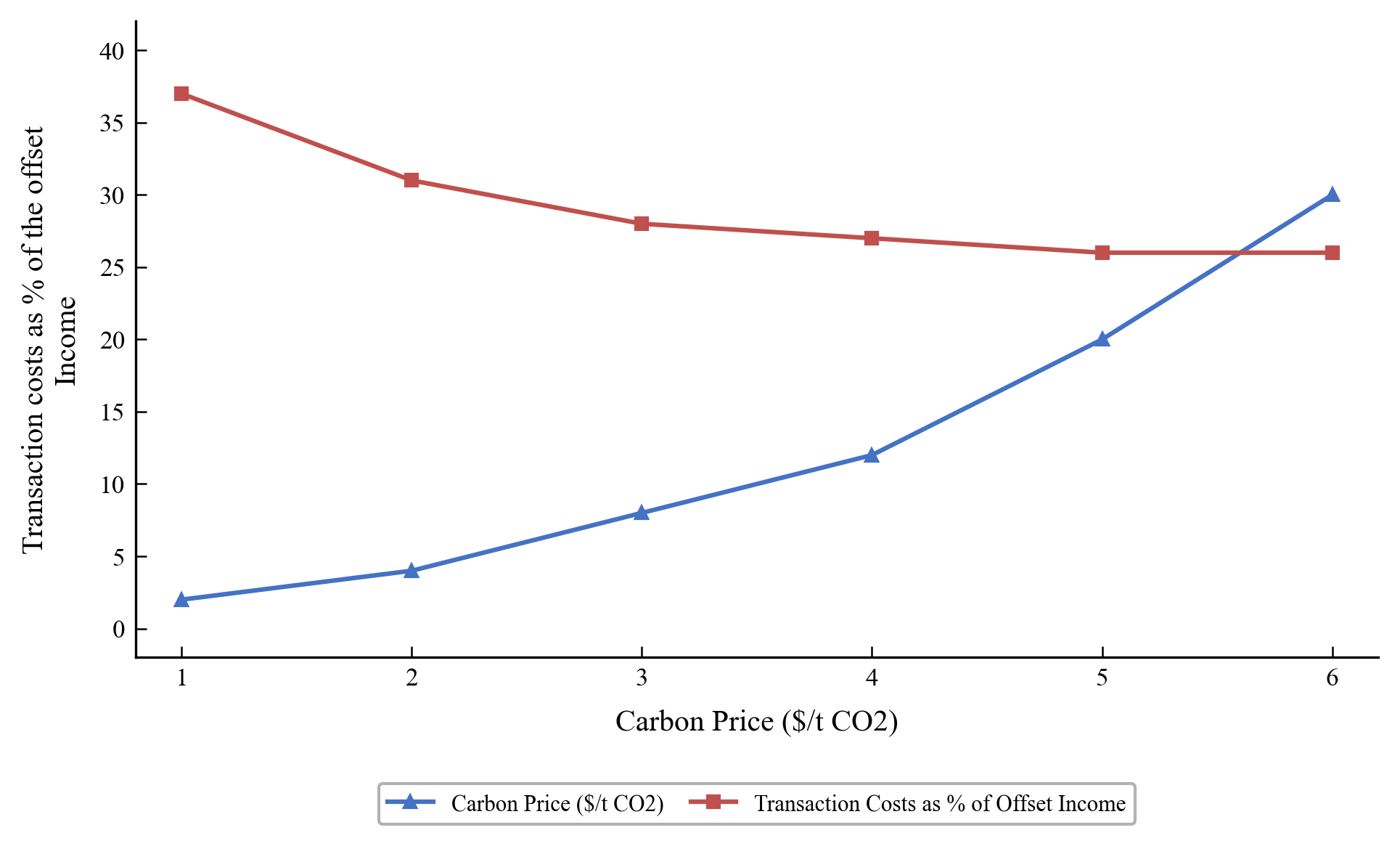}
    \caption{Relationship between carbon price and transaction costs as a percentage of offset income}
    \label{fig:relationship_transaction_costs}
\end{figure}

Transaction costs play a critical role in determining the income earned by project developers and the net benefit of climate mitigation achieved per dollar invested. When the carbon price is relatively low, high transaction costs can significantly reduce the net benefits for carbon credit owners, potentially rendering projects economically unviable. As the price of carbon credits increases and transaction costs represent a smaller proportion of total income, the profitability of the project improves substantially. The relationship between carbon prices and transaction costs is illustrated in Fig.~\ref{fig:relationship_transaction_costs}, which demonstrates how project economics shift across different carbon price scenarios.

\begin{figure}[ht]
    \centering
    \includegraphics[width=0.95\linewidth]{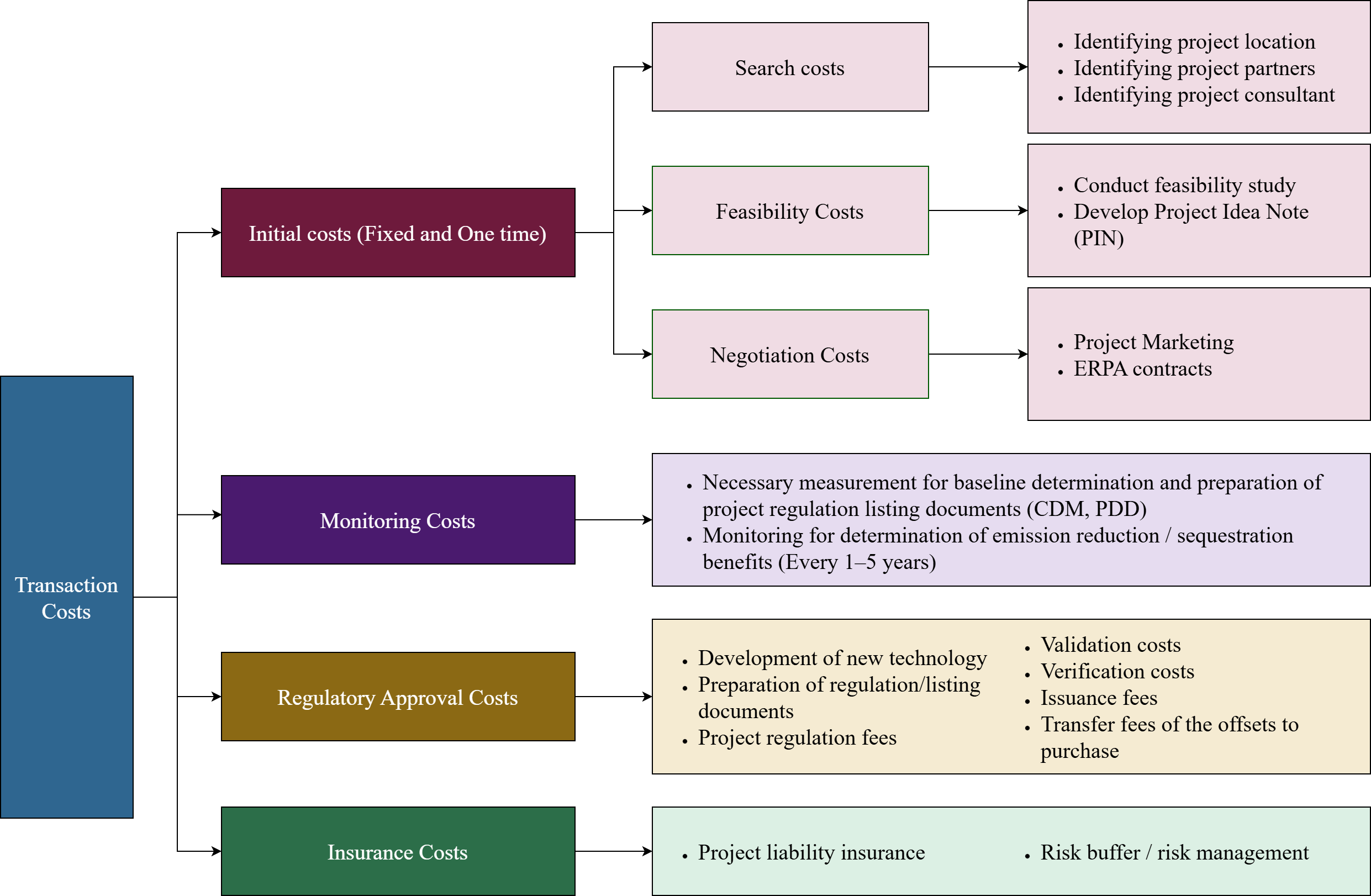}
    \caption{Breakup of costs associated with the REDD Project}
    \label{fig:cost_breakdown_carbon_project}
\end{figure}

The REDD project highlights the substantial influence of transaction costs on the economics of carbon sequestration projects in Southeast Asia. Insurance accounts for nearly 89\% of total transaction costs, making it the single most significant cost component and a critical factor in project design and financial planning. To ensure the long-term viability of REDD projects, transaction costs and carbon credit prices must be carefully balanced. Streamlining regulatory approval processes and reducing insurance costs through mechanisms such as pooled risk arrangements or government backstops could significantly increase the profitability of such projects while maintaining environmental integrity. A comprehensive breakdown of all costs associated with the REDD Project is presented in Fig.~\ref{fig:cost_breakdown_carbon_project}, providing additional details on the financial structure of this forest conservation initiative.

\subsubsection*{Future Prospects}

This initiative offers a scalable and innovative model for climate-resistant agriculture, promoting {sustainable farming} practices and improved livelihoods through community-driven market mechanisms. Despite substantial progress, challenges persist, particularly in terms of farmer adoption, financing, and awareness. To address these ongoing barriers, the project proposes to further develop digital knowledge-sharing platforms that can reach farmers more effectively and provide real-time guidance. Forging collaborations with research institutions will strengthen the scientific foundation of practices and enable continuous improvement based on rigorous evaluation. Enhanced financial support and incentives for farmers will reduce barriers to adoption and ensure that regenerative practices remain economically competitive with conventional approaches. The framework established through this initiative provides valuable information for similar projects around the world, catalyzing the global transition to sustainable agricultural practices.

\subsection{The URVARA Project (India)} 
The URVARA project, implemented by Boomitra \cite{Urvara}, addresses the important issue of soil degradation in Indian agriculture. Conventional farming practices currently contribute to approximately 14\% of India's GHG emissions. By shifting the focus toward inclusive, regenerative agricultural practices, including reduced tillage to optimize fertilizer use, this initiative seeks to simultaneously restore ecosystem health and enhance crop productivity for smallholder farmers.
\subsubsection*{Project Overview}
Spanning across India with initial implementation in six key states (Andhra Pradesh, Karnataka, Kerala, Madhya Pradesh, Maharashtra, and Tamil Nadu), the project leverages the SCM0005 v2.0 methodology to quantify carbon removals. Operating on a 10-year crediting period that commenced in 2021, the project targets an average annual reduction of 16,405 tCO2e, with a total potential of 164,047 tCO2e over the project lifetime.

\subsubsection*{Challenges Faced}

The project areas face multiple interconnected challenges that hinder the widespread adoption of regenerative agricultural practices. Financial constraints pose a significant barrier, as smallholder farmers have limited access to affordable credit and insurance, face high production costs, and often depend on informal lenders for their financing needs. Obtaining government support has certain challenges due to the complexity of available schemes and the lack of knowledge among smallholder farmers about these programs. 

Weak infrastructure, particularly in digital connectivity and the absence of localized content, further impedes the adoption of regenerative agricultural practices across farming communities. Regulatory and investment restrictions, including limitations on foreign investment and complex land tenure systems, discourage large-scale sustainable agriculture initiatives. Socio-cultural resistance remains prevalent, with traditional practices such as burning crop residues persisting alongside challenges posed by fragmented landholdings and generational barriers to change. Additionally, the project's current dependence on carbon finance as the primary funding source for large-scale implementation raises concerns about long-term sustainability, as carbon credit revenue alone may not be adequate to support continued expansion.

\subsubsection*{Solutions Implemented}

To address these multifaceted challenges, the URVARA project has implemented a comprehensive set of solutions. Direct carbon credit payments to farmers help ease the burden of upfront costs associated with transitioning to regenerative agricultural practices. The project provides targeted training programs, localized advisory tools, and demonstration plots to build farmer capacity and confidence. Behavioral change is facilitated through peer learning networks and the active participation of youth in farming communities.

Transparent monitoring, reporting, and verification (MRV) systems coupled with revenue-sharing mechanisms ensure accountability and fair distribution of benefits. The project prioritizes the inclusion of women and marginalized groups in both training and monitoring activities, promoting equitable participation. Furthermore, the initiative creates employment opportunities directly linked to regenerative practices, contributing to local economic development.

\subsubsection*{Best Practices Implemented}

The project promotes a diverse array of regenerative agricultural best practices tailored to local contexts. Soil management techniques include reduced or conservation tillage and improved residue management, with some farmers adopting no-till systems combined with crop residue retention and reincorporation. Fertilizer management focuses on optimized, reduced, or organic fertilizer application, along with the adoption of specialized fertilizers and soil probiotics to enhance soil health.

Water management improvements encompass precision irrigation techniques and alternative wetting and drying methods to optimize water use efficiency. The project encourages diversified cropping systems, including rotational and continuous cropping with cover crops, double cropping, relay cropping, and intercropping practices. Notably, improved agroforestry practices are integrated into commercial crop systems, combining tree cultivation with agricultural production to enhance both productivity and environmental outcomes.

\subsubsection*{Impact of URVARA}

The URVARA Project has delivered measurable environmental, economic, and social benefits across farming communities. These outcomes include improvements in soil and ecosystem health, enhanced resource-use efficiency, and strengthened rural livelihoods, as summarized in Table \ref{tab:carbon_farming_impacts}. The project has also contributed to institutional capacity-building and created new livelihood opportunities through training and monitoring activities.

Environmental impacts include increases in soil organic carbon, reduced GHG emissions, improved soil moisture retention, and enhanced biodiversity on farmlands. Economic benefits manifest through lower input costs due to reduced fertilizer and tillage requirements, revenue generation from carbon credits, long-term productivity gains, and employment generation in local value chains. Social impacts encompass improved livelihoods and income stability, empowerment through capacity building, better food and nutrition security, and strengthened community cooperation and resilience.

\begin{table}[ht]
    \centering
    \caption{Impact of the URVARA project}
    \label{tab:carbon_farming_impacts}
    \begin{tabular}{@{}p{3cm}p{10cm}@{}}
        \toprule
        \textbf{Category} & \textbf{Impact} \\
        \midrule
        Environmental & Increase in soil organic carbon; reduced GHG emissions; improved soil moisture retention; enhanced biodiversity on farmlands. \\
        Economic & Lower input costs due to reduced fertilizer and tillage; revenue from carbon credits; long-term productivity gains; employment generation in local value chains. \\
        Social & Improved livelihoods and income stability; empowerment through capacity building; better food and nutrition security; strengthened community cooperation and resilience. \\
        \bottomrule
    \end{tabular}
\end{table}

\subsubsection*{Future Prospects}

The project envisions several strategic directions for sustained impact and growth. Plans include expanding regenerative practices to scale carbon sequestration across larger geographical areas and farming communities. Integration of advanced remote sensing and digital tools will enable real-time monitoring and more precise management of agricultural activities. Strengthening market linkages for both carbon credits and sustainable produce will enhance economic viability for participating farmers.

The project aims to influence national agricultural policies to support climate-resilient farming at a systemic level. Attracting private sector investment through verified impact metrics will diversify funding sources beyond carbon finance. Finally, enhancing climate adaptation by building soil and ecosystem resilience remains a central objective, ensuring that farming communities are better prepared to face future climate challenges.

\subsection{Grassland Restoration in Northern Mexico}

The Boomitra Northern Mexico Grassland Restoration Project \cite{VerraVCS2887} represents one of the largest soil carbon initiatives globally. It is distinguished by its integration of cutting-edge AI and satellite technology with traditional regenerative grazing practices. The initiative focuses on restoring degraded rangeland ecosystems to improve local livelihoods and sequester significant amounts of atmospheric carbon.

\subsubsection*{Project Overview}

Covering over 1.3 million acres across the Chihuahuan and Sonoran deserts, this project implements rotational grazing and improved grassland management strategies. The initiative operates on a 20-year crediting period, which began in 2018. It targets an estimated annual GHG reduction of 758,269.25 tCO2e for the first monitoring period (2018-21), making it one of the most ambitious regenerative grazing efforts in the world.

\subsubsection*{Challenges Faced}
The project faces numerous interconnected challenges that have historically impeded sustainable land management. Financial barriers present a significant obstacle, with high upfront investment requirements for essential infrastructure such as fencing and water points. 

Knowledge gaps further complicate adoption, as many ranchers lack a technical understanding of sustainable grazing management. In Ejido communities, particularly low literacy levels make conventional training and extension approaches largely ineffective. Social resistance to change has slowed the adoption of new practices, especially in communal lands where collective approval is required for management decisions. Environmental degradation from continuous grazing over decades has resulted in severely compromised land conditions, characterized by shrub encroachment, low vegetative cover, and poor soil fertility. Climate vulnerability has intensified these challenges, with increasing drought frequency driven by El Niño events making ecosystem regeneration harder to achieve without implementing sustainable practices.

\subsubsection*{Solutions Implemented}

The project has implemented a rich set of solutions to address these types of challenges. Carbon finance serves as the primary mechanism for funding critical infrastructure, including fencing, paddock creation, and the water systems necessary to implement rotational grazing. Comprehensive training and capacity-building programs for ranchers are delivered with support from local partners who understand community contexts and needs.

Recognizing literacy constraints, the project has developed peer-led, visual, and audio-based training methods specifically tailored for low-literacy ranchers. Awareness campaigns and participatory planning processes encourage the adoption of new practices by engaging communities in decision-making. AI and satellite-based monitoring systems dramatically reduce the costs associated with soil carbon quantification, making large-scale monitoring economically feasible. Water conservation structures, including keylines, gabions, ponds, and microbasins, have been established to improve water availability and support vegetation recovery.

The best practices implemented through the project are diverse and integrated, as summarized in Table \ref{tab:grassland_practices}. Rotational grazing forms the cornerstone approach, allowing vegetation recovery periods that improve soil health and sequester carbon. Adaptive multi-paddock grazing enables ranchers to adjust livestock numbers based on forage availability, preventing overgrazing while maintaining productive herds. Paddock fencing ensures controlled grazing cycles that protect the recovering vegetation. The water conservation infrastructure improves soil moisture, promotes biomass growth, and enhances carbon storage capacity. Biodiversity monitoring tracks the recovery of native species and habitat improvements. Comprehensive capacity building and training programs equip ranchers with the knowledge and tools needed for sustainable land management in the long term.

\begin{table}[ht]
    \centering
    \caption{Best practices implemented in the grassland restoration project}
    \label{tab:grassland_practices}
    \begin{tabular}{@{}p{4cm}p{9cm}@{}}
        \toprule
        \textbf{Practice} & \textbf{Objective} \\
        \midrule
        Rotational Grazing & Allow vegetation recovery, improve soil health, and sequester carbon. \\
        Adaptive Multi-Paddock Grazing & Adjust livestock numbers based on forage availability. \\
        Paddock Fencing & Prevent overgrazing and ensure controlled grazing cycles. \\
        Water Conservation (keylines, gabions, ponds) & Improve soil moisture, biomass growth, and carbon storage. \\
        Biodiversity Monitoring & Promote native species recovery and habitat improvement. \\
        Capacity Building and Training & Equip ranchers with knowledge and tools for sustainable land management. \\
        \bottomrule
    \end{tabular}
\end{table}

\subsubsection*{Impact}

The Grassland Restoration Project has delivered substantial ecological, economic, and social gains by restoring degraded rangelands, rebuilding forage systems, and strengthening community-led grazing management. These improvements have contributed to healthier ecosystems, more resilient livelihoods, and sustained incentives for regenerative land stewardship, as summarized in Table \ref{tab:grassland_impacts}. The project has also reinforced local capacity through collective planning, knowledge sharing, and inclusive governance structures, particularly within Ejidos.

Environmental impacts have been dramatic, with restoration of degraded lands, biodiversity recovery, improved soil organic carbon levels, and reduced erosion transforming previously barren landscapes. In some areas, native grass species diversity has increased from just 1 species to over 15, demonstrating remarkable ecological recovery. Improved soil fertility and enhanced water infiltration have created positive feedback loops that accelerate regeneration.

Economic benefits extend beyond immediate carbon credit revenue sharing to include increased forage availability for livestock and enhanced long-term ranch productivity. These carbon payments provide sustained incentives for maintaining sustainable practices, supporting rural livelihoods through revenue-sharing arrangements that benefit both individual ranchers and communal landholders. Social impacts have been equally significant, with strengthened community collaboration in Ejidos, enhanced local capacity through training and knowledge transfer, improved livelihood resilience, and reduced poverty vulnerability contributing to more stable rural communities.

\begin{table}[ht]
    \centering
    \caption{Impact of the grassland restoration project}
    \label{tab:grassland_impacts}
    \begin{tabular}{@{}p{3cm}p{10cm}@{}}
        \toprule
        \textbf{Dimension} & \textbf{Impact} \\
        \midrule
        Environmental & Restoration of degraded lands; biodiversity recovery; improved soil organic carbon; reduced erosion. \\
        Economic & Carbon credit revenue sharing; increased forage availability for livestock; enhanced long-term ranch productivity. \\
        Social & Capacity building; improved knowledge sharing; stronger community cooperation; better climate resilience. \\
        \bottomrule
    \end{tabular}
\end{table}

\subsubsection*{Future Prospects}

The project envisions ambitious expansion and refinement of its approach in the coming years. Plans include scaling regenerative grazing and restoration practices across additional degraded grassland regions, extending the proven model to new contexts and communities. Digitization of paddock data collection and grazing management plans will enable improved monitoring, more responsive management decisions, and better documentation of outcomes.

The project seeks to catalyze stronger financial and policy support for sustainable grazing at the national level, working to influence broader agricultural policies and investment frameworks. Expanding carbon revenue-sharing models remains a priority to secure long-term income streams for both individual ranchers and communal landholders, ensuring that regenerative practices remain economically viable and socially equitable into the future.

\subsection{The Ganga Basin Project (India)}

Grow Indigo Pvt. Ltd. \cite{GrowIndigo} spearheaded this project in India to address two critical challenges: mitigating climate change and improving farmer incomes. The study covers four Indian states: Bihar, Madhya Pradesh, Rajasthan, and Uttar Pradesh~\cite{sainath2025} and demonstrates the synergistic benefits of regenerative agriculture in the Ganga Basin.

\subsubsection*{Project Overview}

The {\em Regenerative Agriculture in the Ganga Basin for Farmer Income and Climate Impact\/} project seeks to reduce greenhouse gas (GHG) emissions by adopting sustainable agricultural practices that restore soil organic carbon, improve soil health, and enhance biodiversity. Initiated by Grow Indigo Private Limited, the project operates over a 20-year period from 15 October 2020 to 14 October 2040, targeting an estimated annual GHG reduction of 5,048,425 tons of CO$_2$ equivalent. This ambitious scale reflects the critical need for transformative agricultural interventions in one of India's most agriculturally intensive regions.

\subsubsection*{Challenges in the Project Areas}

Farmers in the areas covered by the project rely largely on conventional agricultural practices characterized by intensive use of chemical fertilizers, frequent and deep tillage, high water consumption, and burning of residues. These unsustainable practices have degraded the agricultural ecosystem over decades, resulting in progressive loss of soil fertility, widespread water pollution, elevated greenhouse gas emissions, and declining farm incomes. The cumulative effect of these practices has created a vicious cycle in which farmers must invest more in inputs to maintain yields on increasingly degraded land, further eroding their economic viability while exacerbating environmental damage.

\subsubsection*{Best Practices Implemented}

The project has adopted various regenerative agriculture practices to address these challenges, as shown in Table~\ref{tab:regenerative_operations}. Minimal tillage approaches minimize soil disturbance and reduce fuel usage, preserving soil structure and reducing carbon emissions from both soil disturbance and machinery operation. Crop residue management practices avoid the burning of agricultural waste by incorporating organic matter back into the soil, building soil carbon while eliminating harmful emissions and air pollution.

Fertilizer management strategies promote precision nutrient management and increased use of organic input, reducing dependence on synthetic fertilizers while improving the efficiency of nutrient use. Innovations in irrigation management include the adoption of Direct Seeded Rice  and alternative Wetting and Drying techniques, which substantially reduce water consumption and methane emissions compared to traditional flooded rice cultivation. Crop rotation and cover cropping systems improve soil structure and health, while decreasing dependence on synthetic fertilizers through natural nitrogen fixation and organic matter addition. Agroforestry practices integrate trees with crops, boosting biodiversity and sequestering additional carbon while providing diversified income sources for farmers.

\begin{table}[ht]
    \centering
    \small
    \setlength{\tabcolsep}{5pt}
    \renewcommand{\arraystretch}{1.3}
    \caption{Best practices implemented by the Ganga basin project}
    \label{tab:regenerative_operations}
    \begin{tabular}{@{}p{2.5cm}p{9cm}@{}}
        \toprule
        \textbf{Practice} & \textbf{Objective} \\
        \midrule
        Minimal Tillage & Minimizes soil disturbance and reduces fuel usage. \\
        Crop Residue Management & Avoids crop residue burning by incorporating organic matter into the soil. \\
        Fertilizer Management & Promotes precision nutrient management and increased use of organic inputs. \\
        Irrigation Management & Adopts Direct Seeded Rice (DSR) and Alternate Wetting and Drying (AWD) to reduce water use and methane emissions. \\
        Crop Rotation and Cover Cropping & Improves soil structure and health while decreasing dependence on synthetic fertilizers. \\
        Agroforestry & Integrates trees with crops, boosting biodiversity and sequestering carbon. \\
        \bottomrule
    \end{tabular}
\end{table}

\subsubsection*{Impact of the Project}

The analysis of the project results reveals environmental, economic, and social benefits, exemplifying the multifaceted impacts of carbon agriculture as shown in Table~\ref{tab:Impact_CF}. This holistic approach improves farmer well-being and overall quality of life in multiple dimensions.

The environmental impacts strengthen soil health through greater carbon sequestration and reduced erosion, reversing decades of degradation. Enhanced water conservation and ecosystem resilience create more stable agricultural systems better adapted to climate variability. Improved biodiversity on farmlands contributes to more robust and self-regulating agricultural ecosystems.

Economic benefits manifest themselves through increased farmer income generated by improved yields, reduced input costs, and revenue from carbon credits. The project has achieved significant savings by curbing excess input use, particularly synthetic fertilizers and irrigation water, which previously represented a major cost burden for farmers. These combined economic improvements improve the financial sustainability and resilience of farming households.

Social impacts extend beyond individual farm economics to contribute to broader development goals. The project promotes the Sustainable Development Goals, notably Zero Hunger (SDG 2) through improved food security, Quality Education (SDG 4) through farmer training and capacity building, and Climate Action (SDG 13) through emission reductions and climate adaptation. Enhanced knowledge and capacity among farmers empower them to make informed decisions about their agricultural practices and improve their ability to adapt to changing conditions.

\begin{table}[ht]
    \centering
    \small
    \setlength{\tabcolsep}{5pt}
    \renewcommand{\arraystretch}{1.3}
    \caption{Impact of the Ganga basin project}
    \label{tab:Impact_CF}
    \begin{tabular}{@{}p{2.5cm}p{9cm}@{}}
        \toprule
        \textbf{Dimension} & \textbf{Key Impact} \\
        \midrule
        Environmental & Strengthens soil health through greater carbon sequestration and reduced erosion. Enhances water conservation and ecosystem resilience. \\
        Economic & Increases farmer income through improved yields, reduced input costs, and revenue from carbon credits. Achieves significant savings by curbing excess input use. \\
        Social & Promotes Sustainable Development Goals (SDGs), notably Zero Hunger (SDG 2), Quality Education (SDG 4), and Climate Action (SDG 13), while enhancing knowledge and capacity among farmers. \\
        \bottomrule
    \end{tabular}
\end{table}

\subsection{Trees for Global Benefit (TGB) Project (Uganda)}

The \emph{Trees for Global Benefit} (TGB) program in Uganda \cite{Masiga2012TFGB,GFC2022,Purdon2022} is a widely cited smallholder agroforestry carbon initiative that appears successful when viewed through project documentation and credit issuance records, but looks very different when examined from the perspective of participating farmers. Developed by ECOTRUST under the Plan Vivo standard, the program encourages smallholders to establish woodlots or agroforestry systems on their land while carbon credits generated from tree biomass are sold to corporate buyers on the voluntary market. Farmers sign long-term contracts, typically lasting around 25 years, during which they are required to maintain tree cover on the enrolled plots in exchange for a sequence of performance-based carbon payments distributed over several years.

\subsubsection*{Project Overview}

TGB operates in several districts in Western and Central Uganda under the stewardship of ECOTRUST, applying the community-based Plan Vivo standard to generate carbon credits from smallholder agroforestry and farm woodlots. Farmers usually join the program by dedicating part of their land—often formerly planted with coffee, bananas, or annual crops—to tree establishment and entering into a long-duration contract with ECOTRUST \cite{Masiga2012TFGB}. Expected carbon revenues are discounted and paid in installments that depend on tree survival and periodic verification. Once these payments are reduced or stopped, farmers are bound to maintain the trees for the rest of the contract period. Participation patterns show that households with larger landholdings are more likely to join, since minimum plot-size requirements and the opportunity costs of diverting cropland to trees deter or disadvantage smaller and poorer farmers \cite{Purdon2022}.

\subsubsection*{Farmer Experience and Key Problems}

Independent research and civil-society investigations reveal a consistent set of difficulties experienced by many participating farmers \cite{GFC2022,Purdon2022}. These concerns, which recur across interviews, focus groups, and formal assessments, are summarized in Table~\ref{tab:tgb_problems}.

\begin{table}[ht]
    \centering
    \caption{Farmer-reported problems under the TGB program}
    \label{tab:tgb_problems}
    \begin{tabular}{@{}p{3.5cm}p{9.5cm}@{}}
        \toprule
        \textbf{Issue} & \textbf{Farmer Experience / Effect} \\
        \midrule
        Payments & Initial payments often fail to cover the costs of establishing trees—including labour and seedlings—and later instalments tend to be small, delayed and uncertain, resulting in limited net financial gains. \\
        Land use \& food security & Land allocated to trees is often land previously used for food or cash crops. As the tree canopy expands, the flexibility to cultivate annual crops diminishes, and some households report declining food production and heightened livelihood vulnerability. \\
        Information and contracts & Contracts are lengthy, technical, and frequently written in English, which restricts understanding. Many farmers have only a partial grasp of payment schedules, penalties, and obligations extending over decades. \\
        Equity and participation & Larger and better-off farmers can more easily meet the minimum land requirements and manage risks, whereas land-poor households face substantially higher opportunity costs and livelihood risks if they participate. \\
        Power imbalance & ECOTRUST and carbon buyers determine contract terms and revenue flows, leaving farmers with limited bargaining power or influence over prices, rules, or the distribution of benefits. \\
        \bottomrule
    \end{tabular}
\end{table}

These issues are not isolated or anecdotal. They recur across multiple studies and form a central reason why TGB is frequently highlighted in discussions on carbon farming, equity, and justice.

\subsubsection*{Lessons from the TGB Project}

Although TGB has successfully issued and sold certified carbon credits and is often presented as a community-based offset model, evaluations centered on farmers’ experiences present a contrasting picture \cite{GFC2022}. Smallholder farmers take on long-term land-use restrictions and climate-related risks in return for payments that are modest, front-loaded, and often insufficient to cover establishment and maintenance costs. Meanwhile, corporate buyers benefit from low-cost carbon credits and associated reputational gains. The shortcomings of TGB therefore stem less from implementation failures than from structural features of its design, which systematically under-reward farmers relative to the obligations they assume. The case illustrates the need for clearer and more accessible contracts, stronger safeguards for food production and land security, and more equitable benefit-sharing mechanisms in future carbon farming policies.

\subsection{Some Observations on the Case Studies}
Taken together, the six case studies underscore that carbon farming is not a monolithic intervention, but a spectrum of project designs shaped by context, incentives, and institutional capacity. All projects essentially rely on broadly similar agronomic best practices. 

A common insight from such case studies is the central trade-off between measurement precision and scalability. Projects relying heavily on field sampling and conservative modeling demonstrate higher credibility but face rising costs and operational complexity, particularly in smallholder landscapes. Projects emphasizing digital MRV and remote sensing achieve scale more rapidly but depend critically on calibration, transparency, and third-party oversight to maintain trust.

The case studies also reveal that carbon credit revenues alone are just not sufficient to drive sustained farmer adoption. Successful projects embed carbon farming within a broader value proposition (which we have extensively described in Section 5, including yield stability, input cost reduction, risk mitigation, and advisory support. Aggregation mechanisms—such as cooperatives, producer companies, and specialized project developers- emerge as indispensable intermediaries, especially in India.

Another important point to note is that policy alignment, market integrity, and MRV credibility are mutually strengthening. Weakness in any one dimension can undermine the entire project value chain. These insights motivate the India-specific analysis in a future section and inform the broader discussion of challenges and remedies presented later in the paper.

Finally, these case studies illustrate that carbon farming is moving beyond theoretical potential into operational reality. Yet, the disparity between the high cost of compliance, as shown by the REDD project’s insurance overhead and the often modest revenue reaching smallholder farmers, identifies the key impediment preventing global scalability. Bridging this gap will require harmonizing technological innovations with robust policy frameworks, ensuring that carbon markets evolve from pilot programs into equitable systems that serve both the planet and the farmers.
\section{Carbon Farming in India: Current Status}\label{}
This section examines carbon farming through the lens of India’s agricultural and policy landscape. It analyses emerging market structures, regulatory developments, and implementation constraints, with particular attention to smallholder farmers, aggregation models, and India-specific opportunities.


As demonstrated by three case studies in the previous section, carbon farming has gained prominence in India as a strategic response to the dual challenges of climate change and sustainable agricultural development. Across diverse agro-ecological zones, multiple initiatives have adopted locally tailored approaches to enhance carbon sequestration and reduce greenhouse gas emissions. This growing momentum aligns with India’s broader transition to a more structured and regulated carbon pricing ecosystem. Against the backdrop of an increasing global emphasis on carbon markets and emissions trading, India is actively developing a rate-based emission Trading System (ETS) in conjunction with voluntary carbon crediting mechanisms \cite{pib2025_carbon_pricing_india}. Reflecting this progress, the World Bank’s \textit{report on State and Trends in Carbon Pricing 2025} \cite{worldbank2025_state_trends} recognizes India’s expanding role among emerging economies in shaping global climate finance and carbon pricing frameworks.


\subsection{Carbon Credit Monetization}
Indian farmers monetize carbon credits mainly by participating in carbon farming initiatives that reward the adoption of sustainable agricultural practices. The mechanisms and market structures supporting this monetization are described below.

\subsubsection*{Participation in Carbon Farming Programs}
Farmers adopt sustainable practices, such as direct-seeded rice (DSR), zero-tillage, and agroforestry, which reduce greenhouse gas emissions and enhance soil carbon sequestration. These interventions are often promoted through government agencies and private organizations-led programs.

\subsubsection*{Carbon Credit Generation} 
Farmers generate carbon credits by implementing practices that reduce or sequester greenhouse gas emissions. Typically, one carbon credit represents reducing or removing one ton of CO$_2$ equivalent. Farmers can sell these credits on carbon markets to companies that want to offset their emissions.

\subsubsection*{Revenue Sharing Models} 
Projects commonly implement revenue-sharing models to ensure farmers receive a significant share of the proceeds from carbon credit sales. These arrangements are designed to directly benefit participants and incentivize the adoption of climate-friendly agricultural practices.

\subsubsection*{Corporate Partnerships} 
Major corporations increasingly buy carbon credits generated from Indian agricultural initiatives. For example, Google has signed an agreement to procure 100,000 tons of carbon credits from Indian farms through Varaha.earth, highlighting growing private sector interest in agricultural emissions offset projects.

\subsubsection*{Carbon Credit Trading Scheme (CCTS)}
The Carbon Credit Trading System (CCTS) is the framework designed by the Indian Government to create a national carbon market that converts emission performance into tradable certificates. CCTS rests on the legal foundation created by the Energy Conservation (Amendment) Act and was notified in 2023, with more detailed operational procedures issued in 2024--25. 

CCTS has a two-track architecture: (1) a \textbf{Compliance Mechanism} that sets sectoral emissions-intensity targets for large emitters, and (2) an \textbf{Offset Voluntary Mechanism} that certifies additional reductions generated across the economy. The Bureau of Energy Efficiency (BEE) is the designated administrator, the Grid Controller/Registry manages the registry, and a National Steering Committee governs policy and sector inclusion.

Compliance-grade Carbon Credit Certificates (CCCs) will be issued to entities that outperform targets; these CCCs can be traded on regulated power exchanges under the oversight of CERC (Central Electricity Regulatory Commission). Early rules allow banking (but not borrowing), do not permit international linkage initially, and anticipate phased sectoral coverage (starting with energy-intensive industries). Full operationalization has been targeted for 2026. 

CCTS, which is still work in progress, aims to replace and broaden India’s earlier PAT (Perform, Achieve, and Trade) approach by directly pricing greenhouse gas reduction performance, but the successful impact will depend on robust MRV, liquidity on exchanges, and a careful target design to deliver real, verifiable emission cuts. 

\subsection{Government Guidelines}

The Indian Government has outlined ambitious climate goals, driven by its commitment to achieve net-zero emissions by 2070. This goal is based on a comprehensive framework of policies, regulations, and initiatives that encompass multiple sectors of the economy.

\subsubsection*{Legislative and Policy Framework}
The Indian government has formulated a legislative and policy foundation for climate action. In 2022, enacting the `Net Zero Emissions" bill and the Energy Conservation (Amendment) Act established the country's legal basis for carbon credit trading. These legislative milestones are supported by a series of strategic roadmaps, including:
\begin{itemize} 
    \item The National Action Plan on Climate Change (NAPCC)
    \item State Action Plans on Climate Change (SAPCC)
    \item The Long-term Low-carbon Development Strategy (LT-LEDS)
\end{itemize}

\subsubsection*{Core Climate Strategy and Targets}

India’s climate strategy targets a 45\% reduction in the intensity of GDP emissions by 2030 (from 2005 levels), 500 GW of non-fossil energy capacity, and meeting 50\% of renewable energy needs - marking a major shift towards clean energy \cite{GOI22}. 

These goals are supported by initiatives like the Perform, Achieve, and Trade (PAT) scheme for industrial efficiency, and investments in grid modernization and energy storage to support renewables. A planned one billion-ton emissions reduction is supported by the National Green Hydrogen Mission and transport electrification. The Net Zero by 2070 target relies on scaling carbon removal technologies, including CCUS. 

\subsubsection*{Carbon Farming and Agricultural Initiatives}

A characteristic of India's climate approach is its emphasis on carbon management within the agricultural sector. The Ministry of Agriculture and Farmers Welfare has taken the lead in developing a comprehensive framework for a {Voluntary Carbon Market} (VCM) explicitly tailored for agriculture. This scheme is designed to provide small and marginal farmers with opportunities to benefit from the generation and sale of carbon credits.

The regulatory landscape for carbon farming has several key components:
\begin{itemize} 
    \item Launch of the Carbon Credit Trading Scheme (CCTS) in 2023
    \item Alignment with the National Agriculture Policy
    \item Detailed guidelines for soil carbon measurement and verification
\end{itemize}

These government actions are crucial in fostering the growth of carbon markets and integrating sustainable agricultural practices into India's broader climate strategy.

\section{Carbon Farming: Key Challenges and Remedies}\label{sec8}
Carbon farming has the potential to transform both farm livelihoods and environmental outcomes, but farmers (especially small and marginal farmers) face several obstacles when trying to participate. These challenges arise in various ways. 
This section presents the principal barriers to large-scale adoption of carbon farming, spanning scientific uncertainty, institutional design, economics, and farmer engagement. The section attempts to outline potential remedies and design principles required to improve credibility, reduce transaction costs, and strengthen long-term viability.

\subsection{Key Challenges and Remedies}
\subsubsection*{Policy and Institutional Barriers}

For farmers to effectively adopt or accept the carbon farming initiatives, the farmers require supportive policies, which may be lacking \cite{ChallengesCF}. Most of the schemes designed are not easily understood by farmers due to unclear procedures. Carbon sequestration projects, such as the REDD$^+$ initiatives described in Case Study 2, faced substantial financial overhead from transaction costs. These costs, which occur during project design, implementation, monitoring, and ensuring compliance with regulatory standards, can significantly undermine the overall profitability and financial viability of these initiatives. These challenges make farmers fear losing eligibility for existing subsidies. Insurance represents a significant component of total project costs. Insurers set high premiums due to the inherent project-specific risks associated with long-term implementation and uncertainties in field conditions \cite{TransactionCosts}. This substantial insurance burden can significantly diminish the net financial benefits obtained from the sale of carbon credits.

\begin{tcolorbox}[Cbox]
    \textbf{Possible Solutions for \textbf{Policy and Institution Barriers}}
    \begin{itemize}
        \item Simplification of guidelines, eligibility criteria, and application procedures, along with extensive training on these processes by specialists and experts through farmer collectives.
        \item Standardizing the approvals and certification protocols in accordance with the farmer categories.
        \item Aligning policies and incentives with farmers through participatory planning
        \item Integrating carbon farming initiatives with existing Governmental schemes. For example, in India, schemes such as the soil health card, PKVY (Paramparagat Krishi Vikas Yojana), and RKVY (Rashtriya Krishi Vikas Yojana) could be embedded into carbon farming initiatives.
        \item Ensuring stability and long-term policy commitments to reduce the uncertainty 
    \end{itemize}
    \end{tcolorbox}

\subsubsection*{Adoption and Economic Barriers}

Adoption of Carbon Farming Initiatives is often constrained by farmers’ limited awareness, fear of productivity loss, and lack of managerial capacity to implement new practices \cite{ChallengesCF}. High upfront costs related to inputs, monitoring, reporting, and verification make participation financially risky, especially for small and marginal farmers \cite{Challenges}. Uncertainty about carbon credit returns, complicated credit approval processes, and perceptions of policy instability further discourage farmers from investing in carbon farming \cite{Challenges17}. These constraints collectively reduce farmers’ enthusiasm to engage with carbon farming initiatives despite their potential benefits.

\begin{tcolorbox}[Cbox]
\textbf{Possible Solutions for \textbf{Adoption and Economic Barriers}}
\begin{itemize}
\item Localized training and awareness programs through Government agencies (for example, KVKs (Krishi Vijnana Kendras) in India), FPOs, and on-field demonstrations to build confidence.
\item Making a clear provision for upfront subsidies, transition grants, and low-interest green loans to reduce financial burden.
\item Implementing minimum carbon price guarantees and risk-sharing mechanisms to address income uncertainty.
\item Transparent benefit-sharing models that clearly communicate how carbon revenues are computed and distributed.
\item Aggregation of farmers through FPOs/cooperatives to reduce per-farmer transaction and verification costs.
\end{itemize}
\end{tcolorbox}

\subsubsection*{Technical and Methodological Barriers}
Many sustainable farming practices, such as Directed Seeded Rice, Alternate Wetting and Drying, minimum tillage, and cover cropping, were adopted long before carbon credit programs, making it difficult to demonstrate “additionality" \cite{CARIAPPA24} (additionality is an important criterion in carbon credit calculation). Small and marginal farmers often lack the technical expertise required for documentation \cite{BARBARTO23}, for soil carbon measurement, and for compliance with verification standards. Diverse agro-ecological conditions further complicate the accurate estimation of soil organic carbon. Further, low digital literacy, uneven smartphone access, and the rapidly evolving landscape of global carbon methodologies make participation challenging for rural farmers \cite{Shen2022}.

\begin{tcolorbox}[Cbox]
\textbf{Possible Solutions for \textbf{Technical and Methodological Barriers}}
\begin{itemize}
\item Development of additionality criteria that specifically reflect local farming realities.
\item Increased adoption of digital MRV tools (remote sensing, AI-based carbon estimation, mobile reporting systems) to reduce complexity.
\item Deployment of field-level carbon technicians to support farmers in documentation and verification.
\item Improving digital literacy and access through rural digital centers and multilingual mobile applications.
\item Establishing a flexible nationwide carbon registry that updates methodologies regularly based on local needs.
\end{itemize}
\end{tcolorbox}

\subsubsection*{Market Access Barriers}
Access to carbon markets remains limited for small and marginal farmers due to fragmented landholdings, lack of aggregation mechanisms, and the absence of a nationwide carbon trading framework. International carbon markets tend to favor large participants with the capacity to meet complex accreditation and compliance requirements \cite{CARIAPPA24}. Price volatility and unclear revenue expectations discourage farmers from viewing carbon credits as a reliable income source. These structural disadvantages reduce the ability of small and marginal farmers to benefit from emerging carbon markets.

\begin{figure}[htbp]
    \centering
    \includegraphics[width=0.8\linewidth]{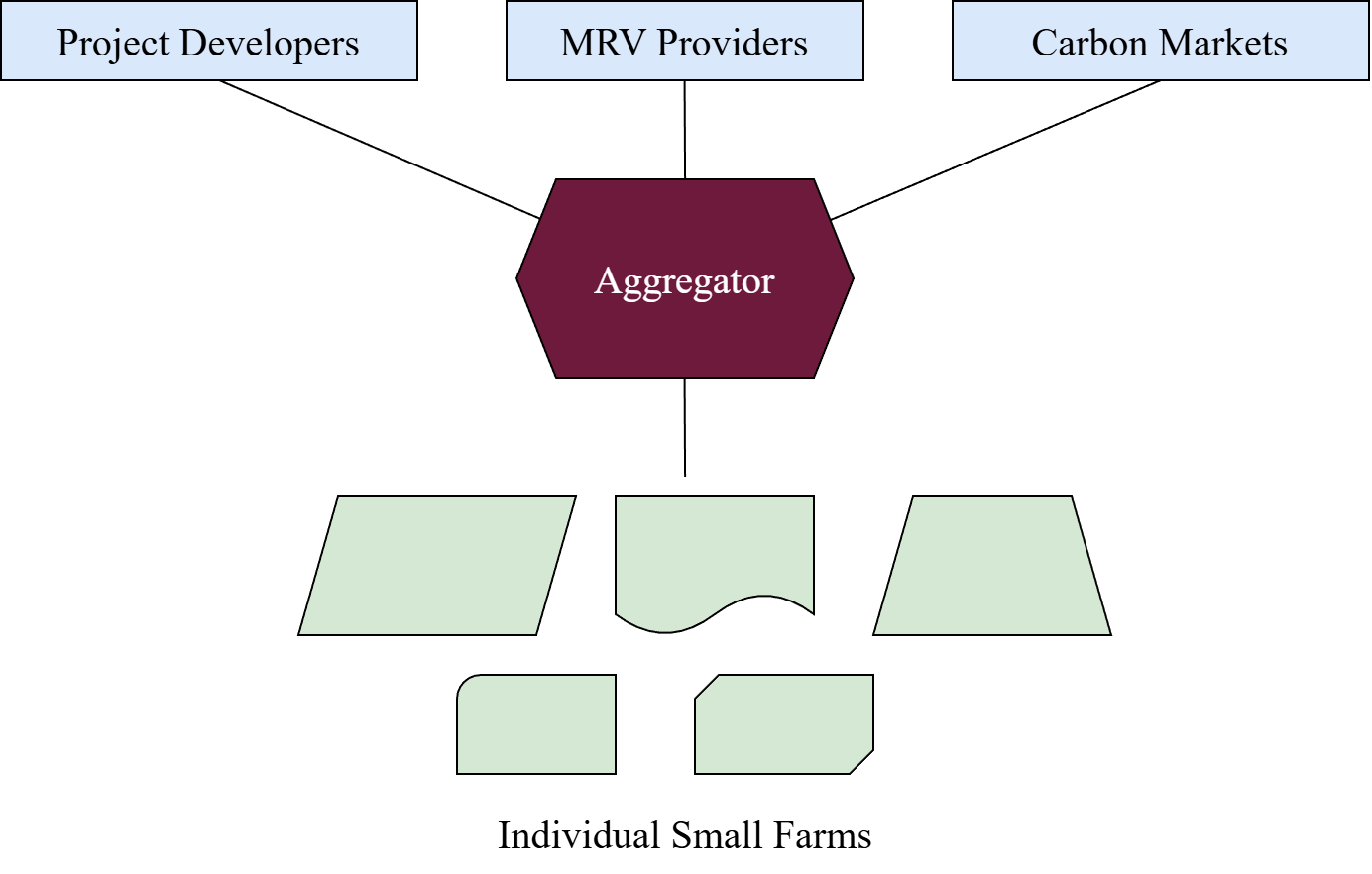}
    \caption{The aggregator model: bridging individual small farms with carbon markets and service providers}
    \label{fig:aggregator_model}
\end{figure}

\begin{tcolorbox}[Cbox]
\textbf{Possible Solutions for \textbf{Market Access Barriers}}
\begin{itemize}
\item Establishment of a nation-wide carbon market with simplified, smallholder-friendly participation rules.
\item Aggregation of farmers through FPOs, cooperatives, and private aggregators to achieve scale and improve bargaining power (see Fig. \ref{fig:aggregator_model}).
\item Simplified accreditation procedures and reduction of transaction costs for smallholders.
\item Enhancing market transparency through regular publication of carbon prices, contract terms, and buyer databases.
\item Facilitating direct linkages between farmer groups and verified carbon credit buyers to improve returns.
\end{itemize}
\end{tcolorbox}

\subsection{Economics of Carbon Farming}
The economic viability of carbon farming hinges on a delicate balance between agronomic benefits, carbon revenues, transition costs, and transaction costs. For farmers, especially smallholder farmers, carbon farming is rarely driven by carbon markets alone. The financial calculus involves a combination of input savings, risk reduction, and modest but uncertain carbon payments.

At the farm level, carbon farming can reduce expenditures on tillage, fertilizers, pesticides, and irrigation. Practices such as minimal tillage, legume-based rotations, cover cropping, and improved water management generate measurable cost savings. In Indian contexts, fertilizer use may decrease appreciably over time, while irrigation savings in rice systems (e.g., AWD or DSR) may reduce costs in a sizeable way. These agronomic gains often exceed carbon-credit income, especially during the early years when SOC changes are small, and credit issuance may be delayed.

The costs of making the transition to carbon farming could be substantial. Farmers incur expenses for new equipment (seeders, residue managers), higher labor for initial cover-crop establishment, opportunity costs from land allocated to non-cash crops, and learning costs associated with new practices. In addition, yield dips can occur during transitional years, raising the short-term risk profile for economically vulnerable households. Without upfront financial support or risk-sharing arrangements, smallholder farmers will be unable or unwilling to adopt new practices despite long-term benefits.

On the revenue side, carbon payments remain modest. In India, for example, soil-carbon projects typically produce 0.5--2.0 tons of CO$_2$/ha/year in tradable credits under conservative assumptions. At current voluntary market prices (\rupee 300 to \rupee 1000 per ton of CO$_2$), farmers receive only \rupee 300 to \rupee 2000/ha/year after deductions. Because aggregators, project developers, MRV providers, and registries all take shares of carbon revenues, the farmer’s portion may fall below 30--40\% of gross credit value. Payment delays of 1–3 years, caused by verification cycles, further reduce economic attractiveness.

From the project developer’s perspective, economics is highly dependent on scale. Soil sampling, modeling, remote sensing, verification, and registry fees create high fixed costs. Projects of less than 20,000 to 30,000 acres may even struggle to break even. Digital MRV promises cost reductions, but reliability and regulatory acceptance remain constraints. Because revenue is tied to credit issuance, developers face cash-flow risks that often necessitate external financing.

At the market level, carbon prices are volatile and influenced by corporate sentiment, regulatory uncertainty, and controversies over credit integrity. This volatility creates uncertainty for both farmers and developers. The long-term viability of carbon farming, therefore, depends on predictable price signals, improved credit quality, and integration with compliance or semi-regulated markets, where prices tend to be higher and more stable.

To summarise, the economics of carbon farming are complex. Although agronomic co-benefits are strong, carbon-revenue streams alone are insufficient to drive widespread adoption. Successful models combine carbon finance with public subsidies, blended finance, concessional credit, ecosystem-service payments, or crop-diversification incentives. A sustainable economic architecture must reduce transaction costs, increase price transparency, improve benefit sharing, and align farmer incentives with long-term climate and soil-health outcomes.

\subsection{The Way Forward for Farmers}\label{sec11}

Effective participation in carbon farming or sustainable agriculture requires careful planning and systematic implementation of regenerative or improved agricultural land management practices. Farmers should establish comprehensive documentation systems to monitor their farming activities, soil carbon levels, and associated environmental impacts. This includes maintaining detailed records of management practices, conducting regular soil tests, and collecting photographic evidence of the adopted methods. This documentation provides essential proof for the verification of carbon credits and facilitates the ongoing optimization of farming strategies.

Farmers should adopt carbon farming practices in a strategic and phased manner. Initially, they should integrate methods compatible with their existing operations and, over time, introduce more advanced or innovative approaches. The selection of appropriate practices must account for local climatic conditions, soil characteristics, and resource availability. A gradual and context-sensitive transition helps minimize potential risks and builds farmers' expertise in sustainable management.

Meticulous financial planning is key to the success of carbon farming initiatives. Farmers should perform detailed cost-benefit analyses considering upfront expenditures and long-term gains. This involves evaluating the subsidies, grants, or support schemes available to offset initial investment costs. The formation of partnerships or joining cooperatives can help distribute costs more effectively and enhance collective bargaining power in carbon markets.

Small and marginal farmers will require considerable guidance and support from the government and other official agencies. Government schemes should be embedded into carbon farming practices, and the same should be informed to the farmers. The government should support NGOs and start-ups to help farmers implement carbon farming practices.

Farmer-Producer Organizations are designed to empower small and marginal farmers through collective action. These organizations, which are usually registered as cooperatives or producer companies, provide comprehensive support services such as input procurement, technical guidance, financial services, and market access. FPOs have shown significant transformative potential, particularly in seed procurement and distribution. Individually, small farmers often face high costs, limited access to quality inputs, and minimal negotiation power. FPOs address these challenges by aggregating farmers' requirements, enabling bulk procurement of seeds and other inputs at reduced costs, thereby improving access and affordability.

This model of collective action can be effectively extended to address the barriers faced by small and marginal farmers in carbon farming and participation in carbon credit markets. FPOs can aggregate carbon credits from member farmers, achieving the scale required for meaningful engagement in carbon markets.

FPOs can collectively invest in the necessary monitoring equipment and certification processes, distributing these expenses throughout their membership. Standardized documentation, managed by FPOs, supports carbon credit verification and simplifies transactions with carbon credit buyers. Furthermore, FPOs can ensure equitable benefit sharing among members while minimizing the transaction costs and technical challenges that often exclude individual smallholders from carbon markets.

FPOs can deliver targeted training programs on carbon-efficient agricultural practices through their organizational capacity, offer shared access to specialized equipment such as laser land levelers, and coordinate the adoption of standardized carbon measurement protocols. These collective measures transform barriers that would be prohibitive for individual farmers into manageable challenges.

In summary, by leveraging collective action, FPOs can help small and marginal farmers access and benefit from carbon farming and carbon credits. 

\section{Artificial Intelligence, Game Theory, and Computation in Carbon Farming}
The success of carbon farming depends critically on the ability to accurately measure carbon outcomes, design effective farm-level interventions, ensure high levels of farmer participation, and enable efficient carbon markets. Artificial Intelligence techniques as well as game theory and mechanism design \cite{NARAHARI2014} will have an important role to play across all these dimensions. Rather than acting as a single solution, AI and game theory will function as a connective layer that integrates data, models, inferences, incentives, and strategic decision-making in carbon farming systems. Besides AI and game theory, there are many computation-oriented issues in the design and practice of carbon farming. The above disciplines do not replace agronomy, soil science, or farmer knowledge. Rather, these disciplines play an augmenting role by enabling scale, integration, and adaptability. 

\subsection{AI for Measurement, Reporting, and Verification} 
We have already seen that one of the most significant barriers to scaling carbon farming is the cost and complexity of MRV.
Traditional soil sampling is expensive.
AI-based techniques are now transforming the digital MRV (dMRV) ecosystem by combining satellite imagery, machine learning, and blockchains for cost-effective measurement and verification at scale.
We now provide several examples of how AI techniques are being used in dMRV. We confine our discussion to startups in India. These examples are by no means exhaustive; however, they provide a glimpse of representative efforts.


\subsubsection*{Methane Reduction in Rice Farming}
Mittilabs \cite{MittiLabs} has created a specialized platform using Synthetic Aperture Radar (SAR) and AI to target rice methane emissions. This detects water levels, soil moisture, and seeding methods using multi-spectral satellite data validated against gas-chamber measurements. This also tracks climate-smart practices (Alternate Wetting and Drying, Direct Seeded Rice) to verify 40--50\% methane reductions and 30\% water savings in real-time.

\subsubsection*{Smallholder-Focused Digital MRV}
Varaha \cite{Varaha} has a comprehensive platform for South Asian smallholder farmers that combines remote sensing and carbon modeling. This platform monitors regenerative agriculture, agro-forestry, and biochar across hundreds of thousands of hectares. The platform aggregates fragmented landholdings into unified projects, maintaining field-level precision while generating Verra VM0042-compliant credits. This significantly reduces measurement costs, relative to traditional soil sampling.

\subsubsection*{Satellite-Based Soil Carbon AI}
Boomitra \cite{Boomitra} has developed an AI system that eliminates the need for extensive soil sampling by combining machine learning with more than 1 million georeferenced soil samples to deliver pixel-level carbon measurements from Sentinel-2, Landsat, and ALOS-2 radar data. This is expected to reduce MRV costs by 90\% while achieving Verra VM0042 accuracy. This system is currently supporting more than 100,000 farmers on more than 5 million acres with annual registry-grade measurements.

\subsubsection*{Blockchain-Enabled Verification}
TraceX has created a comprehensive dMRV platform \cite{TraceX} that automates data collection from satellite imagery, IoT sensors, and mobile applications. A blockchain creates immutable transaction records, preventing double-counting and fraud. The system Integrates geofencing and biogeochemical modeling for GHG tracking and remote verification, streamlining baseline-to-issuance workflows.

\subsubsection*{Scalable MRV for Smallholder Farmers}
GrowIndigo \cite{GrowIndigo} has developed a platform that captures real-time methane and nitrous oxide data through satellite observations integrated with biogeochemical models. The tool uses statistical sampling and farm grouping to reduce costs while maintaining rigor. The tool also provides farmer enrollment, practice tracking, and technical guidance tailored to India's smallholder landscape.

\subsection{AI for Precision Carbon-Positive Farm Management}
Beyond measurement, AI can play a central role in designing and optimizing farm practices that improve carbon sequestration. Carbon outcomes depend on complex interactions among soil type, climate, crops, management practices, and socio-economic constraints. AI models can integrate these variables to provide site-specific recommendations.
We present several ideas here that could potentially be explored. Predictive models can support decisions on crop rotations, cover cropping strategies, residue management, reduced tillage, and the use of organic amendments. Optimization and reinforcement learning methods can explore trade-offs between yield, cost, emissions, and carbon gains, enabling adaptive strategies rather than static prescriptions. For example, AI systems can recommend when to introduce cover crops that maximize biomass input without increasing water stress or risk of yield.

\subsection{Sensor Data, IoT, and Edge AI}
Carbon farming can benefit immensely from data collected from soil sensors, weather stations, flux chambers, and farm machinery. AI enables the fusion of these heterogeneous data sources with remote sensing products to enhance temporal resolution and facilitate the early detection of anomalies \cite{ElJarroudi2024}.
Edge AI, which involves models that can be deployed directly on sensors or local devices, allows real-time monitoring of soil moisture, temperature, and emissions, even in rural environments with limited connectivity. This capability is particularly important for locations dominated by smallholder farmers, where centralized infrastructure may be weak.

\subsection{AI-Based Advisories to Farmers}
Conversational agents driven by AI can translate complex carbon protocols, MRV outputs, and agronomic recommendations into local languages and culturally appropriate guidance. These systems can explain why certain practices improve carbon outcomes, identify trade-offs, and describe how payments are computed. When grounded in curated, locally validated knowledge bases, such systems act as scalable extension services rather than generic chatbots.
In carbon farming, where concepts like additionality, permanence, and uncertainty are non-trivial, explainable, and dialog-based AI becomes especially valuable.

AI-based advisories can transform decision-making in carbon farming by providing highly personalized, timely, and trustworthy guidance. Generative AI could be used to design tools and apps that can converse with farmers in a friendly, natural manner, translating complex agronomic science into simple, actionable advice. Retrieval-Augmented Generation (RAG) can be used to ground these responses in verified sources such as local weather data, soil health cards, crop calendars, and government advisories and schemes. Explainable AI techniques help farmers understand why a recommendation is made by building trust by linking advice to observable factors such as rainfall forecasts or pest pressure. Multimodal AI integrates satellite imagery, drone photos, sensor data, and farmer-uploaded images to diagnose crop stress or disease. Multilingual and voice-based interfaces ensure inclusion, allowing smallholder farmers to access advisories in local languages and dialects, thus improving adoption and impact in the field. There are already several ongoing efforts in this direction for the broad area of agriculture, for example, see the Master's Thesis \cite{MADHURA2025} and the references therein.

\subsection{Incentive Design for Farmers for Participation in Carbon Farming}
Effective incentive design is crucial for motivating smallholder farmers to participate in carbon farming. Carbon revenues are uncertain, delayed, and often modest. Therefore, proactive and risk-reducing incentives are important. Incentives include input subsidies, transition payments during the first two to three years, and assured minimum payments independent of carbon price volatility. Aggregation-based incentives, through FPOs or farmer cooperatives, are an important issue to be looked at, because of lower transaction and MRV costs and improved bargaining power. Incentives should be based on outcomes as well as best practices: farmers are rewarded for measured soil carbon gains as well as adopting verified climate-positive practices (such as reduced tillage, agro-forestry, etc.). Incentives could also be non-monetary, such as priority access to credit, crop insurance discounts, extension support, and access to personalized digital advisory services. Transparent benefit-sharing rules, simple contracts, and timely payouts are essential to build trust, without which smallholder farmers are unlikely to engage in carbon farming projects in the long term. Design of incentives warrants utmost attention and can benefit from data-driven, AI-based methods and other analytical techniques. Mechanism design \cite{NARAHARI2014} offers several scientific techniques for designing incentives in such a way that all stakeholders also behave honestly in terms of implementation and reporting.

\subsection{Cooperative Game Approach to Carbon Farming}
As already stated, carbon farming can attract small farmers through well-designed incentives. Farmer cooperatives and FPOs are critical enablers: they aggregate land and farmers, lower MRV and transaction costs, share risks, build capacity, and negotiate fair carbon prices. Acting as trusted intermediaries, they simplify participation, ensure transparent benefit sharing, and make carbon farming economically viable, scalable, and sustainable for smallholders.
Cooperative game theory \cite{NARAHARI2014} offers a variety of solution concepts (such as Core, Shapley Value, Nucleolus, and Gately Point) that can be used to design fair and robust surplus sharing arrangements among participating farmers. The theory can also be used to determine an optimal size for a group of farmers to come together to maximize their utilities through carbon farming projects. 

\subsection{Mechanism Design for Carbon Markets}
Carbon farming is inseparable from carbon markets, whether compliance-based or voluntary. AI contributes to market integrity and efficiency in several ways. Machine learning models help detect anomalous credit claims, flag potential double-counting, and identify spatial or temporal inconsistencies in reported sequestration.

Mechanism design \cite{NARAHARI2014} is the art of designing a protocol of interactions among strategic players to induce honest participation, maximize collective welfare, and ensure attractive revenues for players. Mechanism design is the science behind market design and is equally applicable to the design of carbon markets, especially voluntary carbon markets. It would be interesting to apply mechanism design to the design of carbon markets, aiming to maximize the welfare of participating farmers and optimize the process. 

\subsection{Simulation for Carbon Farming}
Simulation can play a central role in carbon farming by enabling the systematic exploration of soil carbon dynamics, greenhouse-gas fluxes, and management interventions over long time horizons. Process-based models, such as DSSAT \cite{Jones2003DSSAT} and APSIM \cite{McCown1996APSIM}, simulate interactions among climate, soil properties, crop growth, and management practices, allowing researchers to evaluate sequestration trajectories, emissions trade-offs, and sensitivity to external shocks. Simulations support baseline construction, scenario analysis, and a priori assessment of mitigation potential, particularly where direct measurement is infeasible or expensive. By explicitly representing biophysical processes, simulation provides a transparent and interpretable foundation for carbon accounting, policy design, and risk assessment, complementing empirical measurement and monitoring approaches. Digital twins can extend simulation by creating continuously updated, data-linked virtual representations of farms or landscapes. In carbon farming, digital twins integrate process models with observational data streams such as soil measurements, management records, and remote sensing, to track system evolution in near real time. They enable dynamic assessment of carbon sequestration, detection of deviations from expected trajectories, and evaluation of management changes before physical implementation. Digital twins would be particularly valuable for MRV. 
\subsection{Blockchain Technology for Carbon Farming}
Trust, traceability, and accountability are crucial requirements in carbon farming, and blockchain technology \cite{SHIVIKA2018} will be the perfect technology to ensure these requirements. Carbon credits depend on verifiable records of land practices, measurements, audits, and time-bound commitments. Blockchain-based append-only ledgers will be able to provide tamper-evident storage of carbon claims, measurement reports, and verification events across all stakeholders. Smart contracts in blockchains can encode rules for credit issuance, expiry, and reversal in case of non-permanence. d off-chain/on-chain consistency. A growing number of agritech startups are deploying blockchain technologies to streamline the verification of carbon credits \cite{TraceX}. Informed use of blockchain technology will strengthen confidence in carbon accounting without replacing scientific measurement or regulatory oversight.

\section{Conclusion and Future Directions}
In this paper, we have seen that carbon farming represents the most promising pathway to simultaneously address climate change mitigation, soil degradation, and farmer livelihoods. However, it is also a complex pathway involving numerous tradeoffs and practical challenges. This survey has tried to provide a comprehensive and interdisciplinary synthesis of carbon farming, integrating agronomic best practices, soil organic carbon dynamics, measurement–reporting–verification frameworks, carbon markets, real-world case studies, India-specific issues, and scientific challenges, within a single coherent narrative. By positioning carbon farming alongside sustainable, regenerative, and organic agriculture, the paper clarifies both conceptual overlaps and the distinctive role of quantification and monetization that defines carbon farming.

Through an extensive review of generic and crop-specific practices, the paper has highlighted the significant mitigation potential of agricultural systems when soil carbon sequestration and greenhouse-gas reductions are jointly considered. The detailed analysis of MRV frameworks underscores that credibility, uncertainty management, and cost-effectiveness are central bottlenecks, especially when smallholder farmers have to reap the benefits of carbon farming. We have also examined carbon markets and case studies, and this examination shows that institutional design, aggregation models, and trust are as important as biophysical potential.

Looking ahead, there are many opportunities for future research for improving the scientific robustness and scalability of carbon farming: (1) Long-term validation of SOC permanence (2) Enhancing MRV methodologies (3) Reduction of transaction costs in Carbon Farming (4) Integration of carbon farming with broader water, biodiversity, and livelihood outcomes, (5) Scaling Carbon farming to cover smallholder farmers. We believe some of the directions we have outlined in the previous section on AI, game theory, and computation will have an important role to play in undertaking this research.
\section*{Acknowledgements}
We thank Professor N. Viswanadham and Mr. Krishnaiah Kunkala for many insightful discussions. We thank Dr. Soma Dhavala for numerous helpful comments and the reviewers for their insightful suggestions.
In the process of preparing this document, the authors have utilized various AI language models and assistants, such as GPT5 and Claude, to enhance the writing quality, provide summaries of source materials, gain additional insights and assist in the conceptualization and drafting of specific diagrams. However, the authors diligently verified all information provided by these AI tools to ensure accuracy and factual integrity. The authors thank Sai Vinay Sathvik for helping with the diagrams.

\section*{Conflict of Interest}
On behalf of all authors, the corresponding author states that there is no conflict of interest.

\bibliographystyle{unsrt}  
\bibliography{references}  
\newpage
\section*{Glossary}
\begin{table*}[htbp]
\centering
\caption{Carbon Markets and Policy Concepts}
\label{tab:carbon_markets}
\renewcommand{\arraystretch}{1.2}
\begin{tabular}{p{4cm} p{11cm}}
\toprule
\textbf{Term} & \textbf{Description} \\
\midrule

Additionality & The principle that emissions reductions would not have occurred without carbon credit incentives. \\

Additionality Criterion & Rules and methods used to assess whether a project satisfies additionality. \\

Carbon Credit & A tradable certificate representing one metric ton of CO\textsubscript{2} equivalent emissions. \\

Carbon Market & A platform where carbon credits are traded under regulatory or voluntary systems. \\

Carbon Trading & The mechanism of buying and selling emission permits for cost-effective mitigation. \\

Voluntary Carbon Market & A system where entities voluntarily offset emissions through credit purchases. \\

Compliance Carbon Market & A regulated market where entities must meet emission limits or buy allowances. \\

Carbon Offset & Emission reductions used to compensate for emissions elsewhere. \\

Carbon Registry & A system that tracks issuance and retirement of carbon credits to prevent double counting. \\

MRV & Monitoring, Reporting, and Verification frameworks ensuring transparency in carbon accounting. \\

Carbon Leakage & Shift of emissions-intensive activities to regions with weaker regulations. \\

Permanence & The durability of carbon storage over time. \\

Carbon Footprint & Total greenhouse gas emissions caused directly and indirectly by an entity. \\

\bottomrule
\end{tabular}
\end{table*}

\begin{table*}[htbp]
\centering
\caption{Agricultural Practices and Sustainability Concepts}
\label{tab:agriculture}
\renewcommand{\arraystretch}{1.2}
\begin{tabular}{p{4cm} p{11cm}}
\toprule
\textbf{Term} & \textbf{Description} \\
\midrule

Agroforestry & Integration of trees with crops to enhance biodiversity and carbon storage. \\

Carbon Farming & Practices aimed at increasing carbon storage in soil and vegetation. \\

Cover Crops & Plants grown to protect and improve soil health rather than for harvest. \\

Minimal Tillage & Reduced soil disturbance to preserve soil structure and carbon. \\

No-Till Farming & Planting crops without plowing to minimize soil disruption. \\

Laser Land Leveling & Precision leveling technique improving water efficiency and reducing emissions. \\

Sustainable Agriculture & Farming practices that protect resources and ensure long-term productivity. \\

Sustainable Farming & Practices maintaining productivity while conserving environmental resources. \\

Regenerative Agriculture & Farming focused on restoring soil health and ecosystems. \\

Organic Farming & Agriculture that avoids synthetic inputs and relies on natural processes. \\

Soil Health & The ability of soil to function as a living ecosystem supporting life. \\

Soil Degradation & Decline in soil quality due to physical, chemical, or biological factors. \\

\bottomrule
\end{tabular}
\end{table*}

\begin{table*}[htbp]
\centering
\caption{Carbon Science and Environmental Concepts}
\label{tab:carbon_science}
\renewcommand{\arraystretch}{1.2}
\begin{tabular}{p{4cm} p{11cm}}
\toprule
\textbf{Term} & \textbf{Description} \\
\midrule

Biomass & Organic material from plants and animals used as energy or fuel. \\

Biochar & Carbon-rich material from pyrolysis used for soil enhancement and carbon storage. \\

Methanogenesis & Microbial production of methane under anaerobic conditions. \\

Soil Organic Carbon & Carbon stored in soil, essential for fertility and climate mitigation. \\

Soil Carbon Stock & Total carbon stored in soil, used in carbon accounting. \\

Carbon Sequestration & Capture and long-term storage of atmospheric CO\textsubscript{2}. \\

Carbon Sink & Reservoir that absorbs more carbon than it emits. \\

Carbon Cycle & Natural circulation of carbon across Earth's systems. \\

Greenhouse Gases & Heat-trapping gases such as CO\textsubscript{2}, CH\textsubscript{4}, and N\textsubscript{2}O. \\

\bottomrule
\end{tabular}
\end{table*}

\begin{table*}[htbp]
\centering
\caption{Global Sustainability Frameworks}
\label{tab:frameworks}
\renewcommand{\arraystretch}{1.2}
\begin{tabular}{p{4cm} p{11cm}}
\toprule
\textbf{Term} & \textbf{Description} \\
\midrule

UN Sustainable Development Goals (SDGs) & A set of 17 global goals for sustainable development set by the United Nations. \\

\bottomrule
\end{tabular}
\end{table*}

\newpage
\section*{Appendix}

\appendix

\section{Approaches to Carbon Sequestration}
\label{app:carbon_sequestration}

Carbon sequestration takes place through different approaches, which are described below

\subsubsection*{Biological Sequestration}
This is the process of capturing and storing atmospheric carbon dioxide in living systems. Driven primarily by photosynthesis, it involves plants and microorganisms drawing down CO$_2$ and incorporating it into biomass and then into a complex matrix of organic matter of the soil. Through photosynthesis, plants convert atmospheric CO$_2$ into organic compounds, which are subsequently incorporated into the soil. Biological sequestration provides multiple cobenefits, including improved soil health, enhanced biodiversity, and increased agricultural productivity, and is particularly well-suited for agricultural and forestry practices. 

    \subsubsection*{Ocean Sequestration} 
    The Earth's oceans constitute the planet's largest and most dynamic carbon reservoir, naturally absorbing a significant portion of anthropogenic (resulting from human activities) CO$_2$. Ocean sequestration strategies seek to leverage and accelerate this capacity, either by enhancing natural pathways, such as the biological pump (carbon getting transported from surface to deep ocean), or enhanced techniques like ocean fertilization. Therefore, any pursuit of ocean sequestration demands extreme caution, rigorous research, and robust governance to prevent catastrophic harm to this vital marine ecosystem. 

    \subsubsection*{Terrestrial Sequestration} 
    This focuses on improving carbon storage in terrestrial ecosystems, including forests, grasslands, and wetlands. This approach involves the management of landscapes to maximize carbon accumulation while maintaining vital ecosystem services. The effectiveness of terrestrial sequestration depends on land management practices, climate conditions, and the specific ecosystem, making it a versatile but complex strategy for carbon management. 

    \subsubsection*{Geological Sequestration} 
    Geological sequestration involves the injection of carbon dioxide into deep underground geological formations, such as depleted oil and gas reservoirs, saline aquifers, or unmineable coal seams. This method offers substantial long-term storage capacity and stability, but requires significant infrastructure and precise site selection to prevent leakage. Geological sequestration is primarily geared toward mitigating large-scale industrial emissions and requires advanced monitoring systems for storage verification.

    \subsubsection*{Technological Sequestration} 
    This employs engineered solutions, such as direct air capture (DAC) systems, to actively remove CO$_2$ from the atmosphere. Other technical durable carbon sequestration methods, such as converting biomass carbon into stable forms (e.g., biochar) or storing biomass in anoxic underground conditions, lock carbon into the soil for centuries. The use of biochar in agricultural land increases soil organic carbon (SOC) stocks and improves soil health, while stored biomass helps prevent carbon reemissions. Although technological sequestration offers greater control and precision over the capture process, its implementation is often energy-intensive and is presently limited to small-scale operations. The ongoing research aims to improve efficiency and reduce the costs associated with these technologies.

Among all the sequestration modes, this paper focuses on terrestrial sequestration and explores ways to monetize it through carbon trading.

\section{The EU ETS Market}

The EU ETS was introduced in 2005 in response to the 1997 Kyoto Protocol, creating the first large-scale carbon market. Based on a cap-and-trade framework, the system allocates emission allowances to companies for free or through auctions, thus establishing a market-driven incentive for decarbonization. Since its inception in 2005, the EU ETS has operated as the world’s most extensive cap-and-trade system, regulating emissions across major industrial sectors. Over time, successive regulatory reforms and shifting economic contexts have significantly influenced the price dynamics of carbon credits.

The report \cite{EUETS} analyzes the historical trajectory of carbon prices within the EU ETS, identifies periods of price volatility, and assesses the potential of carbon credits as investment assets. We present some details of the following questions: 

\begin{itemize} 
\item How have carbon credit prices evolved? 
\item What are the key periods of price instability and the underlying causes?
\item What future risks may result from policy credibility, economic crises, and investor sentiment, and how could these factors contribute to price shocks? 
\end{itemize}

\subsection{Evolution of Carbon Credit Prices}

\begin{figure}[ht]
	\centering
	\includegraphics[width=0.9\textwidth]{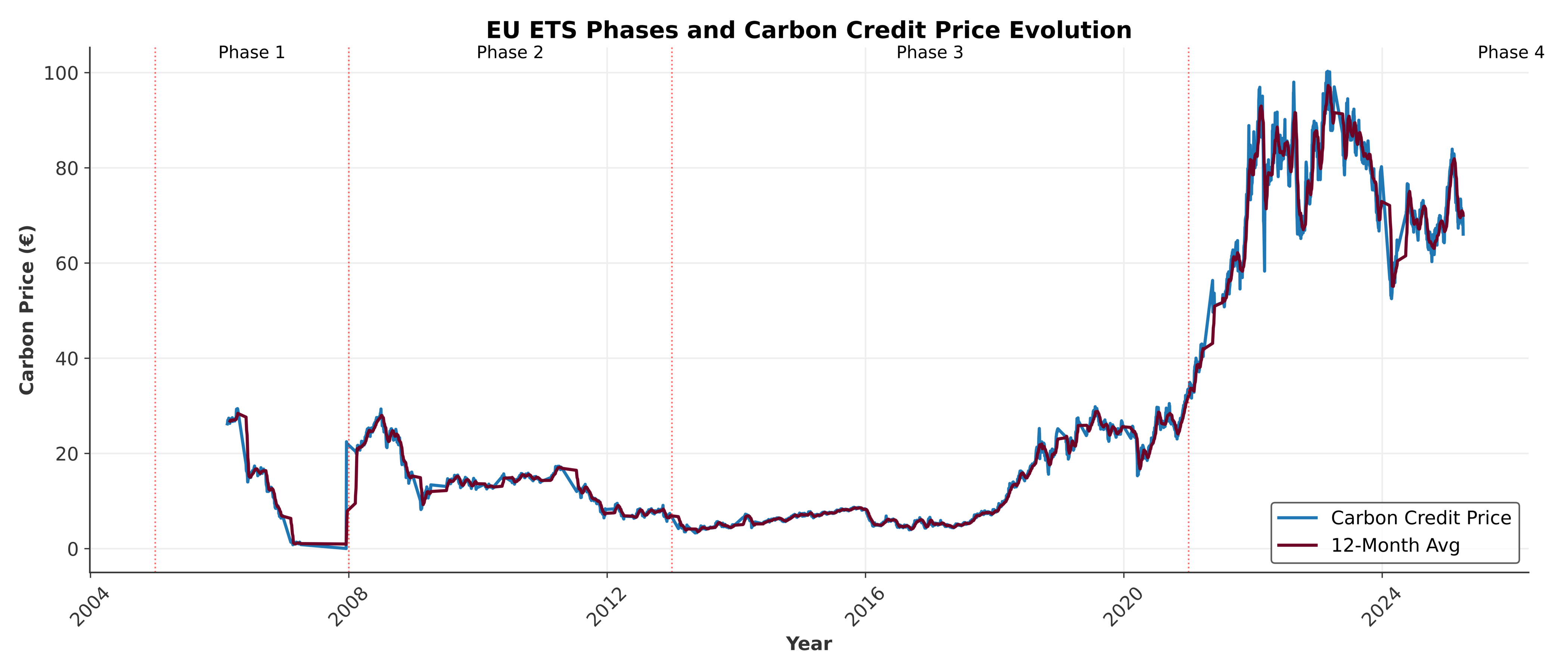}	\caption{EU ETS phases and their correlation with carbon credit price movements over time}
	\label{fig:euets_phases}
\end{figure}

Carbon prices under the EU ETS have progressed through four distinct phases as shown in Table~\ref{tab:eu_ets_phases}. In the initial 2005–2007 pilot phase, free allocation of allowances and the absence of banking provisions led to a severe oversupply, causing prices to collapse. Between 2008 and 2012, stricter emission caps and the inclusion of aviation briefly supported prices, but the global financial crisis triggered another decline. From 2013 to 2020, the shift toward auctioning allowances and the creation of the Market Stability Reserve (MSR) restored confidence and drove gradual price recovery. Since 2021, prices have surged sharply, reflecting the ambitious targets of the EU Green Deal, heightened investor participation, and geopolitical disruptions in energy markets (Fig.~\ref{fig:euets_phases}).

\subsection{Price Trends and Volatility Analysis}
We conducted stationarity tests and rolling volatility analysis to distinguish between policy-driven trends and shock-driven instability. Both the Augmented Dickey-Fuller (ADF) \cite{DickeyFuller1979} and KPSS tests \cite{Kwiatkowski1992} indicate that carbon prices are non-stationary.

Rolling standard deviation analysis (Fig.~\ref{fig:volatility_analysis}) highlights that volatility is often imported from external markets:
\begin{enumerate}
    \item \textbf{2008--2012:} Volatility driven by demand destruction from the Global Financial Crisis (macroeconomic shock).
    \item \textbf{2013--2020:} Relative stability as the MSR absorbed oversupply, decoupling the price somewhat from immediate economic shocks.
    \item \textbf{2021--Present:} Renewed volatility driven by the "financialization" of the market (entry of speculative investment funds) and extreme turbulence in global gas markets following the Russia-Ukraine conflict.
\end{enumerate}
\begin{table}
    \centering
    \small
    \setlength{\tabcolsep}{5pt}
    \renewcommand{\arraystretch}{1.3}
    \caption{Key phases of the EU ETS and their impact on Carbon prices}
    \label{tab:eu_ets_phases}
    \begin{tabular}{@{}p{1.5cm}p{4cm}p{5cm}@{}}
        \toprule
        \textbf{Years} & \textbf{Key Features} & \textbf{Price Impact} \\
        \midrule
        2005--2007 & 
        Pilot phase with free allocation of allowances and no banking between phases. & 
        Oversupply resulted in a collapse of carbon prices. \\
        
        2008--2012 & 
        Stricter emission caps were introduced, including in the aviation sector, and the global economic crisis had a major impact. & 
        The 2008 financial crisis caused oversupply, leading to a sharp decline in prices. \\
        
        2013--2020 & 
        Shift towards auctioning of allowances and the introduction of Market Stability Reserve (MSR) mechanisms. & 
        Improved price stability and a gradual recovery in carbon prices. \\
        
        2021--2030 & 
        Raised emissions reduction target (55\%) aligned with the EU Green Deal. & 
        Significant price surges due to stronger climate ambition and broader market participation. \\
        \bottomrule
    \end{tabular}
\end{table}

\subsection{Determinants of Carbon Credit Prices}

Carbon credit prices are shaped by three primary drivers. First, regulatory reforms, notably the move to auction-based allocation and the MSR, have bolstered confidence and stabilized prices. Second, macroeconomic shocks, such as the 2008 financial crisis \cite{McKibbin2010} and the 2021 energy crisis \cite{Ozili2022}, have introduced significant volatility. Finally, ambitious climate policy goals, including the EU Green Deal and stricter emissions targets \cite{EuropeanCommission2021}, have elevated carbon credits as valuable investment assets. Future pricing trends will hinge on the EU’s ability to balance policy ambition with market stability.

\subsection{Price Trends and Volatility Analysis}
\begin{figure}[ht]
    \centering    
    \includegraphics[width=0.9\linewidth]{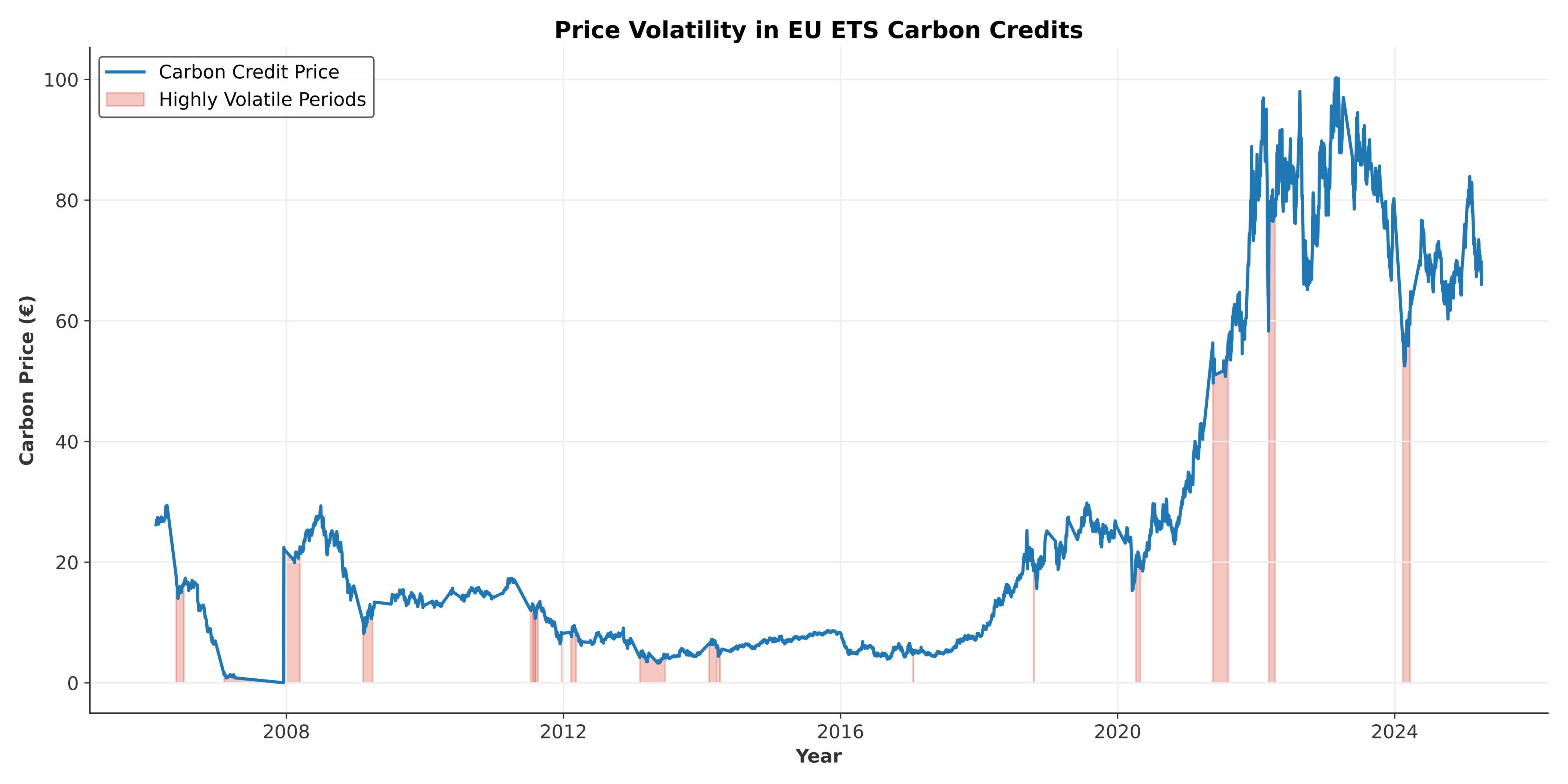}
    \caption{Price instability in the EU ETS market, highlighting periods of non-stationarity}
    \label{fig:volatility_analysis}
\end{figure}

We conducted stationarity tests and rolling volatility analysis to understand market dynamics. Both the Augmented Dickey-Fuller (ADF) \cite{DickeyFuller1979} and KPSS tests \cite{Kwiatkowski1992} indicate that carbon prices are non-stationary and heavily influenced by external shocks. Rolling standard deviation analysis (Fig.~\ref{fig:volatility_analysis}) highlights three distinct volatility periods: extreme fluctuations from 2008 to 2012 due to oversupply and the financial crisis, relative stability between 2013 and 2020 following regulatory reforms, and renewed volatility from 2021 onward, driven by geopolitical uncertainty, energy crises, and ambitious policy measures.

Our conclusions draw on the alignment between the historical price series and these documented events in the literature, and thus follow a descriptive rather than causal-econometric interpretation.

\subsection{Learnings from the EU ETS Market}

Analysis of the EU ETS carbon price is therefore critical for farmers and agribusinesses, since allowance price signals in the energy and industrial sectors indirectly drive input costs (e.g., electricity, fertilizer, feed), and therefore farm-level production decisions.  
In addition, understanding the dynamics of the allowance market informs the design of future policy instruments that ensure consistent mitigation strategies across all sectors.
The lessons of the EU ETS also provide a blueprint for the development of robust monitoring, reporting, and verification (MRV) systems tailored to agriculture’s emissions, an essential prerequisite for any future market-based inclusion \cite{Verschuuren2024}. Finally, the EU ETS experience with price volatility and stability mechanisms offers valuable insights for crafting resilience measures in agricultural markets, thereby supporting long-term investment in low-carbon and sustainable farming practices. The EU ETS has successfully demonstrated the power of market-based decarbonization but remains vulnerable to policy uncertainty, economic downturns, and investor sentiment. Strong regulatory commitments will support high carbon prices, but market shocks or political backlash over energy costs could undermine price integrity. 

Although agriculture is not currently included in the EU ETS, it accounts for roughly 12--13\% of EU greenhouse gas emissions \cite{EEA2024}, and price signals from the system indirectly shape farm-level production costs through energy and input prices. Understanding ETS price dynamics is therefore crucial for designing future policies that integrate agriculture into carbon markets, develop robust MRV systems \cite{Verschuuren2024}, and implement resilience mechanisms to encourage investment in sustainable farming.

\section{Key Carbon Projects in India}
\label{app:key_carbon_projects}
An analysis of India’s carbon credit projects as of December 2024 reveals different types of projects, sectoral distribution, and market performance patterns. The aggregate data, compiled by the Berkeley Policy Carbon Trading Project~\cite{HAYA24}, is drawn from all carbon offset projects listed worldwide by the four major voluntary offset project registries, namely, the American Carbon Registry (ACR), Climate Action Reserve (CAR), Gold Standard, and Verified carbon Standard (VCS) of Verra, with records updated until December 31, 2024.

India hosts 1,720 of the 9,921 registered carbon projects worldwide, making it one of the leading countries in terms of project numbers. It is followed by the United States, which has 1,640 projects, and China, which has 1,509 projects. Among the Indian projects, most are registered under Verra (1,115 of 1,720), while the remaining 605 were listed under the Gold Standard.

\subsection{Sectoral Distribution}

\begin{figure}[ht]
    \centering
    \includegraphics[width=0.9\linewidth]{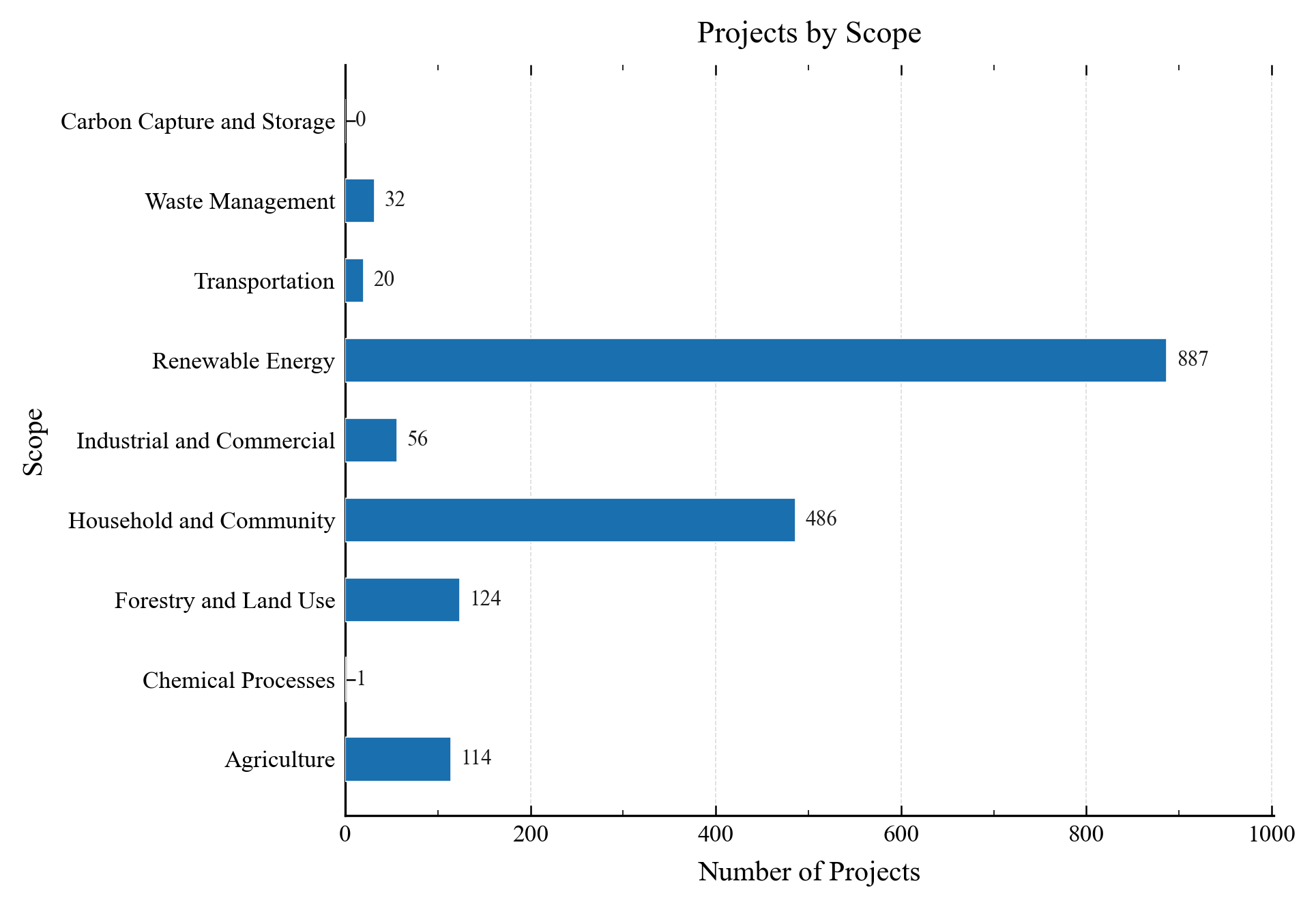}
    \caption{Sector-wise project distribution of carbon credits in India}
    \label{fig:sector-wise-distribution}
\end{figure}

From Fig.~\ref{fig:sector-wise-distribution} we see that renewable energy projects overwhelmingly dominate India’s carbon credit portfolio, representing 88.1\% of all credits issued. In the category of renewable energy, wind energy projects account for 35.9\% of the issuances, while centralized solar projects contribute 26.3\%. Hydropower projects comprise 21.7\%, and the remaining credits come from other renewable sources.

Credits issued in other sectors show a more distributed pattern. Household and community-based projects constitute 6.8\%, while the industrial and commercial sectors represent 3.8\%. Agriculture and forest projects together contribute only 1.0\% of the credits issued.


\begin{table}[ht]
 \caption{Sectoral distribution of issued carbon credits in India (in percentage)}
    \centering
    \begin{tabular}{l r}
        \toprule
        \textbf{Category} & \textbf{Share (\%)} \\
        \midrule
        Agriculture & 0.2 \\
        Chemical Processes & 0.1 \\
        Forestry and Land Use & 0.8 \\
        Household and Community & 6.8 \\
        Industrial and Commercial & 3.8 \\
        Renewable Energy & 88.1 \\
        Transportation & 0.1 \\
        Waste Management & 0.2 \\
        Carbon Capture and Storage & 0.0 \\
        \bottomrule
    \end{tabular}
   
    \label{tab:category_distribution}
\end{table}

\begin{figure}[ht]
    \centering
    \includegraphics[width=0.8\linewidth]{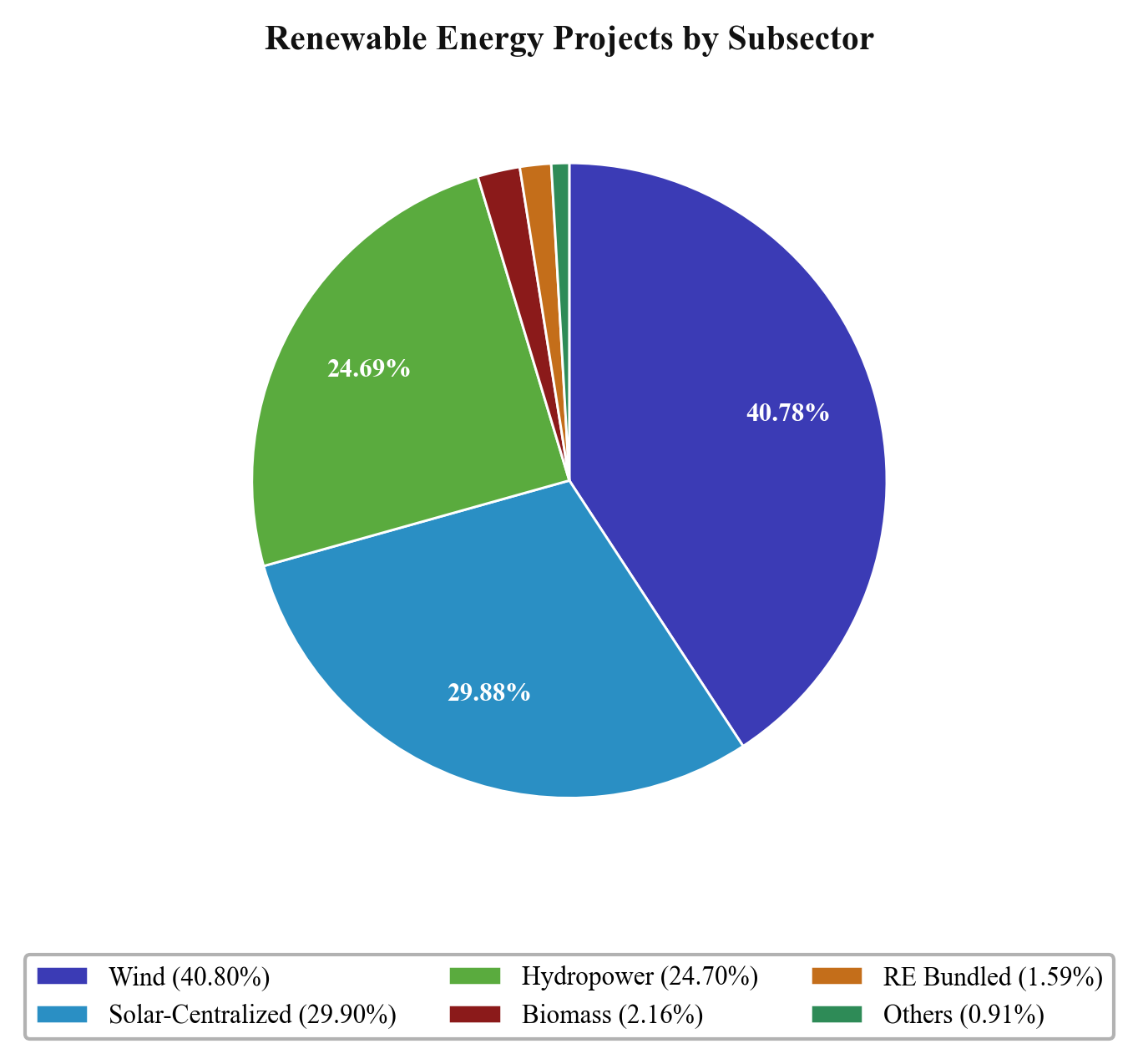}
    \caption{Credits issued across different sub-sectors within renewable energy}
    \label{fig:renewable-energy-split}
\end{figure}

\subsection{Project Efficiency and Approval Rates}

Renewable energy projects comprise 51.5\% of all project submissions (887 projects) but account for 81\% of the total credits issued, indicating a high approval and issuance rate within this sector. In contrast, emerging sectors, such as agriculture, reflect markedly lower success rates, with 114 submissions resulting in only 11 approved projects.

Within the forestry and land use sector, the total credits issued amount to 3.02 million tons of carbon dioxide, with afforestation and reforestation projects contributing 2.46 million tons and other forest initiatives issuing 0.56 million tons. Agricultural projects have issued 0.73 million metric tons, mainly from irrigation management and manure methane digester activities.

\subsection{Market Performance and Trading Efficiency}

The Indian carbon credit market exhibits notable differences between credit issuance and retirement, indicative of underlying market inefficiencies. Between 2008 and 2024, the total issued credits reached 361 million tons of CO$_2$, whereas total retired (traded or consumed) were 207 million tons of CO$_2$. This results in an unsold inventory of 154 million tons CO$_2$, corresponding to 42.7\% of all issued credits.

Even after accounting for a one-year delay in credit trading, approximately 115 million tons of CO$_2$ or 36\% of credits remain unsold. This substantial difference highlights significant liquidity challenges and underscores the need for more robust trading mechanisms and refined market structures to enhance market efficiency and increase credit absorption capacity.

\section{Startup Ecosystem}
\label{app:startups}

There has been a fair amount of startup activity in India in the area of carbon farming and carbon credits. The following is an indicative list of some startups in this space. This list is by no means exhaustive, and the inclusion of a handful of the startups is not to be viewed as any kind of endorsement of these institutions.

\subsubsection*{Boomitra} 

Boomitra \cite{Boomitra} uses satellite and AI techniques to measure, report, and verify soil carbon credits globally. The company has developed a robust data set from more than 1 million soil samples, analyzed in accredited laboratories, to create regional soil models for accurate carbon monitoring. Using satellite imagery, Boomitra can measure soil carbon, plant health, and moisture levels across the globe without additional sampling. Project 'URVARA', a Boomitra project on the issuance of carbon credits to 6000 smallholder farmers in India, received the first issuance of carbon credits in April 2025. The project promotes regenerative practices such as reduced tillage, reincorporation of residues, cover cropping, and improved irrigation.

\subsubsection*{Grow Indigo}
Grow Indigo \cite{GrowIndigo} is an agritech joint venture between Mahyco\cite{Mahyco} and Indigo Ag \cite{IndigoAg}. It develops proprietary microbial and bio-based products, such as seed coatings and soil bio-stimulants, that improve soil health, nutrient uptake, and crop resilience. Through regenerative practices such as direct-seeded rice (DSR) and no-tillage, the company’s carbon farming program allows small farmers to generate voluntary carbon credits. By integrating biological innovations with digital tools and carbon markets, Grow Indigo aims to transform Indian agriculture toward profitability, sustainability, and resilience to climate by 2030.

\subsubsection*{IORA Ecological Solutions}
Iora Ecological Solutions is an environmental advisory firm founded in 2009 with a multidisciplinary team in climate policy, environmental finance, and scientific research. The firm specializes in integrated solutions based on nature-based climate mitigation and resilience, which include the development of REDD (Reducing Emissions from Deforestation and forest degradation in developing countries) methodology, forest and biodiversity conservation, sustainable agriculture, and water resource management across India, Asia, and Africa. Using geospatial analysis and remote sensing, Iora has built decision support tools such as the IFSDM (IEMaC Fuelwood System Dynamics Model) and contributed to India's NDC (National Development Council) roadmap, state climate action plans and the IUCN (International Union of Conservation of Nature) Global Biodiversity Compensation Strategy. Flagship programs include IEMaC (Innovations in Ecosystem Management and Conservation) for gender‑sensitive fuelwood sustainability and the Indian Organic Waste Management Programme (IOWMP) for community biogas generation and methane‑reduction carbon credits.

\subsubsection*{Mitti Labs} 
Mitti Labs \cite{MittiLabs} focuses on sustainable rice farming using technology to reduce methane emissions, water consumption, and stubble burning. By integrating AI with satellite imagery, Mitti Labs addresses specific issues in rice farming, including water management and fertilizer use, to reduce rice GHG emissions by more than 40\%. The company aims for scalable solutions with high-resolution data to improve environmental impact while maintaining productivity.

\subsubsection*{Varaha.earth} 
Varaha.earth \cite{Varaha} has pioneered a digital MRV (Measurement, Reporting, and Verification) platform that combines remote sensing, carbon modeling, and an in-house proprietary carbon quantification tool. The company focuses on critical environmental projects, including regenerative agriculture, afforestation, reforestation, revegetation, and biochar production. With these initiatives, Varaha.earth aims to sequester 1 billion tons of carbon by 2030, demonstrating how technology-driven solutions can accelerate climate action. 
Google has recently signed a deal with Varaha to purchase 100,000 tons of carbon credits from their biochar projects, marking Google's first foray into India's carbon removal sector and one of the largest biochar credit purchases to date. 

\subsubsection*{The/Nudge Institute}

The Nudge Institute \cite{TheNudgeInstitute} is an action institute that works toward a poverty-free India within our lifetime through partnerships with governments, markets, and civil society. It operates flagship programs in rural development, such as the Transforming Agriculture for Small Farmers (TASF) initiative, which conducts field research on natural agricultural impacts, studies of direct-seeded rice (DSR) to reduce methane emissions, and surveys smallholder farmers’ climate-adaptation practices. Under TASF, reports highlight the adoption of crop rotation, organic manure, and water-efficient practices among farmers in six states. The institute also publishes reports on Agri-IKIGAI solutions, outlining 13 scalable, environment- and farmer-friendly practices for sustainable agriculture.“Asha Kiran,” its Economic Inclusion Program, promotes backyard poultry and goatry to build sustainable livelihoods for rural women.

\subsubsection*{TraceX} 
TraceX \cite{TraceX} provides advanced digital measurement, reporting, and verification solutions to address critical challenges in food traceability, sustainability, carbon management, and regulatory compliance, 
including EUDR, CSRD, and CS3D.
The company emphasizes natural solutions, such as agroforestry, which integrates vegetation components such as trees and shrubs into agricultural landscapes. This approach helps sequester atmospheric carbon, mitigating climate change while enhancing soil fertility, biodiversity, and ecosystem health. TraceX empowers enterprises to meet sustainability targets, align with global regulations, and drive meaningful environmental impact by combining technology with sustainable practices. With a focus on digital transformation, TraceX is committed to fostering climate resilience and promoting sustainable development in the agricultural sector.

\subsubsection*{VNV Advisory}
VNV Advisory is an unfunded, privately held advisory firm. It operates as a provider of carbon emission reduction programs and sustainability services. As part of the Value Network Ventures group, a project developer with more than 15 years of experience in carbon finance, VNV Advisory leverages community-centric, nature-based solutions in Asia and East Africa. Its core services span social forest, mangrove restoration, climate-smart agriculture, energy access and clean cooking, waste management, water access, sustainability reporting, and CSR strategy. The firm integrates digital MRV platforms such as TraceX to monitor alternative wetting and drying (AWD) irrigation in paddy cultivation, reducing methane emissions, and facilitating robust carbon credit generation.

\end{document}